\documentclass[twocolumn,preprint,trackchanges]{aastex63}

\usepackage{lineno}
\usepackage{graphicx}
\usepackage{booktabs}
\usepackage{tabularx}
\usepackage{amsmath,amssymb}
\usepackage{xcolor} 
\usepackage{colortbl} 
\usepackage{CJK}
\usepackage{natbib}
\usepackage{multirow}

\hypersetup{linkcolor=cyan,citecolor=magenta,filecolor=yellow,urlcolor=blue}

\hypersetup{linkcolor=magenta, citecolor=cyan, filecolor=yellow, urlcolor=blue}

\received{ddmmyyyy}
\revised{\today}
\accepted{ddmmyyyy}
\published{ddmmyyyy}

\submitjournal{ApJ}

\begin{document}

\title{GRB 220101A: a most energetic $10^{54}$ erg long GRB triggered by two supernovae $3.5$ seconds apart}

\author[0000-0003-0829-8318]{R.~Ruffini}
\affiliation{ICRANet, Piazza della Repubblica 10, 65122 Pescara, Italy}
\affiliation{ICRA, Dip. di Fisica, Sapienza Universit\`a  di Roma, Piazzale Aldo Moro 5, I-00185 Roma, Italy}
\affiliation{Universit\'e de Nice Sophia-Antipolis, Grand Ch\^ateau Parc Valrose, Nice, CEDEX 2, France}
\affiliation{INAF,Viale del Parco Mellini 84, 00136 Rome, Italy}

\author[0000-0001-5717-6523]{Y. Aimuratov}
\affiliation{Fesenkov Astrophysical Institute, Observatory 23, 050020, Almaty, Kazakhstan}
\affiliation{INAF -- Osservatorio Astronomico d'Abruzzo, Via M. Maggini snc, I-64100, Teramo, Italy}

\author{L.~M.~Becerra}
\affiliation{ICRANet, Piazza della Repubblica 10, 65122 Pescara, Italy}
\affiliation{ICRA, Dip. di Fisica, Sapienza Universit\`a  di Roma, Piazzale Aldo Moro 5, I-00185 Roma, Italy}
\affiliation{Centro Multidisciplinario de Física, Vicerrectoría de Investigación, Universidad Mayor, Santiago de Chile 8580745, Chile}

\author[0000-0003-2624-0056]{Chris L., Fryer}
\affiliation{CCS-2, Los Alamos National Laboratory, Los Alamos, NM 87545, USA }

\author[0000-0002-1343-3089]{Liang Li}
\affiliation{ICRANet, Piazza della Repubblica 10, 65122 Pescara, Italy}
\affiliation{ICRA, Dip. di Fisica, Sapienza Universit\`a  di Roma, Piazzale Aldo Moro 5, I-00185 Roma, Italy}
\affiliation{INAF -- Osservatorio Astronomico d'Abruzzo, Via M. Maggini snc, I-64100, Teramo, Italy}

\author{G. J. Mathews}

\affiliation{Department of Physics and Astronomy, Center for Astrophysics, University of Notre Dame, 46556, IN, USA}

\author[0000-0001-9277-3366]{M.~T.~Mirtorabi}
\affiliation{ICRANet, Piazza della Repubblica 10, 65122 Pescara, Italy}
\affiliation{ICRA, Dip. di Fisica, Sapienza Universit\`a  di Roma, Piazzale Aldo Moro 5, I-00185 Roma, Italy}
\affiliation{INAF -- Osservatorio Astronomico d'Abruzzo, Via M. Maggini snc, I-64100, Teramo, Italy}
\affiliation{Department of Fundamental Physics, Faculty of Physics, Alzahra University, Tehran, Iran}

\author[0000-0002-2516-5894]{R.~Moradi}
\affiliation{ICRANet, Piazza della Repubblica 10, 65122 Pescara, Italy}
\affiliation{ICRA, Dip. di Fisica, Sapienza Universit\`a  di Roma, Piazzale Aldo Moro 5, I-00185 Roma, Italy}
\affiliation{INAF -- Osservatorio Astronomico d'Abruzzo, Via M. Maggini snc, I-64100, Teramo, Italy}
\affiliation{Key Laboratory of Particle Astrophysics, Institute of High Energy Physics, Chinese Academy of Sciences, Beijing 100049, People's Republic of China}

\author{F. Rastegarnia}
\affiliation{ICRANet, Piazza della Repubblica 10, 65122 Pescara, Italy}
\affiliation{ICRA, Dip. di Fisica, Sapienza Universit\`a  di Roma, Piazzale Aldo Moro 5, I-00185 Roma, Italy}
\affiliation{INAF -- Osservatorio Astronomico d'Abruzzo, Via M. Maggini snc, I-64100, Teramo, Italy}
\affiliation{Key Laboratory of Particle Astrophysics, Institute of High Energy Physics, Chinese Academy of Sciences, Beijing 100049, People's Republic of China}

\author{J. A. Rueda}
\affiliation{ICRANet, Piazza della Repubblica 10, 65122 Pescara, Italy}
\affiliation{ICRA, Dip. di Fisica, Sapienza Universit\`a  di Roma, Piazzale Aldo Moro 5, I-00185 Roma, Italy}
\affiliation{Dip. di Fisica e Scienze della Terra, Universit\`a degli Studi di Ferrara, Via Saragat 1, I--44122 Ferrara, Italy}

\author{C. Sigismondi}
\affiliation{ICRANet, Piazza della Repubblica 10, 65122 Pescara, Italy}
\affiliation{ICRA, Dip. di Fisica, Sapienza Universit\`a  di Roma, Piazzale Aldo Moro 5, I-00185 Roma, Italy}
\affiliation{INAF -- Osservatorio Astronomico d'Abruzzo, Via M. Maggini snc, I-64100, Teramo, Italy}

\author{S.~S.~Xue}
\affiliation{ICRANet, Piazza della Repubblica 10, 65122 Pescara, Italy}
\affiliation{ICRA, Dip. di Fisica, Sapienza Universit\`a  di Roma, Piazzale Aldo Moro 5, I-00185 Roma, Italy}

\author[0000-0001-7959-3387]{Yu Wang}
\affiliation{ICRANet, Piazza della Repubblica 10, 65122 Pescara, Italy}

\affiliation{INAF -- Osservatorio Astronomico d'Abruzzo, Via M. Maggini snc, I-64100, Teramo, Italy}

\email{ruffini@icra.it, liang.li@icranet.org, yu.wang@icranet.org}

\begin{abstract}

GRB 220101A is a long GRB, with a total energy exceeding $10^{54}$ erg with a redshift $z = 4.61$ and one of the largest ever high-quality multi-wavelength observational coverage, from a large number of space-based and ground-based telescopes. We interpret this source in a doubly Binary driven peta nova (BdP-N) model. The progenitor is composed of a massive CO core of $\sim 10\,M_\odot$, highly magnetized with $B \sim 10^{6}$ G, associated to a neutron star (NS) and a white dwarf (WD) with orbital periods from minutes to hours. The large GRB luminosity is explained by a sequence of 7 episodes:   episode 1 is triggered by a new kind of pair supernova (HB) which originates from the collapse of the strongly magnetized CO core. Accretion of the HB supernova ejecta (the ejecta)  onto the white dwarf companion triggers after 3.5 sec the episode 2: the second supernova emitting neutrinos and creating a new neutron star ($\nu$NS). The ejecta, interacting with the magnetosphere of the binary NS companion originate the episode 3:  the Ultra relativistic Prompt Emission (UPE) emission by far the most energetic episode of this GRB, with  the formation of a powerful  jet normal to the plane of the GRB. Following  the UPE energy loss, the accretion of the ejecta on  the NS companion leads to the episode 4: the formation of a black hole (BH) of $2.3 \  M_\odot$ leading  to the observed GeV afterglow emission. Further accretion of the ejecta spin up the $\nu$NS to a period of $1.3$ ms which gives origin to the episode 5: the birth of a pulsar. The interaction of this milli-second pulsar with the remnants lead to the Episode 6: the synchrotron emission observed in the X-ray, optical and radio,  The episode 7 is a 56.7 ms pulsar, as observed $10^{10}$ s after the first burst in the crab nebula.

\end{abstract}

\keywords{gamma rays: bursts --- gamma-rays: individual (GRB 220101A) --- supernovae: general --- binaries: close}

\section{Introduction} \label{sec:intro}

GRB~220101A provides an unprecedented opportunity to study multiple fundamental processes in a new class of gamma-ray bursts (GRBs) with isotropic energies exceeding $10^{54}$ erg. This source appears to be associated with a new type of supernova (HB supernova), distinct from the traditional core-collapse SNe observed in the BdHN model, typically associated with Type Ic events. In those systems, the isotropic energy is of the order of $10^{52}$ erg. This difference indicates the need to extend the original BdHN model to a much larger energy domain. We propose that a HB supernova triggers these most energetic systems, and an unprecedented process occurs $\sim 7$ seconds after the SN trigger. We give evidence that the SN ejecta trigger the gravitational collapse of a white dwarf (WD), a third component of the progenitor system. In the traditional BdHN model, the progenitor is a binary system composed of a massive $\sim 10 M_\odot$ CO core and a companion neutron star (NS). The orbital period is from minutes to hours. The additional presence of the WD appears essential for these more energetic sources. The induced gravitational collapse of the WD leads to the formation of a new NS ($\nu$NS) with a characteristic spin period of $\sim 1$--$10$~ms and an observed emission of $10^{52}$~erg. Part of the remaining SN ejecta accretes onto the companion NS, generating an emission of $10^{54}$~erg during the UPE phase. As soon as this process is completed, a black hole (BH) from the NS companion collapse, and three afterglows are observed: (1) a GeV emission from the BH, (2) an X-ray emission from synchrotron radiation by the expanding SN remnant, and (3) an optical emission by the $\nu$NS millisecond pulsar. All these unprecedented processes are documented below using an extensive set of observational data from the space observatories AGILE, Swift, and Fermi, as well as a vast set of ground-based optical observations, including Xinglong, CAHA, NOT, and the Liverpool Telescope. This is a unique observational campaign from space and ground. We are going to present specific signatures that characterize a handful of the most energetic GRBs previously observed, now clearly identified thanks to GRB~220101A, including GRB~130427A \citep{2019ApJ...886...82R,2019ApJ...874...39W}, GRB~180720B \citep{2022ApJ...939...62R,2022EPJC...82..778R}, and GRB~240825A (Ruffini et al., in preparation).

The paper is organized as follows. In Section~\ref{sec:obs}, we present the main observational properties of GRB~220101A and briefly review previous articles on this GRB, emphasizing the features that require a different physical picture. Section~\ref{sec:simulation} introduces the traditional BdHN framework together with the previous numerical simulation.In Section~\ref{sec:newmodel} the new additional scenario is presented. In Section~\ref{sec:observation}, we analyze multi-wavelength datasets spanning optical, X-ray, MeV, and GeV from various ground-based telescopes and space satellites. In Section~\ref{sec:unique}, we examine the conditions that make GRB~220101A a favorable and unique case for studying an energetic GRB at high redshift, in particular, the detection of multiple emission episodes over a broad energy range. Section~\ref{sec:episodes} interprets the event within the modified BdHN scenario by identifying its seven episodes and deriving the global properties of the system. Section~\ref{sec:upe-bh-gev} addresses the formation of the BH, and its associated GeV emission, which is used to obtain constraints on the BH mass and spin. In Section~\ref{sec:details}, we examine the role of overcritical electromagnetic fields during the prompt phase. In  Section~\ref{sec:Summary}, \ref{sec:mainresults} and  \ref{sec:perspectives} we summarized the main result of this study and provide new perspective for further investigation. Final conclusion is given in Section \ref{sec:conclusion}.

\section{Observations and Review of Previous Work}
\label{sec:obs}

We were immediately aware of the extraordinary importance of GRB 220101A as soon as the redshift was detected at $z=4.615$, indicating that it was then the most powerful GRB observed. It exemplifies the potential to deepen our understanding of GRBs via the cosmological time-dilation effect enabled by its location in the early universe \citep{2024ApJ...966..219B}. The multiwavelength observational campaign of GRB 220101A, from GeV, MeV to optical and radio, reveals a complex light curve and energy spectrum \citep{2022GCN.31347....1T,2022GCN.31348....1T,2022GCN.31388....1C,2022GCN.31360....1L,2022GCN.31354....1U,2022GCN.31353....1F,2022GCN.31350....1A,2022GCN.31433....1T,2022GCN.31357....1P,2022GCN.31359....1F}. We identified the SN rise predicted by the BdHN model in \citet{2022GCN.31465....1R}. We also identified the X-ray afterglow and, shortly thereafter, the GeV emission. The total isotropic energy released from the keV to the GeV range is $6 \times 10^{54}$ erg, as reported in \citet{2022GCN.31648....1R}. Instead of applying these results within the traditional BdHN model, we incorporate additional observational results obtained in the meantime to improve the understanding of the BdHN model itself, extending it to this new, more energetic family of GRBs with isotropic energy $\sim 10^{54}$ erg.

A vast campaign of observations provided additional data on this source. These observations were interpreted by the authors within a ``traditional'' GRB model, assuming a single Kerr BH as the GRB energy source and the observed emissions arising from a postulated jet launched from the BH. No model for the origin of the BH nor of the jet was advanced.

\begin{figure*}[!t]
\centering
\includegraphics[angle=0, scale=0.8]{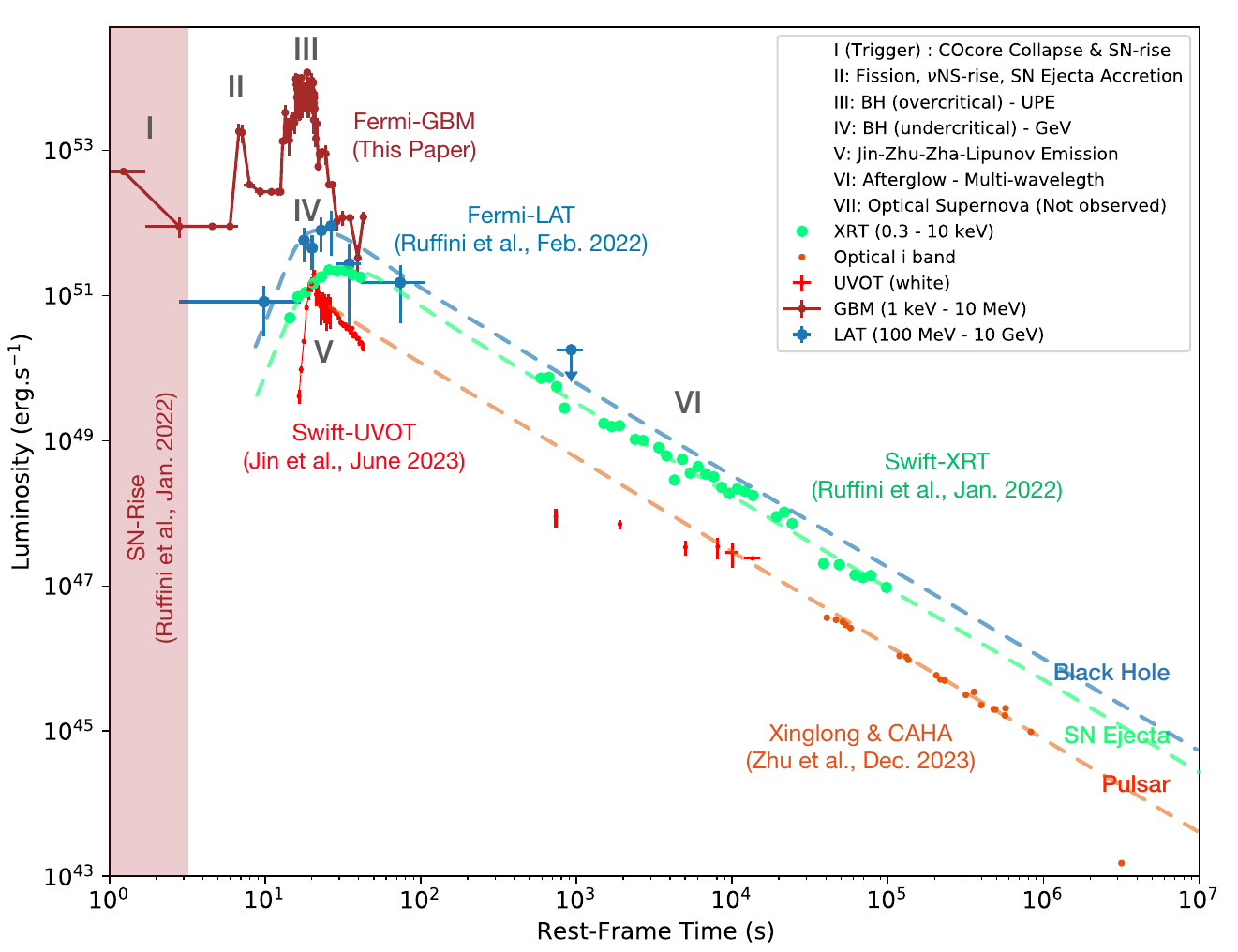}
\begin{picture}(0,0)
    \put(-90,135){\makebox(0,0)[b]{\large \textbf{VII}}}
\end{picture}
\caption{Comprehensive multi-wavelength luminosity light curve of GRB 220101A spanning from trigger time up to more than $10^6$~s in the rest frame. This plot combines data from Fermi-GBM ($1$ keV--$10$ MeV, red points), Fermi-LAT (100 MeV - 10 GeV, blue points), Swift-XRT ($0.3$--$10$ keV, green points), Swift-UVOT (white band, bright red points), and ground-based optical telescopes (Xinglong \& CAHA, i-band, orange points). The $x$-axis is rest-frame time (log scale), and the $y$-axis is isotropic, k-corrected luminosity (log scale). Roman numerals (I-VI) denote the  BdHN Episodes. Dashed lines illustrate the distinct rising and the long-term power-law decay behaviors associated with the three main emission components attributed to the BH (GeV), the SN ejecta (X-ray), and the Pulsar (Optical). Citations indicate initial reports or key analysis papers for different datasets/Episodes. Details are in the entire article.}\label{fig:lc}
\end{figure*}
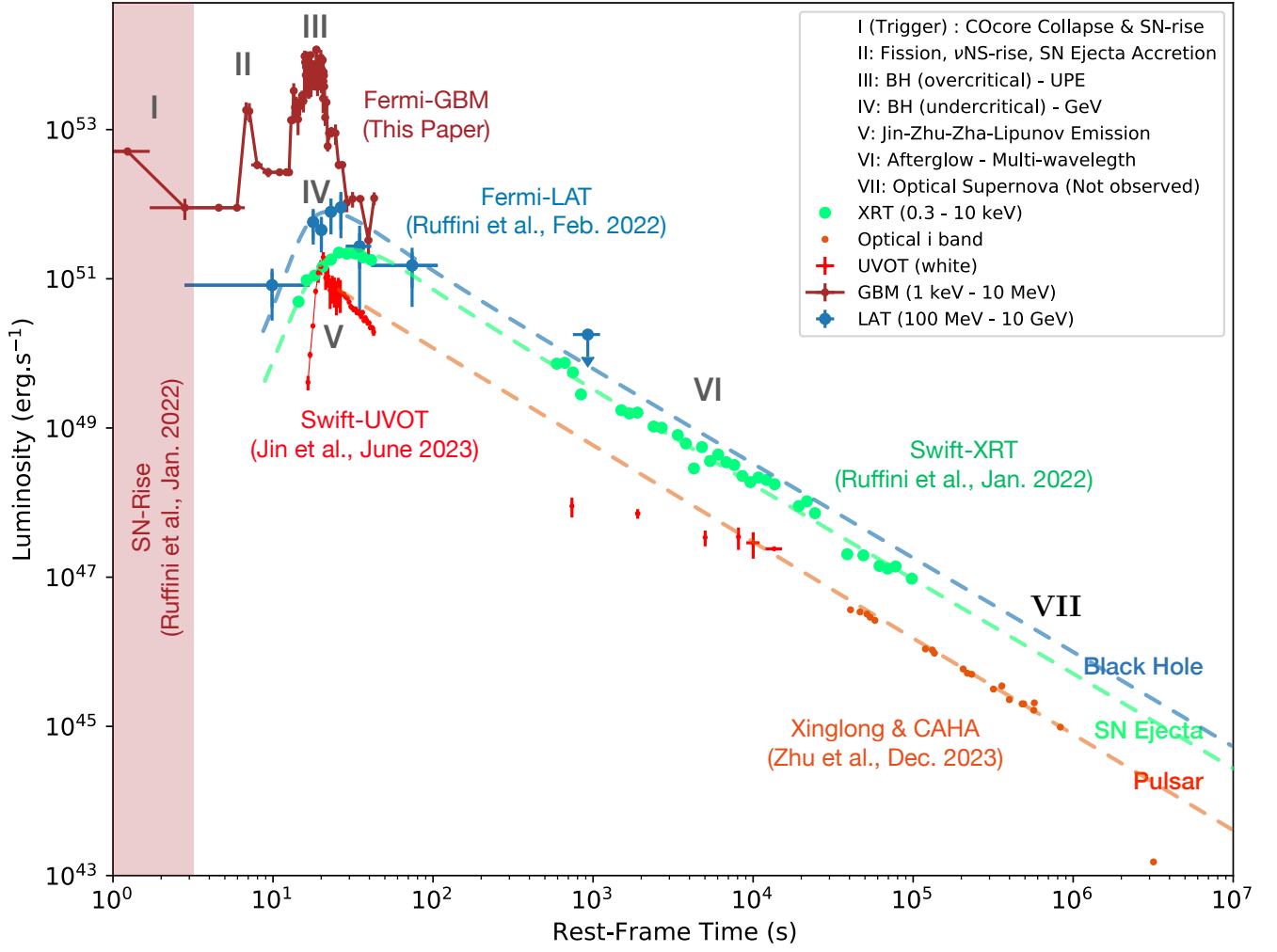

We use these data and interpret them within a modified BDHN model comprising seven episodes, identified from the data analysis presented in this article and summarized in Figure~\ref{fig:lc}. We first recall the relevant observational data acquired in the meantime and their astrophysical significance within the BdHN model.

A fundamental input comes from the optical observations by \citet{2023NatAs...7.1108J}, which revealed an outstanding initial optical emission detected by Swift-UVOT and an exceptionally bright ultraviolet flare within the first $150$~s. This previously unpredicted result was corroborated by ground-based observations from the Xinglong $2.16$~m Telescope \citep{2023ApJ...959..118Z}, as well as XL2.16/BFOSC and NOT/ALFOSC, which provided optical and near-infrared data. In the time interval from $80$ to $120$~s following the prompt emission, these observations show one of the most luminous optical afterglows ever detected. The optical afterglow shows a lower luminosity than the X-ray and GeV components, while following the same power-law dependence in time (see Figure~\ref{fig:lc}). In this article, we use these data to derive the spin-up phase of the $\nu$NS leading to the pulsar emission.

The observations of the JWST for the Crab Nebula \citep{2024ApJ...968L..18T} followed. They evidenced synchrotron emission and allowed us to return to our earlier hypothesis to identify the X-ray afterglow observed by Swift originating from the energy injection by the SN explosion into the SN remnant \citep{2018ApJ...869..101R,2020ApJ...893..148R,2022ApJ...939...62R}. Here we extend to the multi-band synchrotron emission triggered by the new neutron star interacting with the remnant.

The Agile observations from \citet{2022ApJ...933..214U} revealed a unique spectral evolution in the high-energy MeV-GeV region, with notable hardening during the prompt emission. In the traditional model, they inferred a wind-like density medium for the propagating jet, while we use these data to determine the GeV emission from the BH.

Similarly, \citet{2023ApJ...956..101S} analyzed both the Fermi-GBM and Fermi-LAT data, relating them to a single jetted emission within the traditional model. In the BdHN model (see Figure \ref{fig:lc}), we show how the Fermi-GBM and LAT data refer to two different BdHN Episodes: the MeV data from Fermi-GBM refer to the ultra-relativistic prompt emission (UPE) and the GeV data from Fermi-LAT refer to the formation of the BH and its emission (see details in Sections \ref{sec:observation} and \ref{sec:unique}).

\section{Traditional BdHN scenario and simulations}\label{sec:simulation}

In the BdHN model, long GRBs consist of seven episodes. Each episode is characterized by specific energetics and angular properties, and each contributes to the overall interpretation of the burst \citep{1999A&A...350..334R, 2000A&A...359..855R,2012ApJ...758L...7R, 2014ApJ...793L..36F, 2015ApJ...812..100B, 2019ApJ...886...82R, 2020EPJC...80..300R, 2021A&A...649A..75M, CAMPION2021136562,2012A&A...548L...5I, 2012A&A...543A..10I, 2015ApJ...798...10R, 2019ApJ...874...39W, 2021MNRAS.504.5301R,2023ApJ...955...93A}.

\begin{figure*}
  \centering
  \includegraphics[width=0.67\hsize,clip]{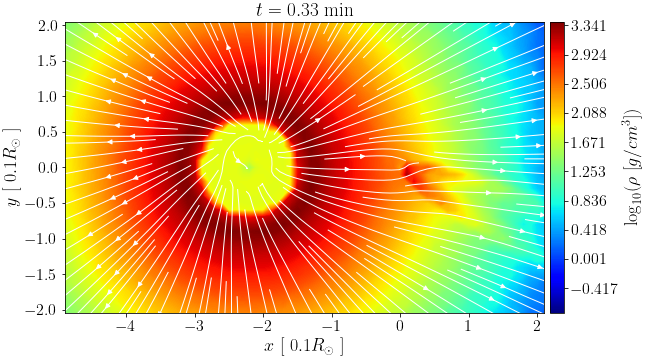}
  \includegraphics[width=0.67\hsize,clip]{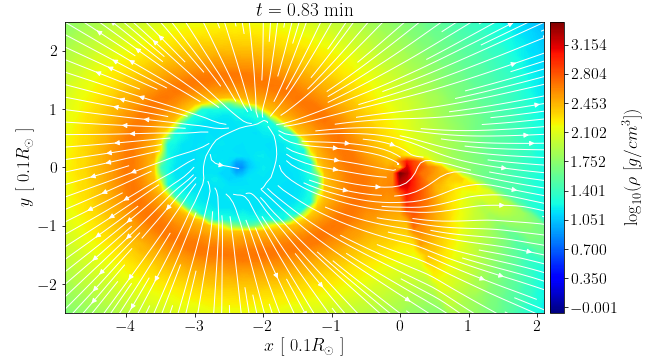}
  \includegraphics[width=0.67\hsize,clip]{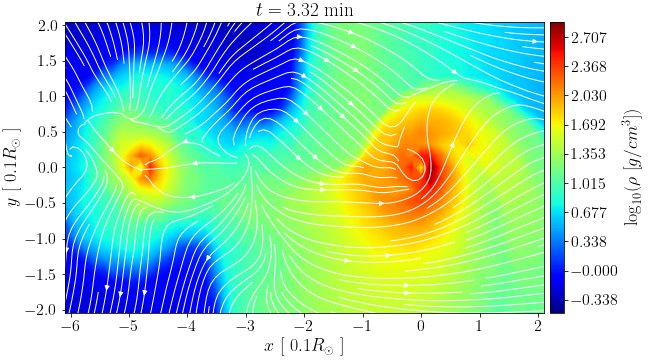}
  \caption{Binary density color map at three selected times from the SN breakout: $t=0.32$ min ($19.2$ s, upper panel), $t=0.83$ min ($49.8$ s, middle panel), when the NS reaches the critical mass point, and a post-collapse time, $t=3.17$ min ($190.2$ s, lower panel). The white curves show the velocity field of the ejected material from the SN. In this simulation, the pre-SN binary consists of a $9 M_\odot$ CO core and a $2.0 M_\odot$ NS companion. The binary orbital period is $6$ minutes, and the kinetic energy of the SN is $7.85\times 10^{50}$ erg. The angular momentum of the NS companion at the beginning of the simulation is $J = 2.6 G M_\odot^2/c$.}
  \label{fig:simulation}
\end{figure*}

Figure \ref{fig:simulation} shows the results of a numerical simulation of a traditional BdHN scenario, which starts with a binary progenitor composed of a CO core of $\sim 10 M_\odot$ and a NS companion of $\sim 2 M_\odot$ with a binary period ranging from minutes to hours \citep[see][and references therein]{2024ApJ...976...80B}. 

These smoothed-particle-hydrodynamics (SPH) simulations are performed with the \textit{SNSPH} code adapted to the BdHN progenitor \citep[see][for details]{2019ApJ...871...14B}. The three-dimensional (3D) Lagrangian code calculates the evolution of the position, momentum (linear and angular), and thermodynamic properties (pressure, density, and temperature) of the pseudo-particles, which are assigned a mass $m_i$ according to the mass-density distribution of the SN ejecta. The SPH simulation begins ($t=0$) when the SN shock front reaches the CO star surface, when we transition from a 1D core-collapse supernova simulation to a 3D-SPH configuration \citep{2018ApJ...856...63F}. At this moment, the collapse of the CO star has formed a $\nu$NS of $1.75~M_\odot$ at the center, and around $7.14~M_\odot$ are ejected by the SN explosion. In the simulation, the $\nu$NS and the NS companion are modeled as point masses and interact only gravitationally, with the SN particles and with each other. We allow such point particles to increase their mass by accreting other particles from the SN material following the algorithm described in \citet{2019ApJ...871...14B,2022PhRvD.106h3002B,2024ApJ...976...80B}.

Figure \ref{fig:simulation} shows snapshots of the density color map, including the velocity field lines of the SN ejected material, at three selected simulation times. In this simulation, we adopt that the pre-SN binary comprises a CO core of $9 M_\odot$ with a $2.0 M_\odot$ NS companion, with an initial angular momentum $J = 2.4 G M_\odot^2/c$ (i.e., $J\approx 0.6 G M^2/c$). The binary orbital period is $P_{\rm orb}\approx 6$ min and the SN kinetic energy $7.85\times 10^{50}$ erg. In this BdHN, the NS companion reaches the critical mass for BH formation at $t = 0.83$ min $\approx 49.8$ s (middle panel). 

Depending on the binary parameters, especially the orbital period, the BdHN model predicts different energy releases that reflect the underlying physics, leading to the identification of three types of BdHNe. In BdHNe I, the most energetic with $E_{\rm iso}\gtrsim 10^{52}$ erg, with orbital periods of a few minutes, the accretion onto the NS companion brings it to the critical mass and forms a BH. A prototype of this class has been GRB 190114C \citep{2021A&A...649A..75M,2022ApJ...929...56R}. BdHNe II emit energies $E_{\rm iso} \sim 10^{50}$--$10^{52}$ erg and have orbital periods from a few tens of minutes. The NS mass does not increase sufficiently to collapse into a BH. The energy division between BdHNe I and II at $\sim 10^{52}$ erg is set by the energy needed to bring the NS to the critical mass \citep{2016ApJ...832..136R,2018ApJ...859...30R}. A source analyzed within this family is GRB 190829A \citep{2019ApJ...874...39W}. The BdHNe III, with $E_{\rm iso}\lesssim 10^{50}$ erg, have orbital periods of hours, and the accretion is negligible. A typical example is GRB 171205A \citep{2023ApJ...945...95W}. A summary of the traditional BdHN model and the analysis of specific sources can be found in \citet{2023ApJ...955...93A}.  We here  moved to a specific, accurate description of GRB 220101A, an energetic source of $E_{iso} \sim 10^{54}$ erg, introducing new astrophysical process characterizing the seven episodes. In doing this we have realized indeed there is an entire family of these new class of Binary driven Peta Nova and will apply the consideration of this article also to GRB 180720B, GRB 221009A and GRB 240825A.

\section{The new scenario}\label{sec:newmodel}

We here extend our basic BdHN model proved for GRBs with  $E_{iso}$ of up to $10^{52}$ erg in \cite{2023ApJ...955...93A} to GRB 220101A to the new class of more energetic sources of $E_{iso} > 10^{54}$ erg., typical of a Peta nova.  Their progenitors, generalize the ones of the binary BDHN composed of a massive CO Core and a binary NS companion, to a  three body system composed of a highly magnetized CO-core, again of approximately 10 Solar masses with magnetic field of $10^6$ Gauss, surrounded by two companions: a binary NS and a binary WD. Again, as in the previous case of the BDHN, their  binary orbital periods of minutes to hours. In view of their larger energies we call these systems Binary Driven Peta Novae (BdP-N). 

In order to attempt the completion of the self consistent result attempts on simplified systems addressing specific time scales and expected emissions have been presented in \cite{2025arXiv250906243R}, and the second Ruffini et al. (2026, submitted to JHEA) to be used as tutorial of illustrating the procedure needed to obtain the final results contained in this final paper.

In this choice we have been inspired by a recent proposal by Michel Mayor \citep{Mayor2020, Mayor2024Plurality} based on  the classic works of Struve \citep{1952Obs....72..199S} and Fred Hoyle \citep{Hoyle1947}. Mayor  proposes  a change of paradigm from the traditional works on highly rotating single objects, see e. g.  Jacobi \citep{Jacobi1834}, James Jeans  \citep{Jeans1919Ellipsoidal} and Chandrasekhar \citep{Chandrasekhar1969}, {\it Ellipsoidal Figures of Equilibrium}. We have been following ourself in many papers the single fast rotating systems we instead for the case of the BdP-N. We consider the Mayor's approach more satisfactory and intuitive: there we are in presence of multiple systems composed of planets with large angular momentum around a massive non-rotating star. In the present case we consider a Massive CO Core of $10 M_\odot$, surrounded by  a combination of neutron stars, white dwarfs and possibly even  magnetars,  highly rotating and carrying very high angular momentum. Each combination may lead to a specific  GRB type.  In the case of GRB 220101  at the center is a massive magnetized CO core of 10 solar masses, surrounded by a neutron star and a white dwarf with a binary period as short as minutes or hours. The collapse of the CO core leads to the HB supernova a specific example of BdP-N (Episode I). 

What makes the results reported in Fig \ref{fig:lc} and Fig. \ref{fig:lc-linear} truly exceptional is that all observational instruments, from space and from the ground, were able to detect the sources in the equatorial plane defined  the ejecta of the HB supernova and described by the seven episodes 

We show in this article and accompanying papers, how the evolution of these progenitor system can lead to an HB supernova of $10^{54}$ erg, the ejecta of such SN interacting with the magnetosphere of the companion NS, leads to the UPE jetted emission normal to the orbital plane of the system which originate a jetted emission  almost ninety degree from the equatorial plane of the GRB. Again the accretion of the HB supernova ejecta on the white dwarf,originate the second SN and the spin up of the newly created NS $\nu$NS  becomes a  pulsar, giving origin to the optical afterglow. The same HB supernova ejecta accreting on the binary NS companion lead to the formation of the BH originating the GeV afterglow emission. The synchrotron emission created by the fast rotating msec NS interacting with the remnant gives origin to the Synchrotron emission observed in the X-ray, optical and Radio details in the following. Having determined all the energetics of the seven Episodes of the GRB 220101, a last study is needed to verify that the interaction of both the highly spinning NS companion and white dwarf binary companion can trigger the observed energetic of the HB SN. On this preliminary results have been obtained by Wang Yu and additional work has started lead by Chris Freyer at Los Alamos.

\section{Data Analysis}\label{sec:observation}

GRB 220101A is a long GRB that occurred at 2022-01-01 05:11:13 (UT). The event triggered multiple satellites, including Swift \citep{2022GCN.31347....1T,2022GCN.31348....1T}, Fermi \citep{2022GCN.31360....1L,2022GCN.31350....1A}, AGILE \citep{2022GCN.31354....1U}, and Konus-Wind \citep{2022GCN.31433....1T}. Optical observation by the Xinglong-$2.16$m telescope \citep{2022GCN.31353....1F} revealed a broad absorption feature in the spectrum indicating the presence of Lyman-$\alpha$ absorption, as well as other absorption lines from which the redshift was determined to be $z=4.61$, confirmed by the Liverpool telescope \citep{2022GCN.31357....1P} and NOT \citep{2022GCN.31359....1F}. The burst exhibited a bright and complex multi-peaked time profile within the first $\sim 150$~s. The estimated isotropic equivalent energy is $E_{\rm iso} \sim 4 \times 10^{54}$~ erg and a peak luminosity is estimated to be $L_p \sim 9 \times 10^{53}$~erg~s$^{-1}$, making GRB 220101A  one of the most luminous GRBs ever observed \citep{2022GCN.31360....1L,2022GCN.31433....1T}. 

\subsection{Fermi-GBM MeV Data Analysis}\label{sec:time-resolved}

\begin{figure*}
\includegraphics[angle=0, scale=0.60]{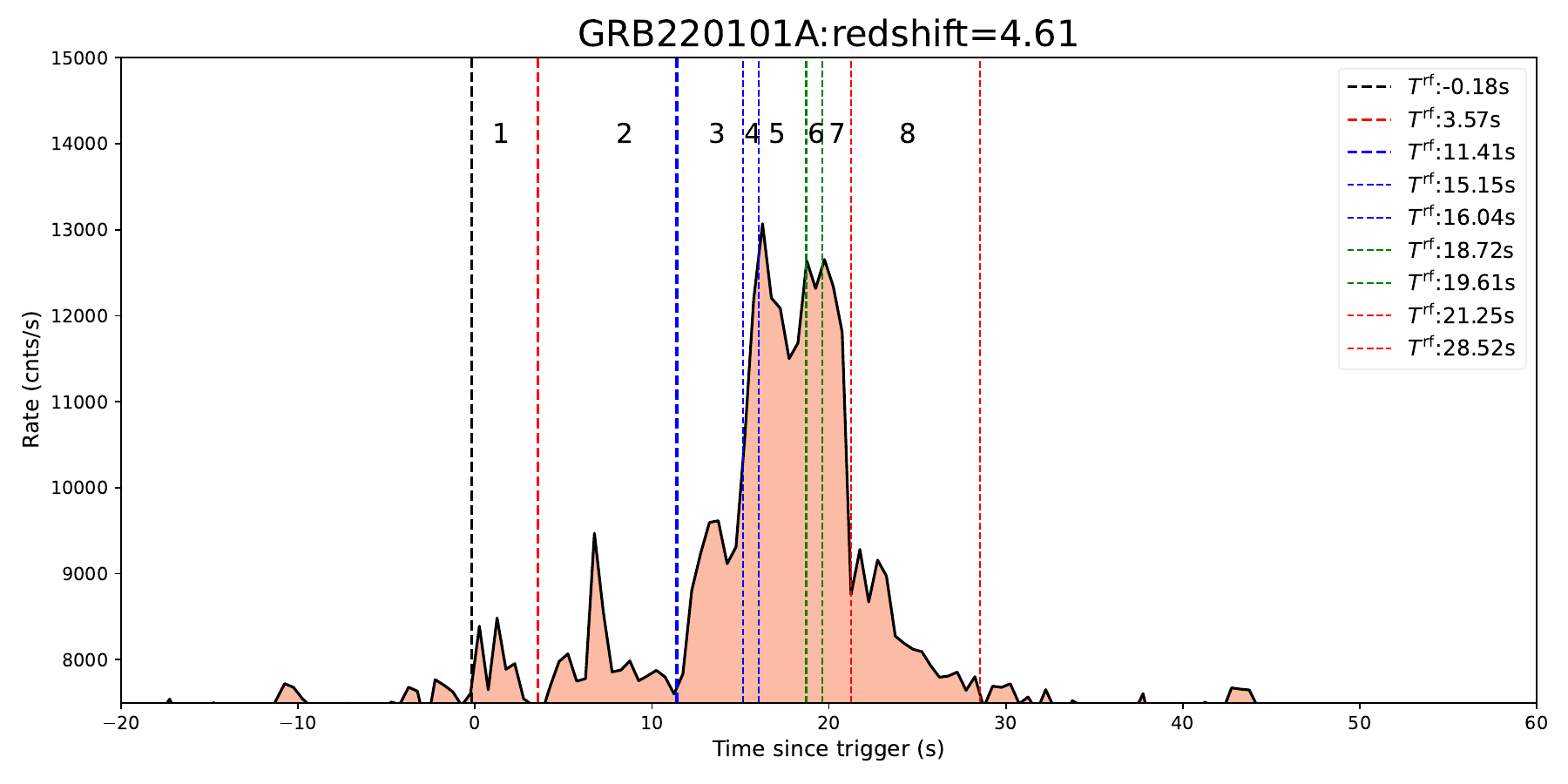}
\caption{Count light curve of \emph{Fermi}-GBM of GRB 220101A with $8$ slices. Binning is performed based on the morphology of the light curve and the fitting functions of the spectra. In Section \ref{sec:episodes}, we summarize the newly identified physical processes, corresponding to each Episode.}
\label{fig:lightcurve}
\end{figure*}

\begin{figure*}
\includegraphics[angle=0, scale=0.45]{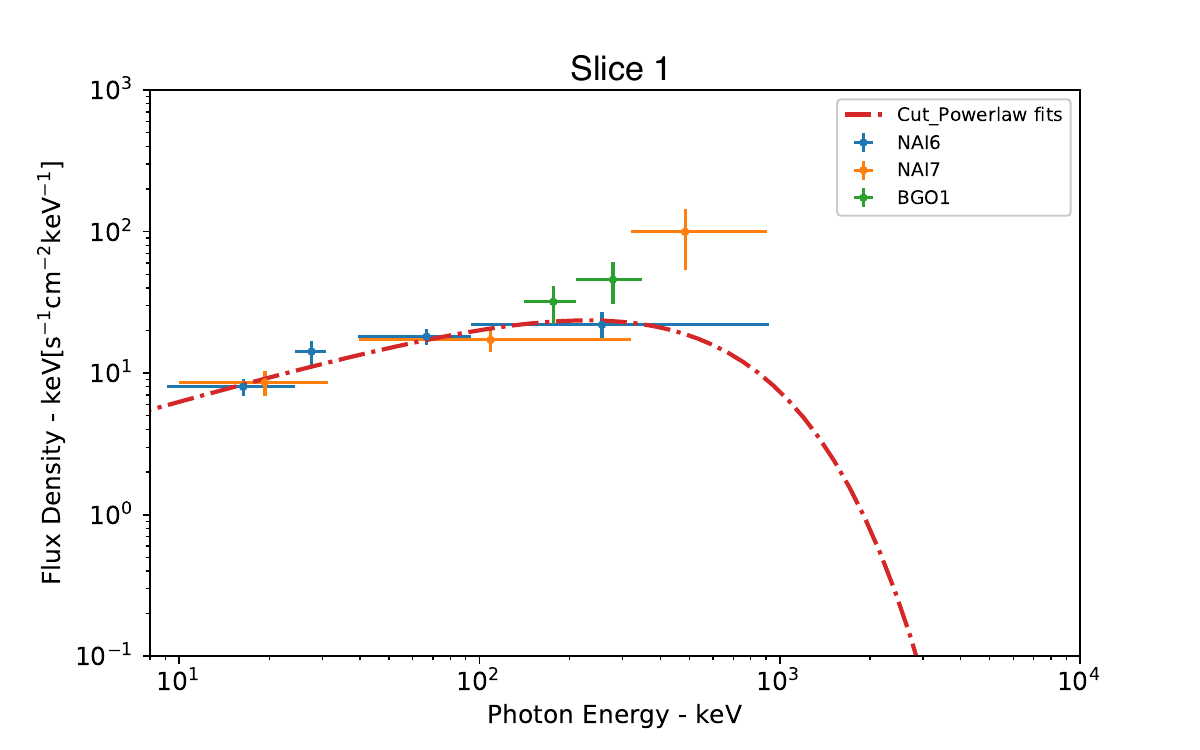}
\includegraphics[angle=0, scale=0.45]{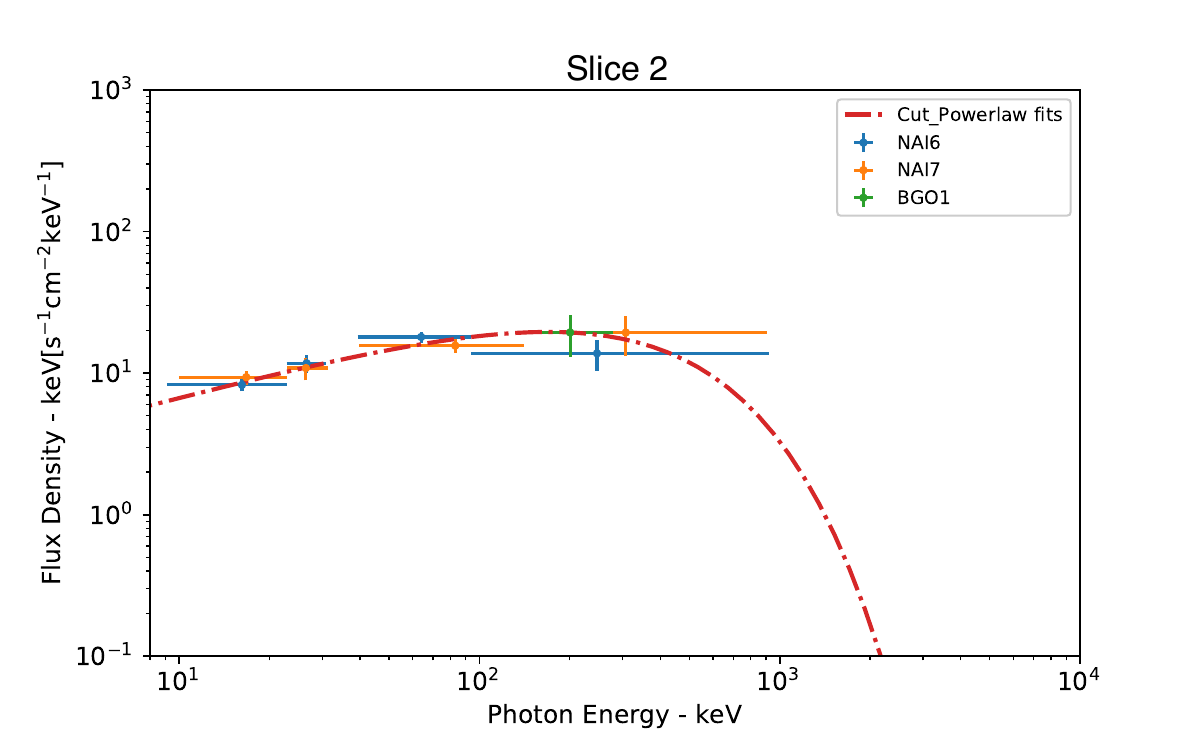}
\includegraphics[angle=0, scale=0.45]{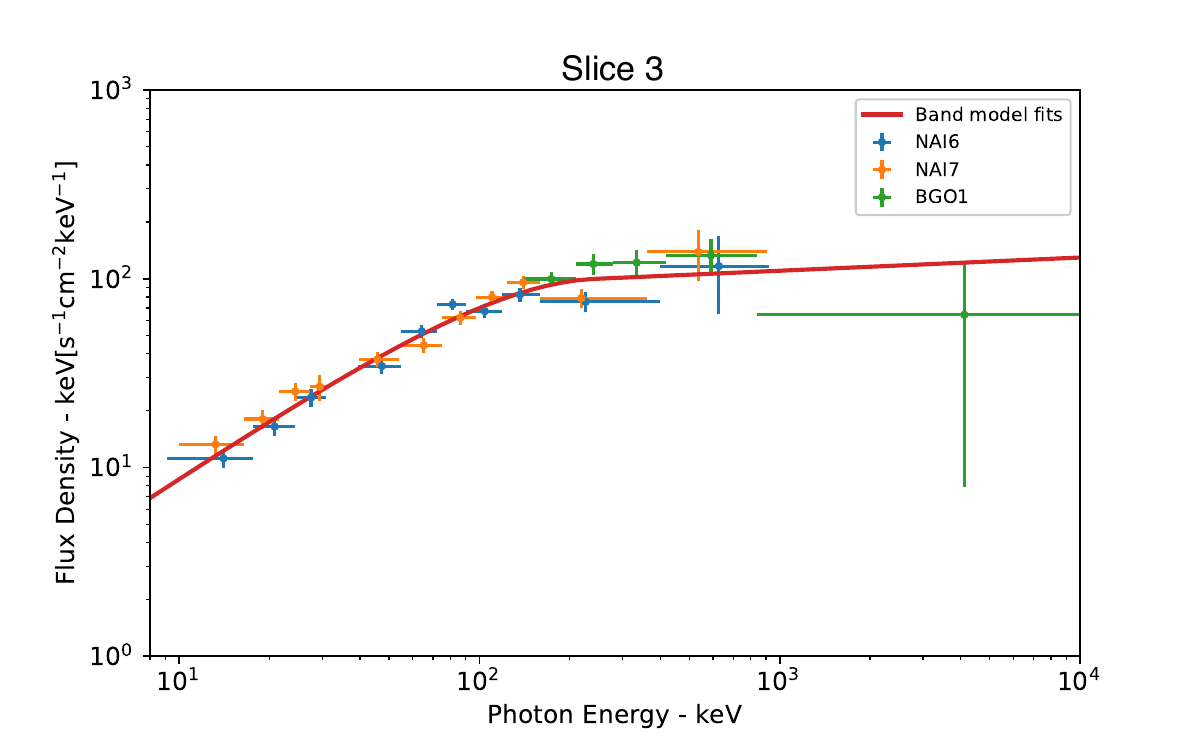}
\includegraphics[angle=0, scale=0.45]{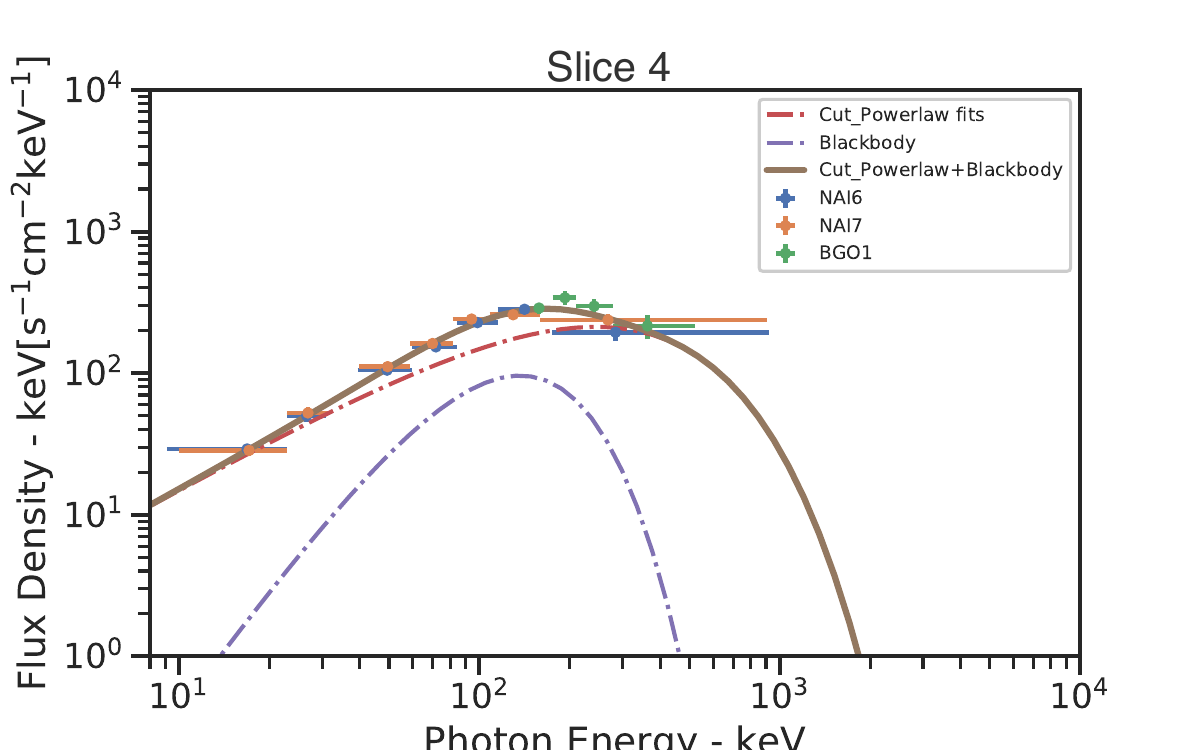}
\includegraphics[angle=0, scale=0.45]{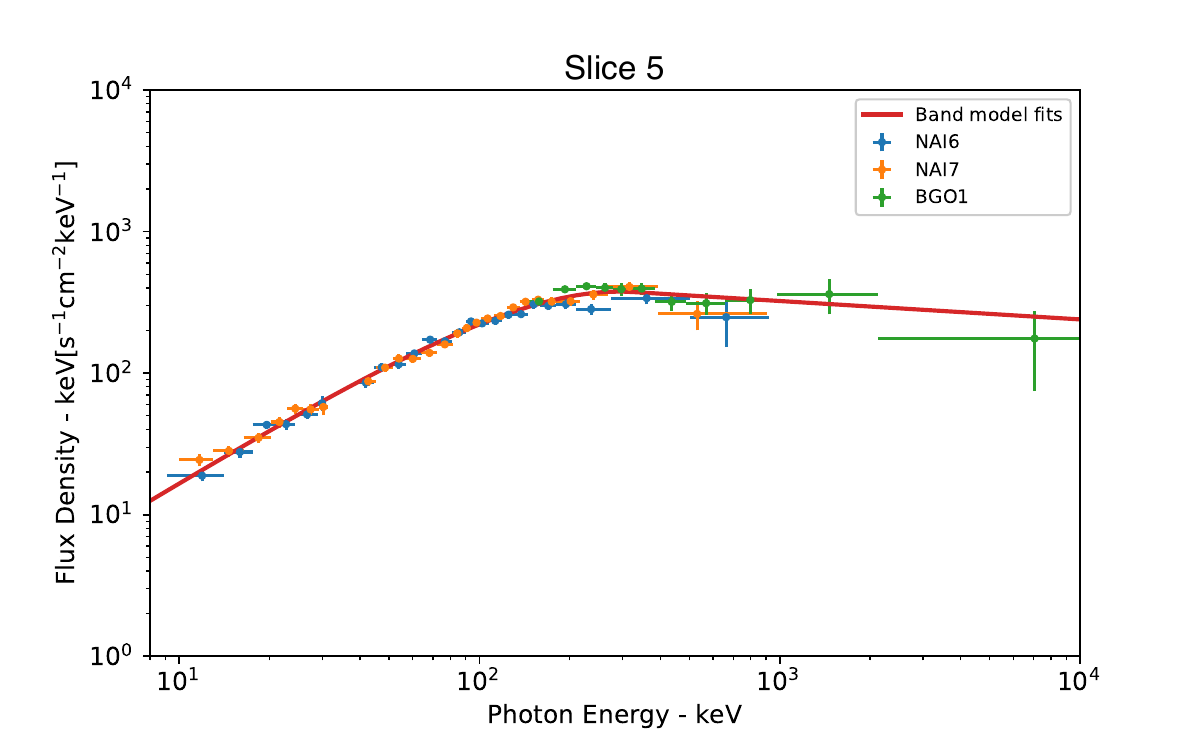}
\includegraphics[angle=0, scale=0.45]{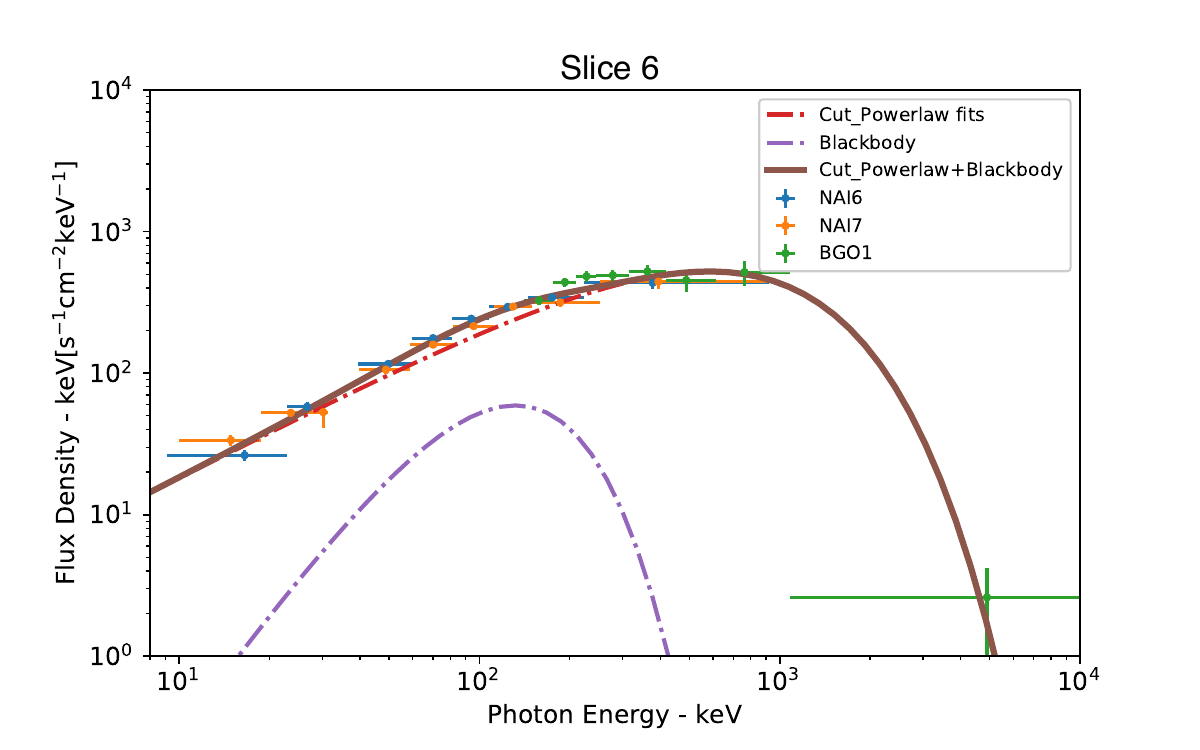}
\includegraphics[angle=0, scale=0.45]{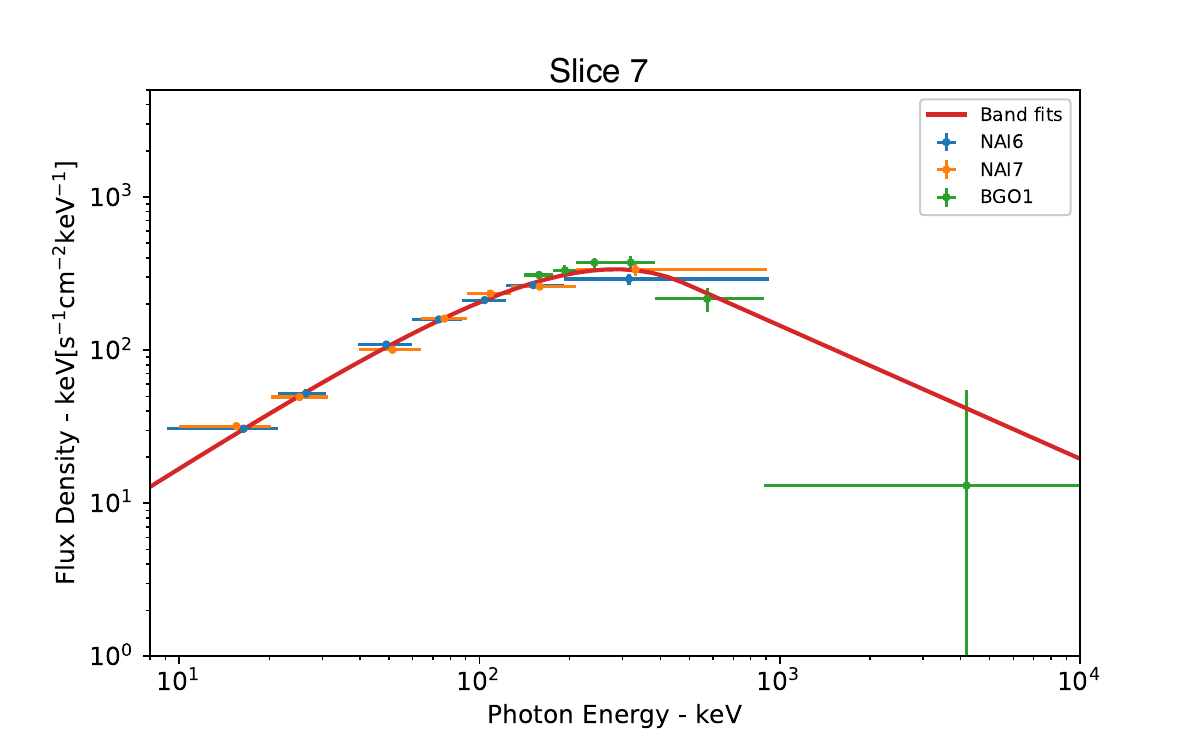}
\includegraphics[angle=0, scale=0.45]{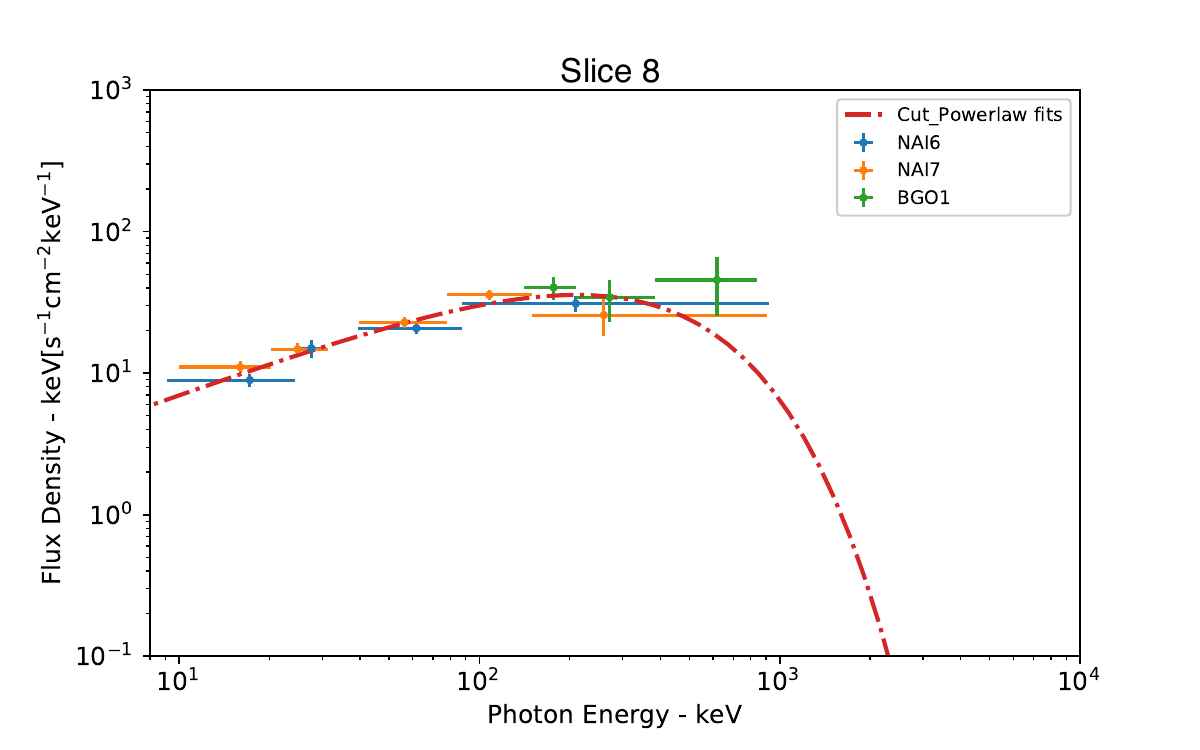}
\caption{The spectra of the Slices 1-8 from Fermi-GBM.}
\label{fig:specra}
\end{figure*}

The  Fermi Space Telescope is designed to observe energetic events in the keV-MeV-GeV energy range in the sky and provide information about their location, duration, energy, and spectra. The Fermi Gamma-ray Burst Monitor (GBM) aboard the Fermi satellite comprises $14$ scintillation detectors that cover the entire sky and detect gamma rays in the $8$ keV-$40$ MeV energy range. The GBM has a wide field of view and can detect GRBs from any direction in the sky. 

The spectral analysis in the keV-MeV energy range provides a crucial clue for diagnosing different physical components within the BdHN model. In this study, we use a pure Python package, {\tt the Multi-Mission Maximum Likelihood Framework} ({\tt 3ML} \citealt{Vianello2015}), to perform detailed time-integrated and time-resolved spectral analyses. Based on 3ML and following standard practices \citep{Li2019b,Li2019c,Yu2019,Burgess2019,Li2020,Li2021a,Li2021b}, supplied by the \emph{Fermi} team, we apply a Bayesian analysis and Markov Chain Monte Carlo (MCMC) iteration to determine their best-fit spectral parameters. 

Fermi observations have revealed that GRBs have diverse spectral properties. Some GRBs can be well-fitted with a typical GRB spectrum, the so-called Band function \citep{Band1993}, while others may require the addition of a thermal component to achieve acceptable fits. We first use the Band function to fit the observed time-resolved spectra of GRB 220101A.

We conduct a detailed time-resolved spectral analysis to search for an additional thermal component. We perform our spectral analysis in four time iterations, each with increasingly shorter time bins. Initially, the entire burst duration is divided into several slices. We then fit the spectral data for each slice with both the Band and CPL+BB models and select the favored model based on the deviance information criterion (DIC; \citealt{Spiegelhalter2002, Moreno2013}). We found a weak thermal component in the intervals [$80$--$90$ s] and [$100$--$110$ s]. These results are confirmed when a shorter iteration is involved (see Table \ref{tab:Summary1}). Next, we perform a spectral analysis of the mini-pulses from GRB 220101A, which may be associated with different physical processes. As shown in Figure \ref{fig:lightcurve}, the GBM light-curve can be divided into $8$ interesting slices. We then check their spectral properties in detail. The results of this refined spectral analysis can be summarized as follows. Each slice is categorized into an Episode of a specific physical process (see Table \ref{tab:Summary1} and Section \ref{sec:episodes}). We use the cosmological redshift of the source $z = 4.61$.

\begin{itemize}
    \item \textbf{Slice 1 ($-0.18$ to $3.57$ s)}: The initial pulse of the bursts best fits with a CPL model, featuring a cutoff energy $E_{\rm c} = 228^{+297}_{-132}$ keV and a power-law index $\alpha = -1.39^{+0.19}_{-0.18}$. The time-averaged energy flux is $1.42^{+1.19}_{-0.63} \times 10^{-7}$ erg s$^{-1}$ cm$^{-2}$. The isotropic luminosity is $L_{\rm iso} = 3.0^{+3.0}_{-1.0} \times 10^{52}$ erg s$^{-1}$ and the energy released is $E_{\rm iso} = 1.2^{+1.0}_{-0.5} \times 10^{53}$ erg.

    \item \textbf{Slice 2 ($3.57$ to $11.41$ s)}: The following pulse is best modeled by a CPL with cutoff energy $E_{\rm c} = 172^{+119}_{-73}$ keV and power-law index $\alpha = -1.42^{+0.13}_{-0.13}$. The time-averaged energy flux is $1.20^{+0.58}_{-0.41} \times 10^{-7}$ erg s$^{-1}$ cm$^{-2}$. The isotropic luminosity is $L_{\rm iso} = 3.0^{+3.0}_{-1.0} \times 10^{52}$ erg s$^{-1}$ and the energy release is $E_{\rm iso} = 2.1^{+1.0}_{-0.7} \times 10^{53}$ erg.

    \item \textbf{Slice 3 ($11.41$ to $15.15$ s)}: This slice is best represented by a Band model, with a peak energy $E_{\rm p} = 258^{+49}_{-48}$ keV and power-law indices $\alpha = -0.93^{+0.09}_{-0.08}$ and $\beta = -1.93^{+0.12}_{-0.12}$. The average energy flux measures $5.97^{+1.31}_{-1.19} \times 10^{-7}$ erg s$^{-1}$ cm$^{-2}$. The isotropic luminosity is $L_{\rm iso} = 3.0^{+0.3}_{-0.3} \times 10^{53}$ erg s$^{-1}$ and the energy release is $E_{\rm iso} = 5.0^{+1.1}_{-1.0} \times 10^{53}$ erg.

    \item \textbf{Slice 4 ($15.15$ to $16.04$ s)}: Fitted best by a CPL+BB model, this slice has a peak energy $E_{\rm p} = 263^{+48}_{-102}$ keV, a power-law index $\alpha = -0.57^{+0.21}_{-0.28}$, and a blackbody temperature $kT = 21.9^{+15.8}_{-18}$ keV. The average energy flux is $1.22^{+0.45}_{-0.57} \times 10^{-6}$ erg s$^{-1}$ cm$^{-2}$. The isotropic luminosity is $L_{\rm iso} = 2.7^{+10.2}_{-1.3}\times 10^{53}$ erg s$^{-1}$ and the energy release $E_{\rm iso} = 2.4^{+9.1}_{-1.1} \times 10^{53}$ erg. The blackbody component accounts for $\sim 30\%$ of the total flux.

    \item \textbf{Slice 5 ($16.04$ to $18.72$ s)}: A Band model with peak energy $E_{\rm p} = 291^{+17}_{-17}$ keV, and power-law indices $\alpha = -0.70^{+0.04}_{-0.04}$ and $\beta = -2.13^{+0.09}_{-0.09}$ fits this slice optimally. The time-averaged energy flux is $1.85^{+1.77}_{-1.56} \times 10^{-6}$ erg s$^{-1}$ cm$^{-2}$. The isotropic luminosity is $L_{\rm iso} = 4.2^{+0.4}_{-0.4}\times 10^{53}$ erg s$^{-1}$ and the energy released is $E_{\rm iso} = 1.12^{+0.11}_{-0.09} \times 10^{54}$ erg.

    \item \textbf{Slice 6 ($18.72$ to $19.61$ s)}: Best fitted by a CPL+BB model, with peak energy $E_{\rm p} = 589^{+157}_{-165}$ keV, power-law index $\alpha = -0.88^{+0.09}_{-0.09}$, and blackbody temperature $kT = 32.2^{+7.1}_{-7.8}$ keV. The average energy flux is $2.31^{+1.05}_{-0.64} \times 10^{-6}$ erg s$^{-1}$ cm$^{-2}$. The isotropic luminosity is $L_{\rm iso} = 5.2^{+2.4}_{-1.5}\times 10^{53}$ erg s$^{-1}$ and the energy release is $E_{\rm iso} = 4.6^{+2.1}_{-1.3} \times 10^{53}$ erg. The blackbody component contributes to $\sim 25\%$ of the total flux.

    \item \textbf{Slice 7 ($19.61$ to $21.25$ s)}: It fits best with a Band model, peak energy $E_{\rm p} = 284^{+17}_{-17}$ keV, and power-law indices $\alpha = -0.74^{+0.04}_{-0.04}$ and $\beta = -2.87^{+0.42}_{-0.40}$. The time-averaged energy flux is $1.42^{+0.16}_{-0.14} \times 10^{-6}$ erg s$^{-1}$ cm$^{-2}$. The isotropic luminosity is $L_{\rm iso} = 3.2^{+0.4}_{-0.3} \times 10^{53}$ erg s$^{-1}$ and the energy release is $E_{\rm iso} = 5.3^{+0.6}_{-0.5} \times 10^{53}$ erg.

    \item \textbf{Slice 8 ($21.25$ to $28.52$ s)}: The last slice of the Fermi-GBM fits a CPL model with cutoff energy $E_{\rm c} = 211^{+86}_{-59}$ keV and power-law index $\alpha = -1.22^{+0.10}_{-0.10}$. The average energy flux is $1.83^{+0.65}_{-0.50} \times 10^{-7}$ erg s$^{-1}$ cm$^{-2}$. The isotropic luminosity is $L_{\rm iso} = 4.0^{+1.0}_{-1.0} \times 10^{52}$ erg s$^{-1}$ and the energy release is $E_{\rm iso} = 3.0^{+1.1}_{-0.8} \times 10^{53}$ erg.
\end{itemize}

Moreover, we checked if the thermal component can be measured in shorter time intervals. In Iteration III of Table \ref{tab:initialchecking},  Slices 4 and 6 were divided into two parts each, and in Iteration IV, Slice 6 into four parts. In all of these shorter intervals the CPL+BB model is unconstrained. The reason is that the BB and the CPL cover almost the same energies, so with fewer photons the fit can no longer distinguish them. In fact, we see this already in the full slices, where $k_{\rm B}T_{\rm obs} = 21.9^{+15.8}_{-18.0}$ keV for Slice 4. Dividing an interval into two parts lowers its significance by a factor $\sqrt{2}$, which makes the separation harder. Therefore, slices 4 and 6 are the shortest intervals in which we can measure the thermal component with \textit{Fermi}-GBM.

\begin{figure*}
\centering
\includegraphics[angle=0, scale=1.0]{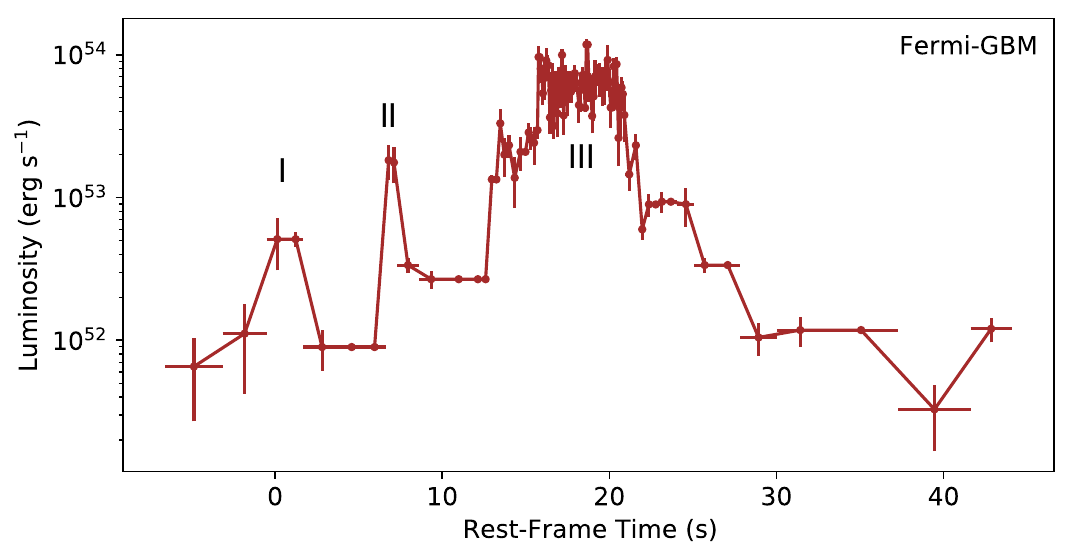}
\caption{Isotropic luminosity light-curve from Fermi-GBM of 1 keV to 10 MeV. Including the three Episodes of I, II, and III. }\label{fig:gbm}
\end{figure*}

To convert the count rate light curve from Figure \ref{fig:lightcurve} to the luminosity light curve of Figure \ref{fig:gbm}, we apply Bayesian blocks binning \citep{2013ApJ...764..167S}. This enables a finer temporal resolution beyond the initial 8 slices, allowing the binning to adapt to the data variability and effectively capture rapid flux changes. For each time bin, we fit the data using the spectral functions identified across the 8 slices, including the Band and CPL models as needed. We then integrate these fitted spectra over the energy range from $1$ keV to $10$ MeV to obtain the energy flux, $f_{\rm obs} $, and apply the observed redshift $z = 4.61 $ to calculate the $k$-correction \citep{2001AJ....121.2879B}. Using a standard cosmological model, we calculate the luminosity distance $D_L(z) $ from the redshift. By combining these terms and assuming a $4 \pi$ solid angle, we convert the energy flux into the isotropic luminosity according to $ L = 4 \pi k D^2_L(z) f_{\rm obs}$, leading to Figures \ref{fig:lc} and \ref{fig:lightcurve}--\ref{fig:gbm}. 

\subsection{Fermi-LAT GeV Data Analysis}\label{sec:fermilatgevdata}

The Large Area Telescope (LAT) is another instrument onboard the Fermi satellite. The Fermi-LAT has a large field of view, covering about $20\%$ of the sky at any given time, and can detect gamma rays from $20$ MeV to $300$ GeV \citep{2012ApJS..203....4A}.

The LAT data is retrieved from the FSSC LAT Data server \footnote{\url{https://fermi.gsfc.nasa.gov/cgi-bin/ssc/LAT/LATDataQuery.cgi}} by specifying the location of RA=1.35372 and Dec=31.76882, which is a refined position provided by the Swift UVOT\footnote{\url{https://www.swift.ac.uk/xrt_positions/01091527/image.php}}. A time window of one hour before and three days after the GBM trigger time is chosen, along with an energy range of $30$ MeV to $300$ GeV and a selection radius of $20$ degrees. The data was analyzed by Fermitools\footnote{\url{https://github.com/fermi-lat/Fermitools-conda/wiki}}. The study utilizes the event selection of {\tt P8\_TRANSIENT020E}. The recommended data cuts, such as {\tt zmax = 100 degrees, skybinzsize=0.2, thetamax=180 degrees}, are implemented. The background consisted of the galactic diffuse emission template, the isotropic emission template, and approximately 128 point sources within a 15-degree radius of the GRB, except for the 4FGL J2311.0+3425, for which the TS value is 3; other contributions are negligible. 

The time-integrated LAT spectrum can be fitted by a single power-law of index $-2.54\pm0.258$; the total isotropic energy from 100 MeV to 10 GeV is $E_{\rm LAT} = 6.65\pm1.69 \times 10^{53}$~erg.  The LAT light curve is produced by fitting multiple time-resolved spectra using the power-law functions. Each spectrum's time bin has a TS value of at least 20. The luminosity light curve depicted in Figure \ref{fig:lc} was calculated for the integration of spectra of energy from $100$ MeV to $10$ GeV. Figure \ref{fig:lat-photon} shows the probability of each photon belonging to the GRB 220101A. The probability is estimated by the {\tt gtsrcprob} command. The first LAT photon of probability more than $80\%$ comes at time $2.8$~s, and the highest rest-frame energy photon is $\sim 5$~GeV.

The entire GeV observation by the Fermi-LAT is named the 11th slice of the GRB (see Figure \ref{fig:lat-photon}):

\begin{itemize}
\item \textbf{Slice 9 ($2.8$ to $100$ s):} It encompasses high-energy emissions from 100 MeV to 10 GeV observed by Fermi-LAT. A single power-law with an index of $-2.54\pm 0.26$ is the optimal model for this slice. Taking into account the redshift of $z=4.61$, the isotropic LAT energy for this slice is estimated to be $E_{\rm LAT} = 6.65\pm1.69 \times 10^{53}$~erg.
\end{itemize}

\subsection{Swift-UVOT, Xinglong, and CAHA Optical Data}\label{sec:swiftuvotxinglongcahaoptical}

Swift-UVOT is a UV/optical telescope onboard the Swift satellite.  It is designed to observe the afterglows of GRBs and other transient sources in the UV and optical wavelengths. It has a $30$~cm aperture and can be observed in six different filters, covering a wavelength range from $170$ to $600$ nm. The UVOT can provide precise positions for GRBs and other transients, enabling the study of their spectral and temporal properties. Additionally, it can observe and study the host galaxies of GRBs, SNe, and other transients \citep{2005SSRv..120...95R}.

The Swift-UVOT data are analyzed and presented in detail by \citet{2023NatAs...7.1108J}. We adopt their UVOT products in this article. The UVOT detection of GRB 220101A was significant. The spike has an exceptionally high absolute AB magnitude of $-39.4 \pm 0.2$, indicating extreme brightness in the optical-ultraviolet range. The UVOT instrument observed the burst in the V, B, U, W1, M2, W2, and White bands over several epochs in image mode. The first white-band exposure in event mode began approximately 90 seconds after the trigger and lasted about $150$ s ($\sim15$--$30$~s in the source's rest frame). The optical-ultraviolet emission did not trace the gamma-ray activity. It displayed a smooth light curve, lacking the variability typically seen in gamma-ray emissions, as shown in Figures \ref{fig:lc}. The spectral energy distribution (SED) analysis showed that the optical emission during the rise and fast-decline phases was significantly above the extrapolated high-energy spectrum, suggesting different physical origins for the optical and high-energy radiation.

The observations of GRB 220101A from the Xinglong $2.16$ m and CAHA $2.2$ m telescopes, especially details in the I/i band, provided critical insights into its time evolution and brightness \citep{2023ApJ...959..118Z,2022GCN.31388....1C}. The Xinglong $2.16$ m Telescope, equipped with the Beijing Faint Object Spectrograph and Camera (BFOSC), captured the afterglow's brightness in the I band at $17.76 \pm 0.01$ ($0.219$ days after burst). Over time, the I-band brightness gradually declined, with measurements of $19.95 \pm 0.03$ ($2.227$ days after the burst) and $20.80 \pm 0.04$ (4.219 days after the burst). By $7.221$ days after the burst, the I band brightness had decreased to a magnitude of $21.3 \pm 0.1$. Similarly, the CAHA 2.2 m Telescope, using the Calar Alto Faint Object Spectrograph (CAFOS), documented the fading brightness in the i band, showing $18.30 \pm 0.03$ ($0.544$ days after the burst) and later $21.61 \pm 0.17$ ($6.611$ days after the burst). The optical afterglow of GRB 220101A is one of the most luminous ever detected, highlighting the exceptional energy output of this burst.

The rising part of the optical light curve is identified as \textbf{Slice 10} ($16.54$~s to $21$~s), with an isotropic energy of $(2.56 \pm 0.54) \times 10^{51}$~erg. The subsequent decaying phase is \textbf{Slice 11} ($21$~s to $10^6$~s), with an isotropic energy $E_{\rm iso} = (5.31 \pm 1.75) \times 10^{52}$erg (see Figure \ref{fig:optical-lightcurve}).

\subsection{Swift-XRT X-ray Data Analysis}\label{sec:swiftxrtdatanalysis}

\begin{figure*}
\centering
\includegraphics[angle=0, scale=0.85]{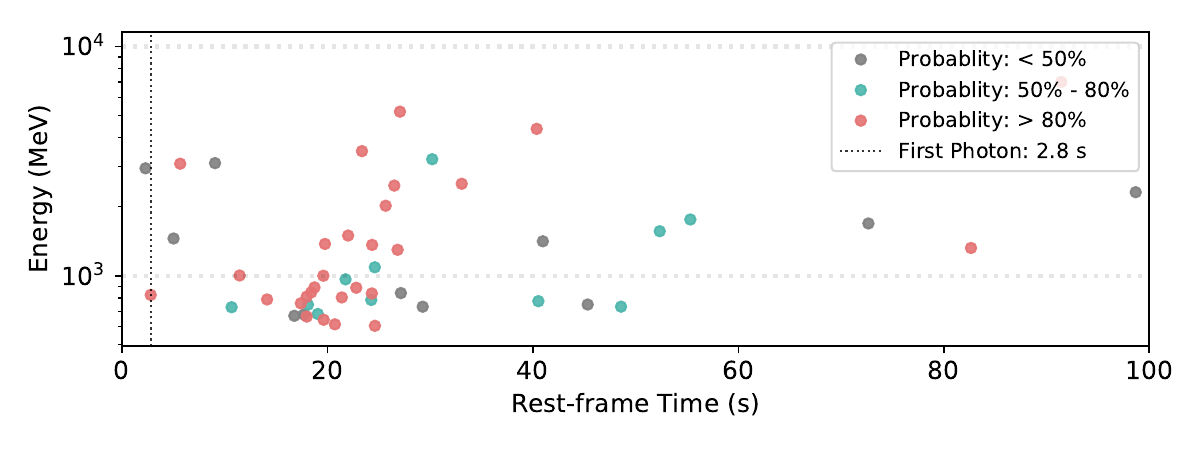}
\includegraphics[angle=0, scale=0.37]{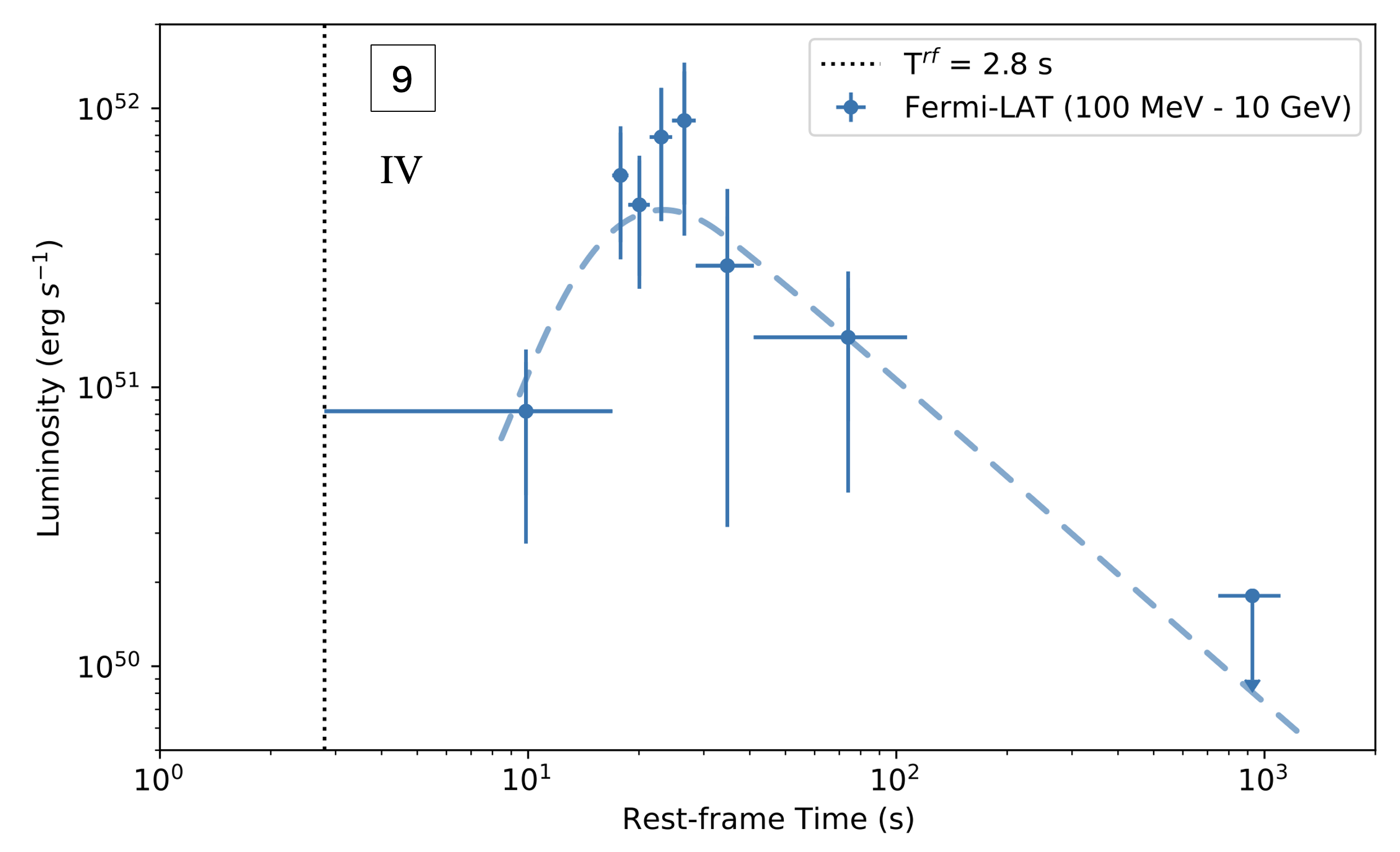}
\setlength{\unitlength}{1cm}
\caption{\textbf{Upper:} Photon arrival times (rest-frame, seconds) versus photon energy (MeV, log scale) detected by Fermi-LAT. Points are color-coded based on the probability of the photon belonging to GRB 220101A, estimated using gtsrcprob: grey ($<50\%$), green ($50\%$--$80\%$), red ($>80\%$). The first photon with $>80\%$ probability arrives at $2.8$ s. The highest energy photon detected has an energy of $\sim 5$ GeV in the rest frame.\textbf{Lower: } GeV luminosity light curve ($100$ MeV--$10$ GeV) derived from Fermi-LAT data. The x-axis is rest-frame time in seconds (log scale), and the y-axis is isotropic luminosity in erg s$^{-1}$ (log scale). The data points (blue) represent luminosities in time bins with a detection significance TS > 20. The light curve shows emission starting around 2.8 s and lasting up to ~1000 s (rest frame), following a decaying trend approximated by the dashed line. This GeV emission constitutes Episode III in the BdHN model, powered by the spin-down of the newly formed BH. The entire LAT observation period is designated as Slice 9.}\label{fig:lat-photon}
\end{figure*}

\begin{figure*}
\label{fig:optical-lightcurve}
\centering
\includegraphics[angle=0, width=0.82\hsize,clip]{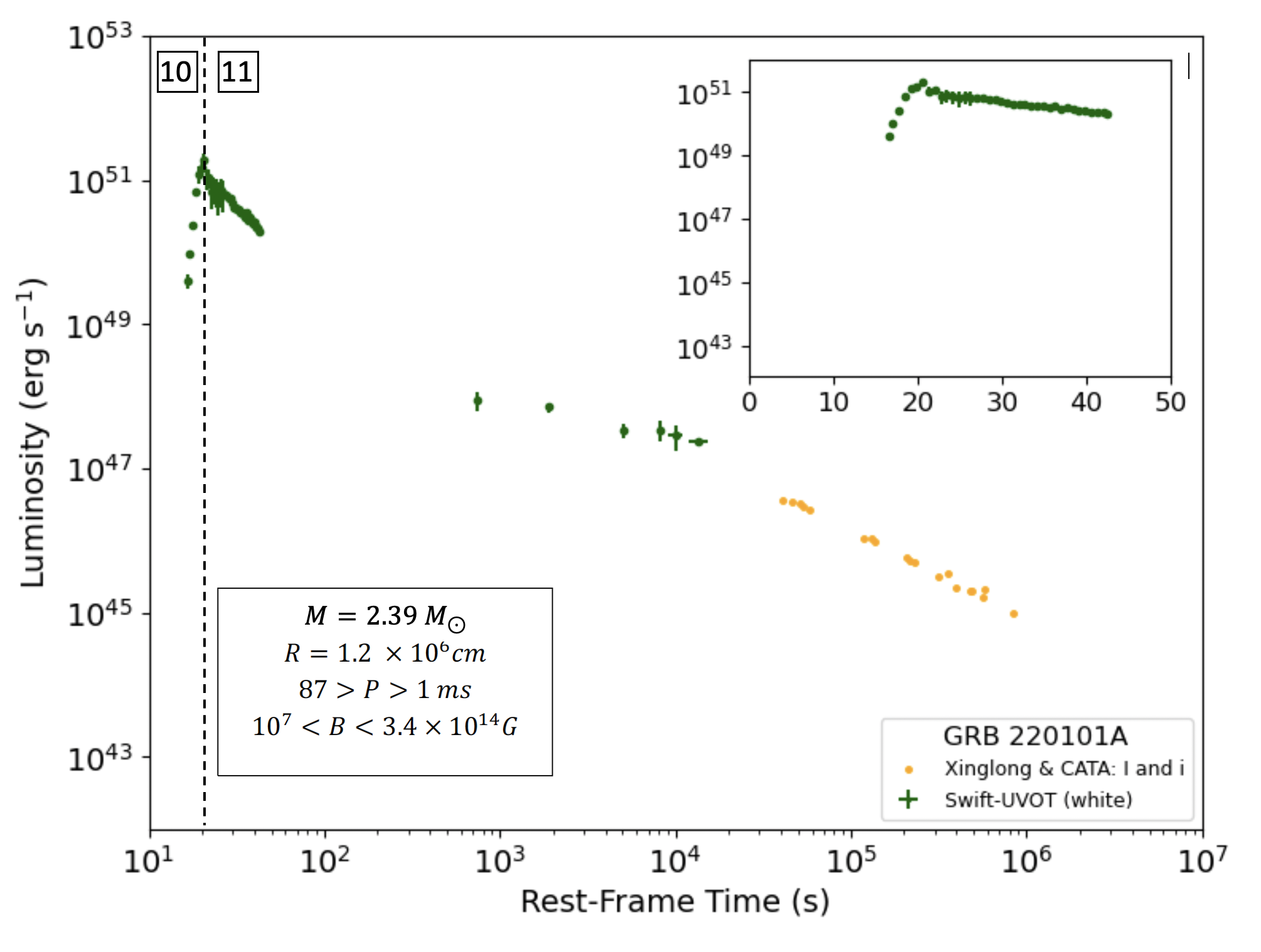}
\includegraphics[angle=0, width=0.8\hsize,clip]{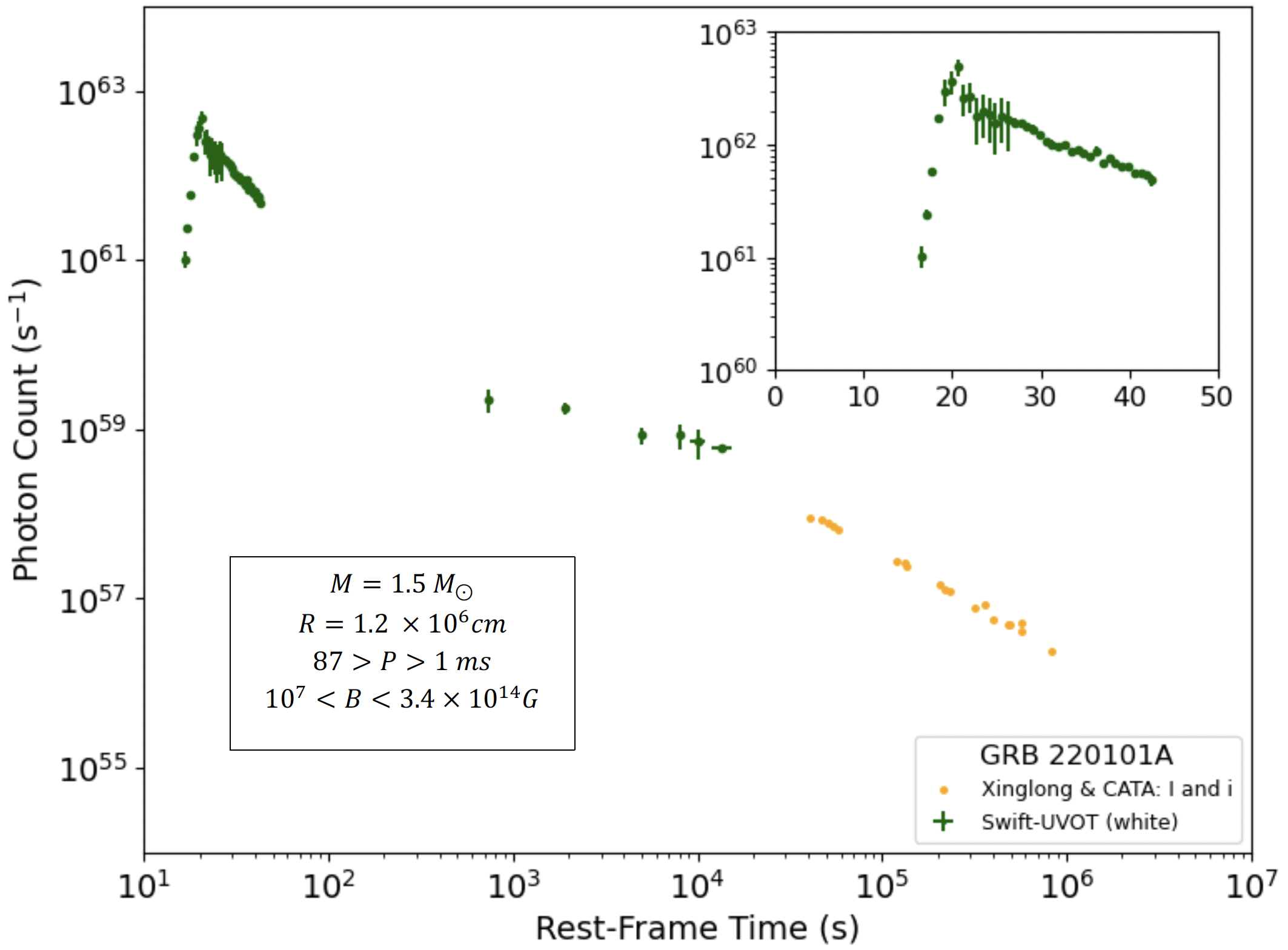}
\setlength{\unitlength}{1cm}
\begin{picture}(0,0)
    \put(-5.5,5.5){\makebox(0,0)[b]{\Large \textbf{V}}}
\end{picture}
\caption{The optical luminosity light-curve, composed of two slices 10 and 12, observed by the Swift-UVOT, Xinglong, and CAHA telescopes. The inset graphs are enlargement of early 50 second of the light curve. The rising part is the spin-up of the $\nu$NS, while the part after $20$~s corresponds to $\nu$NS spin-down. }
\end{figure*}
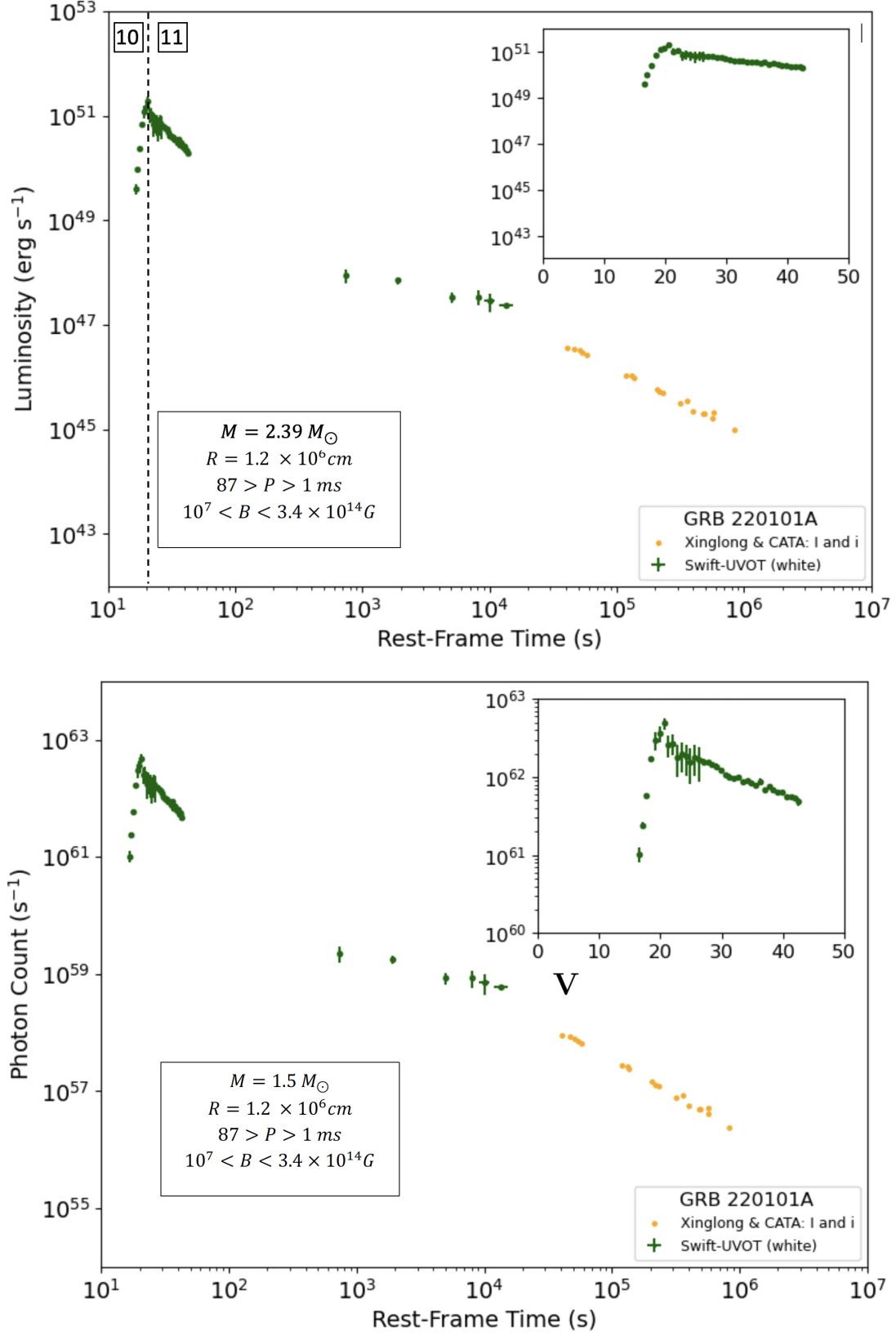

\begin{figure*}
\centering
\includegraphics[angle=0, scale=0.90]{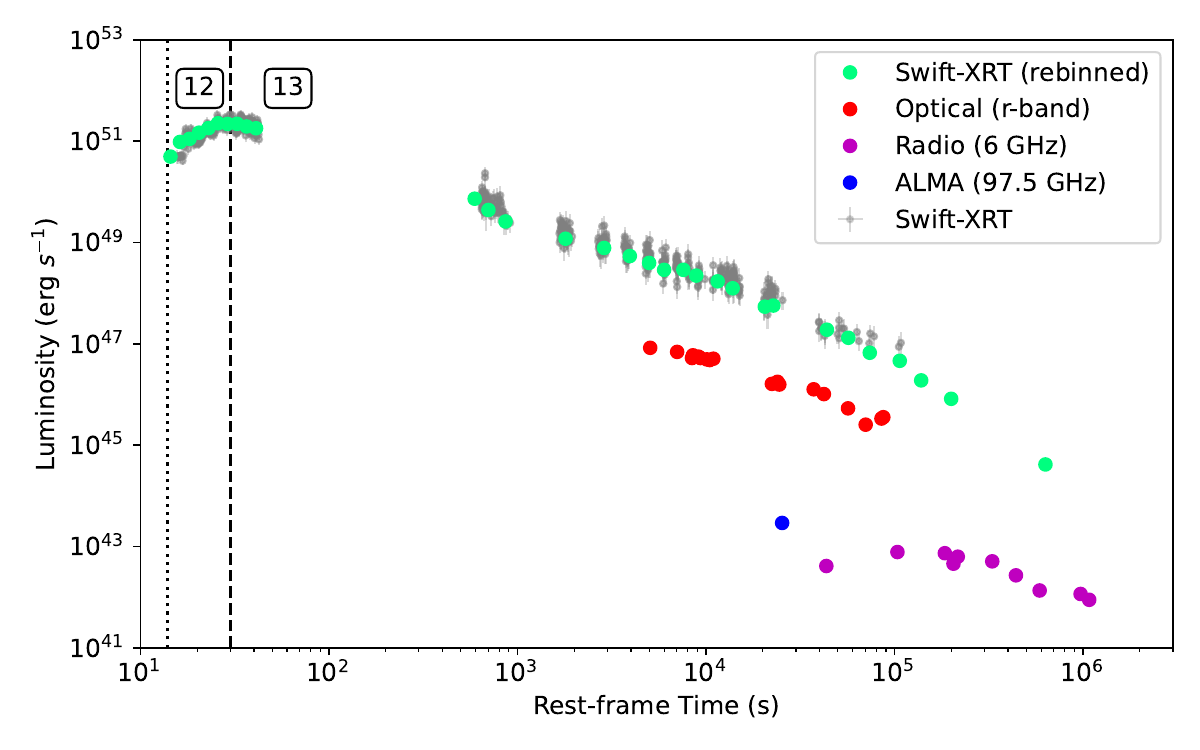}
    \setlength{\unitlength}{1cm}
    \begin{picture}(0,0)
        \put(2,5){\makebox(0,0)[b]{\large \textbf{VI}}}
    \end{picture}
\caption{The luminosity light-curve of the Swift-XRT, the time ranges are from $14$~s to $30$~s and from $30$~s till the end of the Swift-XRT observation ($\sim 10^5$)~s.  These are labeled as time slices 11 and 13, respectively.}\label{fig:xrt-lightcurve}
\end{figure*}

The Swift X-Ray Telescope is designed to observe the X-ray afterglows of GRBs and provide rapid follow-up observations of other high-energy astrophysical sources. The XRT is a focusing telescope that uses grazing-incidence mirrors to focus X-rays onto a detector. It has a field of view of approximately $23$ arcminutes and can detect X-rays in the energy range of $0.3$--$10$ keV.

The standard Swift analysis software included in NASA's Heasoft 6.31\footnote{\url{https://heasarc.gsfc.nasa.gov/docs/software/heasoft/}}, along with relevant calibration files, is used to analyze the XRT data obtained from the UK Swift Science Data Center (UKSSDC)\footnote{\url{https://www.swift.ac.uk}}. The data consisted of Windows Timing (WT) mode ranging from $87$~s to $3788$~s and Photon Counting (WT) mode ranging from $3807$~s to $1107247$~s. The spectral fitting and flux calculation are performed using \textit{XSPEC}. An absorption model multiplied by a power law is adopted as the spectral function. We notice that the early afterglow at the observer's time $\sim 100$~s has a bump structure, of which the photon index changes rapidly from $\sim 0.5$ to $\sim 2$. The variation of the photon index induces the change of the $k$-correction and consequently affects the shape of the light curve. In Figure \ref{fig:xrt-lightcurve}, the X-ray light curve has duly considered the time-resolved $k$-correction. 

Swift's observations are divided into two slices, the 9th and 10th slices (see Figure \ref{fig:xrt-lightcurve}):

\begin{itemize}
\item \textbf{Slice 12 ($14$ to $30$~s):} It represents the rising phase of X-rays. A single power-law model with index $1.22\pm0.02$ best fits the slice emission. Taking into account the redshift of $z=4.61$, the isotropic energy for soft X-rays released during this period is calculated as $E_{\rm iso}=(1.56\pm0.04) \times 10^{52}$ erg. See Section \ref{sec:earlyX} for the early soft X-ray emission.

\item \textbf{Slice 13 ($30$ s to $2\times10^5$ s):} It encompasses the declining phase of the multi-wavelength afterglow. The optimal model for the X-ray observations in this slice is also a power-law, of index $1.68\pm0.04$, along with a hydrogen column density of $B$. Considering the redshift $z=4.61$, the estimated isotropic energy of the soft X-rays in this slice is $E_{\rm iso}=(3.8\pm1.28) \times 10^{53}$ erg. 
\end{itemize}

\begin{figure*}[!t]
\centering
\includegraphics[angle=0, scale=0.8]{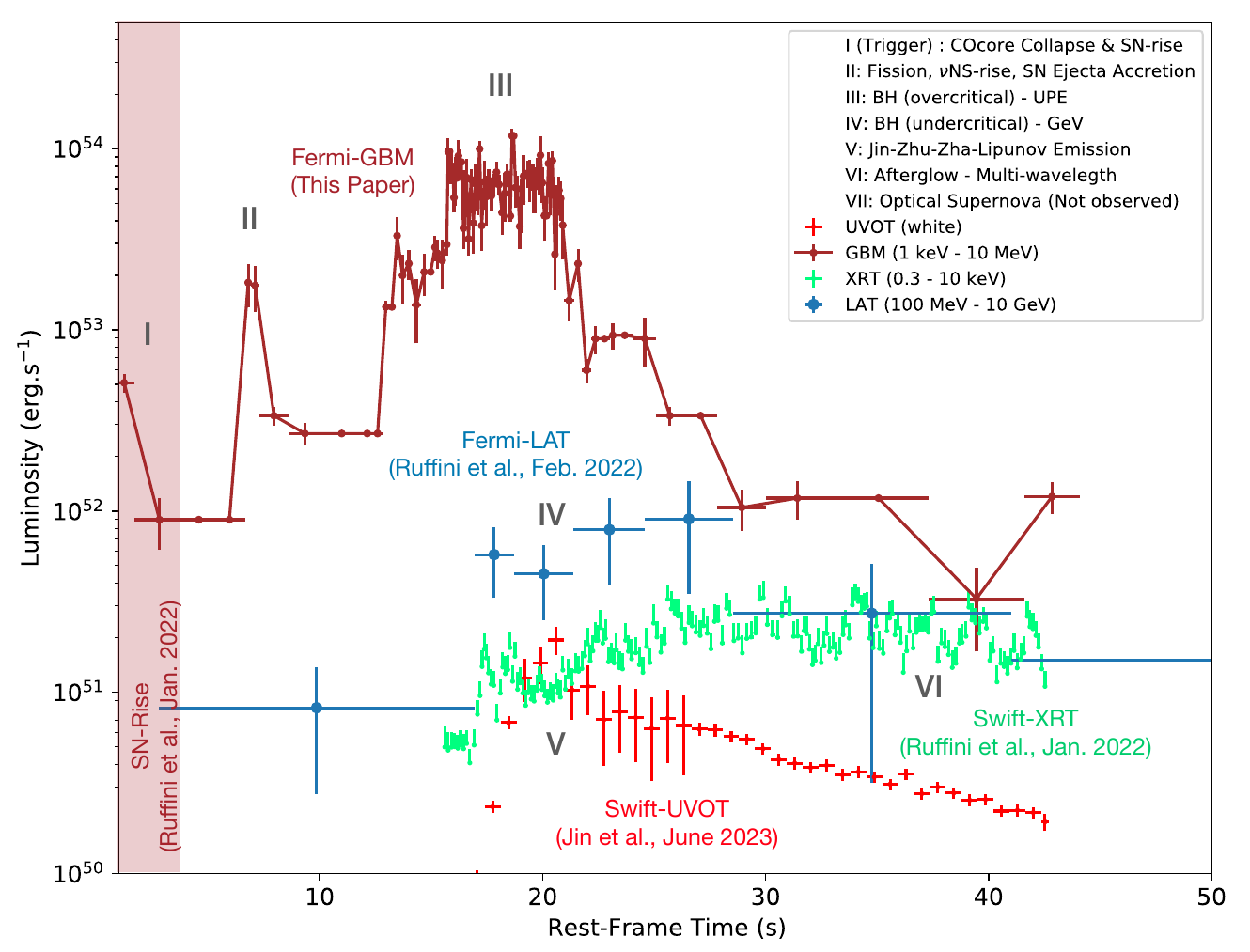}
\caption{Zoomed-in view of the multi-wavelength luminosity light curves for the first $50$ s (rest-frame time) of GRB 220101A. This plot overlays data from Swift-UVOT (white band, optical, bright red points), Swift-XRT ($0.3$--$10$ keV, X-ray, green points), Fermi-GBM ($1$ keV -- $10$ MeV, gamma-ray, red line with error bars), and Fermi-LAT ($100$ MeV -- $10$ GeV, high-energy gamma-ray, blue points with error bars). Luminosities are isotropic and k-corrected. Roman numerals (I-VI) indicate the approximate time periods corresponding to the first six emission Episodes identified within the BdHN model. Citations indicate the initial GCN reports or papers where specific components were identified or analyzed. The shaded pink area is the earliest SN-rise phase.}\label{fig:lc-linear}
\end{figure*}

\subsection{Other Satellites}

We also utilize data from multiple satellites and telescopes, as reported in available articles. \citet{2022ApJ...933..214U} presents the AGILE observations. Using data acquired by the onboard SuperAGILE, Anti-Coincidence, and MCAL, the AGILE satellite provides a broad-band overview of the temporal and spectral behavior of this GRB, spanning from a few tens of keV to tens of MeV and lasting over $100$ s within the AGILE field of view. The fittings in the $18$ keV -- $50$ MeV range do not require extra components beyond single Band or cutoff power-law models. \citet{2023ApJ...959..118Z} conducts the optical and infrared data analysis of a large collection of data, including the Xinglong $2.16$ m Telescope \citep{2016PASP..128k5005F}, the Ningbo Bureau Of Education and Xinjiang Observatory Telescope (NEXT), the Nordic Optical Telescope (NOT), the Centro Astronomico Hispano en Andalucıa (CAHA) $2.2$ m telescope, the Telescopio Nazionale Galileo (TNG), the Tautenburg $1.34$ m Schmidt Telescope \citep{2016SPIE.9908E..4US}, the $10.4$ m Gran Telescopio CANARIAS (GTC). All the data are duly calibrated, and the extinctions are corrected. The obtained light curve spans the g, r, i, z, J, H, and K bands, and covers an approximate time window from $0.2$ to $30$ days.

\section{Unique Opportunity of Observing Energetic GRB at High Redshift}\label{sec:unique}

\subsection{Fermi-LAT Detection of Highest Redshift}\label{sec:fermi-lat}

From the statistics of $10$-year Fermi-LAT GRBs \citep{2019ApJ...878...52A}, $169$ bursts detected photons above $100$~MeV, a small fraction ($\sim 5\%$) of the total $3381$ detected by Fermi-GBM, Swift-BAT, INTEGRAL, and IPN. Within these $169$ bursts, only $\sim 20\%$ bursts have observed photons more than $1$ GeV, and the photon number is usually as low as $<10$ photons in each GRB.

For a GRB at a very high redshift, e.g., GRB 090423 at $z=8.1$ \citep{2009Natur.461.1258S}, the $\sim 100$~MeV and $\sim 1$~GeV photons in the rest-frame are shifted to $\sim 10$~MeV and $\sim 100$~MeV in the observer's frame, respectively. Their fluxes are reduced by one to two orders of magnitude due to the distance (e.g., for the same burst energetics, the $z=8$ to $z=2$ received flux ratio is $0.04$). Therefore, first, most of the intrinsic $\sim 100$~MeV photons will be shifted to $\sim 10$~MeV and beyond the Fermi-LAT energy band ($100$ MeV to $100$ GeV); second, the intrinsic $> 1$~GeV photons will have a meager chance to be detected because of the very low fluence. The lack of high-energy photon observations of GRB 090423 could be due to either of these two reasons. The redshifted very high-energy photons may fall within the Fermi-GBM energy range, but Fermi-GBM has an effective area that is more than 100 times smaller than that of LAT. As a result, Fermi-GBM is nearly incapable of detecting those redshifted photons.

GRB 220101A also has a high redshift of $z=4.61$. Still, compared to GRB 090423, this redshift is not extreme enough to shift most of the very high-energy photons out of the observational bandwidth of the Fermi-LAT, but rather at the lower energy edge. Moreover, the isotropic energy of GRB 220101A is about 70 times that of GRB 090423. Typically, GRBs with higher isotropic energy produce more high-energy photons, which increases the probability that Fermi-LAT will detect them. Indeed, LAT did observe several tens of photons. The calculated energy in the range of 100 MeV to 10 GeV reached $E_{\rm LAT} = (6.65\pm 1.69)\times 10^{53}$ erg, which is more than an order of magnitude greater than typical GRBs.

In summary, firstly, the high-energy redshifted photons of GRB 220101A are still within the observable energy range of Fermi-LAT. Secondly, its extremely high energy ensures sufficient flux despite its far distance. Statistically speaking, the second catalog of Fermi-LAT GRBs includes $34$ redshift samples \citep{2019ApJ...878...52A}, and the redshift of $4.61$ for GRB 220101A is higher than all of these $34$ samples.

\subsection{Fermi Detection of Complete Episodes}\label{sec:complete-episodes}

For Fermi-GBM observations in the keV-MeV range, the source's observed flux is reduced at high redshifts, which may affect detection. During periods of weak emission, background photons can overshadow the signal, leaving only the more intense emission phases detectable. As shown in \citep{2013ApJ...765..116K}, the observed $T_{90}$ decreases for redshifts above $z \sim 4$. However, GRB 220101A is an exception. With a redshift $z=4.61$, its luminosity distance is about three times that of a typical GRB at $z=2$, which would generally reduce its flux by nearly an order of magnitude. Despite this, its energy is two orders of magnitude higher than that of a typical GRB, compensating for the reduced flux due to the distance and thus making it a bright outburst. Further, due to the cosmological time dilation, the observed duration of GRB 220101A is stretched by a factor of $1 + z$, allowing emissions from various origins during different Episodes to be distinguished and analyzed \citep{2024ApJ...966..219B}, summarized in Table \ref{tab:Summary1}.

\subsection{Early Soft X-ray Observation}\label{sec:earlyX}

Due to operational constraints, Swift-XRT is typically unable to detect X-ray emission during the first few tens of seconds after a GRB trigger. This limitation arises because the Swift satellite requires approximately 10 to 20 seconds to detect a Swift/BAT trigger, calculate the source coordinates, evaluate the feasibility of slewing to those coordinates, and begin slewing. The slewing itself takes an additional 20 to 75 s \footnote{See, e.g., E. Troja, The Neil Gehrels Swift Observatory Technical Handbook Version 17.0: \url{https://swift.gsfc.nasa.gov/proposals/tech_appd/swiftta_v17.pdf}.}. Consequently, X-ray emission occurring within the first 40 s following the GRB trigger remains largely unobserved by Swift/XRT.

The cosmological time dilation effect is crucial in this context. A time interval $\Delta t$ in the source frame corresponds to a longer interval $(1 + z) \Delta t$ observed on Earth. Thus, short intrinsic rest-frame durations (typically $\sim 20$~s) at high redshifts appear as longer observed durations (up to $\sim 100$~s). Consequently, high-redshift GRBs offer a unique opportunity to study transient X-ray activity, as the observed duration can exceed the Swift-XRT slewing time. For GRB 220101A, with a high redshift of $z = 4.61$, XRT began collecting data at $87.47$ s after the trigger. Correspondingly, in the rest frame, it started observing X-rays at $15.59$ s, as shown in Figure \ref{fig:xrt-lightcurve}. 

The soft X-ray observation by Swift-XRT could be the least affected by the high redshift. Because the afterglow X-ray spectrum follows a power-law spanning from $\sim 0.1$ keV to tens of keV, the spectrum is still within the XRT energy bandwidth ($0.3$--$10$ keV) after the redshift expansion. Second, XRT is very sensitive, and it is capable of resolving the flux that is lowered several orders of magnitude from typical GRBs till $10^{-14}$ erg~cm$^{-2}$s$^{-1}$ \citep{2007A&A...469..379E}, as the XRT observation of GRB 220101A extends to $\sim 10^6$~s in the observer frame. 

Observing soft X-rays in the prompt emission provides information about the properties of the GRB's central engine and helps constrain the GRB emission mechanism. In the BdHN model, the rise of the soft X-ray emission indicates the rise time and the $\nu$NS properties. 

 \section{Seven Episodes of BdP-N Model}\label{sec:episodes}
 
As noted in the Introduction, the original BdHN model was introduced and tested on a sample of 25 GRBs associated with core-collapse SNe that evolved through 7 distinct episodes. The progenitor of such a BdHN is a binary composed of a $10\,M_{\odot}$ CO core and a companion NS with a characteristic orbital period from minutes to hours. The GRB is triggered by the collapse of the CO core in the SN event, leading to the formation of a $\nu$NS. The further evolution of the system involves three components: the expanding CO core ejecta, the companion NS undergoing accretion and collapsing into a BH, and the $\nu$NS, which accretes matter and evolves into a fast-spinning NS. As summarized in Table~1 of \citet{2023ApJ...955...93A}, the isotropic energy $E_{\mathrm{iso}}$ of these GRBs spans from $10^{47}$ erg up to typical values of $10^{52}$--$10^{53}$ erg. The 7th episode is associated with the emission of nickel in the expanding SN remnant.

The case of GRB~220101A was markedly different from the outset, with an isotropic energy exceeding $10^{54}$ erg. While such extreme energetics are not unique, they appear to be shared by events such as GRB~130427A and GRB~240825A. One of the main objectives is therefore to identify distinctive signatures of this emerging high-energy class, including its specific physical processes and evolutionary pathways, and to systematically classify these events into a coherent family. We aim to minimize the number of additional components required by the new observations. At the same time, we retain the binary nature of the companion system, while revising the nature of the third component in the BdHN scenario: instead of an NS, we consider a WD of $\sim 1.4\,M_\odot$. We proceed by estimating the properties of the CO core and its gravitational collapse, which triggers the GRB event. As in the previous BdHN model, the GRB emission is determined by the accretion process following the first SN explosion onto the companion and other system components. However, in this revised framework, a key difference emerges from the presence of the WD. Following the explosion of the CO core, the energy released by the core-collapse SN (SN~I) is $\sim 1.2 \times 10^{53}$ erg, significantly larger than the typical $\sim 10^{52}$ erg expected for non-rotating cores. A crucial aspect of this scenario is accretion onto the WD, which triggers a new phenomenon: the WD's induced gravitational collapse, leading to a second SN (SN~II). This second explosion is less energetic than SN~I and results in the formation of a rapidly rotating $\nu$NS, which is further spun up by the ejecta from SN~I. The onset of accretion onto the WD, beginning at $\sim 1$ s after SN~I, represents a defining characteristic of this more energetic class and leads to profound changes in the observational signatures. While the number of emission episodes remains consistent with the previous BdHN model, they are now governed by different physical mechanisms and energetics.

As shown by the data presented in the previous paragraphs, there is a clear difference between GRB~220101A and the previous BdHNe originating from core collapse SNe. A major difference is the type of the pair SN triggering these most energetic GRBs. The acceleration process acting in GRBs is the pair self accelerating process \citep{1973PhRvL..31.1362R,Ruffini:1975ne} and the basic process being the Homi Baba scattering. (see eg. Wikipedia). in his memory we call these new type of most powerful SN  an HB SN., which points to a new high-energy electro-dynamical process that we propose originates from a magnetized, fast-rotating core endowed with a magnetic field of $10^{6}$~G; see, for example, Ruffini et al. (2026) (submitted). We proceed by building on the successful BdHN model to extend its applicability to this energetic system with a minimal number of new components, as requested by the new observational results.  In view of the above, this new kind of SN is driven by a highly magnetized CO core of $10 M_\odot$. We are proceeding to identity the properties of this system from direct observations, and quantify their energetic requirements, the collapse of the magnetized CO core does not leave any remnant and generate energetic relativistic ejecta accreting onto all remaining components of the BdHN process.

The ejecta from the first SN play a central role in driving the subsequent evolution: its impact induces the collapse of the WD of $1.4$~M$_\odot$ at $3.57$ s, leading to the second SN with an energy of about $10^{53}$ erg in photons, and possibly more in neutrinos, and resulting in the formation of a $\nu$NS at $\sim 17$~s. The same ejecta are subsequently accreted onto the $\nu$NS, spinning it up from $\sim 60$~ms to the $\sim 1$~ms. Simultaneously, part of the ejecta is also accreted by the companion NS, releasing a large amount of energy and giving rise to the UPE. This continued accretion drives the companion NS beyond its critical mass, leading to its collapse into a BH, which subsequently produces GeV emission.

\begin{table*}[hbtp]
\centering
\caption{Episodes and afterglows of GRB 220101A. This table reports the Episode name (Ep.), time slices (Slice), start -- end in seconds, best-fit spectrum, isotropic energy ($E_{\rm iso}$ in erg), and underlying astrophysical processes for GRB 220101A. The redshift is $z=4.61$, and the total isotropic energy is $E_{\rm iso}=4 \times 10^{54}$ erg.}\label{tab:Summary1}
\small
\setlength{\tabcolsep}{4pt}
\begin{tabular}{l c l l c c l}
\toprule
Ep. & Slice & Event & Duration (s) & Spectrum & $E_{\rm iso}$ (erg) & Physical Phenomena \\
\midrule
I & 1 & 1st Explosion & -0.18 -- 3.57 & CPL & $(1.2 \pm 1.0) \times 10^{53}$ & Collapse of CO Core\\
\midrule
II & 2 & WD Collapse & 3.57 -- 11.41 & CPL & $(2.1 \pm 0.9) \times 10^{53}$ & Second SN rise\\
   & 3,5,7,8 & Accretion& 11.41 -- 28.52 & Band & $(2.4 \pm 1.5) \times 10^{54}$ & \\
\midrule
III & 4 & UPE I & 15.15 -- 16.04 & CPL+BB & $(2.4 \pm 5.1) \times 10^{53}$ & Energy from self-similar \\
    & 6 & UPE II & 18.72 -- 19.61 & CPL+BB & $(4.6 \pm 1.7) \times 10^{53}$ & and QED process \\
\midrule
IV & 9 & BH-rise \& Growth & $\sim$10 -- $\sim$20 & PL & $(2.1 \pm 0.3) \times 10^{53}$ & Accretion on companion NS \\
   & 9 & GeV Afterglow & $\sim$20 -- 100 & PL & $(4.6 \pm 1.4) \times 10^{53}$ & BH spindown \\
\midrule
V & 10 & $\nu$NS Spinup & 17 -- 21 & -- & $(2.6 \pm 0.5) \times 10^{51}$ & $\nu$NS as a millisecond pulsar \\
\midrule
VI & 12, 13 & X-ray Afterglow & 16 -- $2 \times 10^{5}$ & PL & $(4.0 \pm 1.3) \times 10^{53}$ & Powered by SN ejecta \\
\midrule
VII  & 11 & Optical Afterglow & 21 -- $3 \times 10^{5}$ & -- & $(5.3 \pm 1.8) \times 10^{52}$ & Pulsar \\
\bottomrule
\end{tabular}
\end{table*}

We now proceed to a more detailed description of the seven episodes characterizing this more energetic family of BdHNe, and summarized in Table~1.

\noindent \textbf{Episode I: First SN explosion (Slice 1)}  
The event begins with the collapse of the magnetized, rapidly rotating CO core, whose evolution is strongly affected by tidal interactions with the NS companion in a short binary period. The trigger of the GRB with an initial pulse releasing the initial MeV pulse, which is more energetic than the SN of the traditional BdHN. This episode is observed in Slice~1, lasts $3.75$~s ($-0.18$--$3.57$~s), is best fitted by a CPL spectrum, and releases an isotropic energy of $(1.2 \pm 1.0)\times10^{53}$ erg.

\noindent \textbf{Episode II: WD collapse and SN ejecta accretion (Slices 2, 3, 5, 7, 8)}  
Following the first SN explosion, the ejecta interact with the remaining components of the GRB system. Slice~2 ($3.57$--$11.41$~s) is associated with the collapse of the WD, a process dominated by neutrino emission, observed with a CPL spectrum and an energy of $(2.1 \pm 0.9)\times10^{53}$ erg, characterizing the second SN. The subsequent slices~3, 5, 7, and 8 trace the continued accretion of the first SN ejecta onto the $\nu$NS formed from the WD collapse. These phases, described by Band and CPL spectra, correspond to the progressive transfer of mass and angular momentum, with isotropic energies of $(5.0 \pm 1.1)\times10^{53}$ erg, $(1.1 \pm 0.1)\times10^{54}$ erg, $(5.3 \pm 0.5)\times10^{53}$ erg, and $(3.0 \pm 1.0)\times10^{53}$ erg, respectively. 

\begin{figure*}[!t]
\includegraphics[angle=0, scale=0.35]{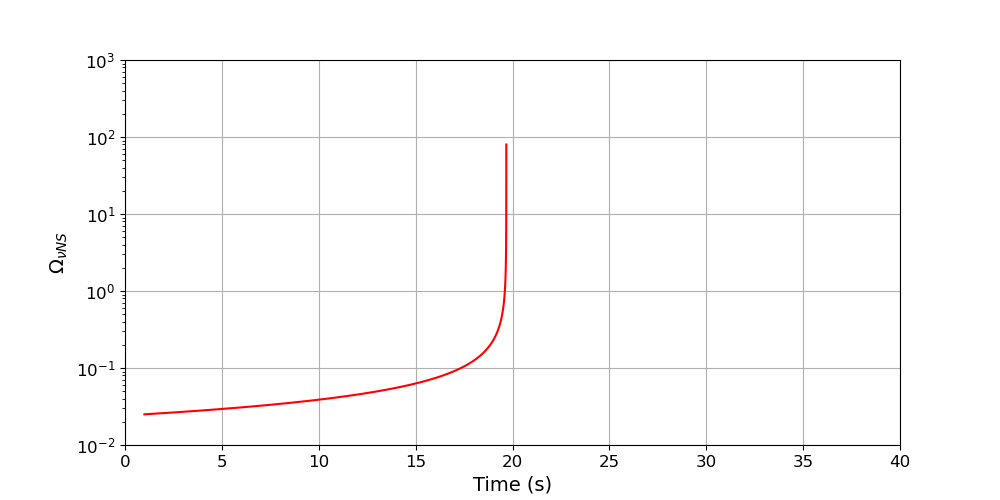}
\includegraphics[angle=0, scale=0.35]{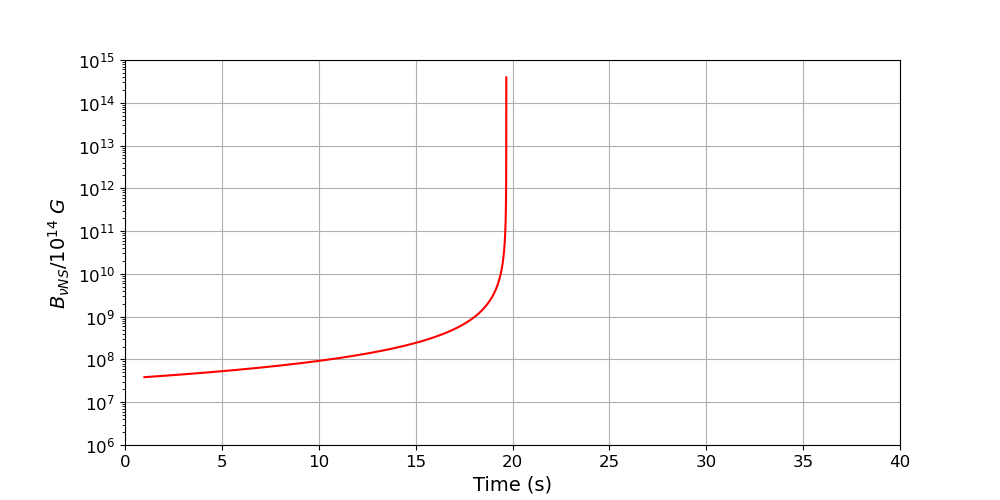}
\caption{Angular velocity ($\Omega_{\nu \rm NS}$) and  magnetic field ($B_{\nu \rm NS}$),  of 1.4 $M_\odot$ WD collapsing into a $\nu$NS vs. collapse time.}
\label{fig:specra2}
\end{figure*}

\begin{figure*}[!t]
\includegraphics[angle=0, scale=0.35]{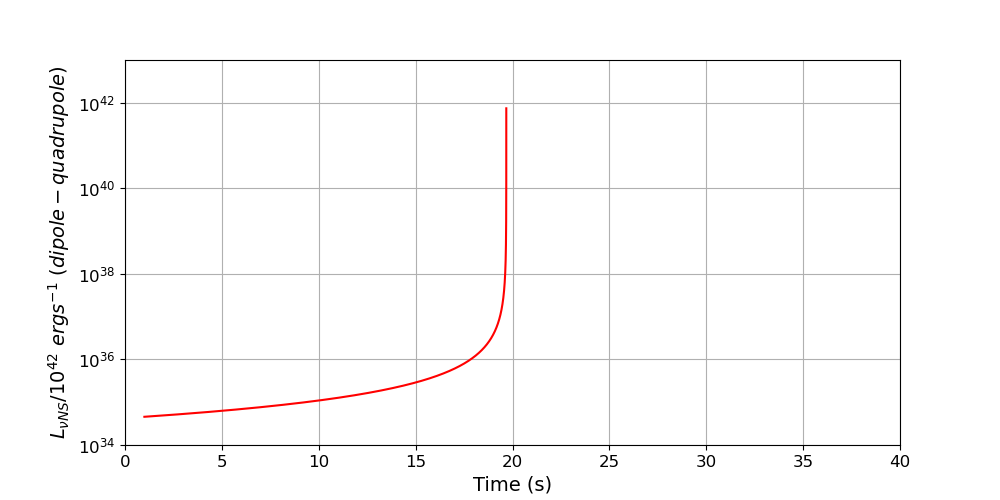}
\includegraphics[angle=0, scale=0.35]{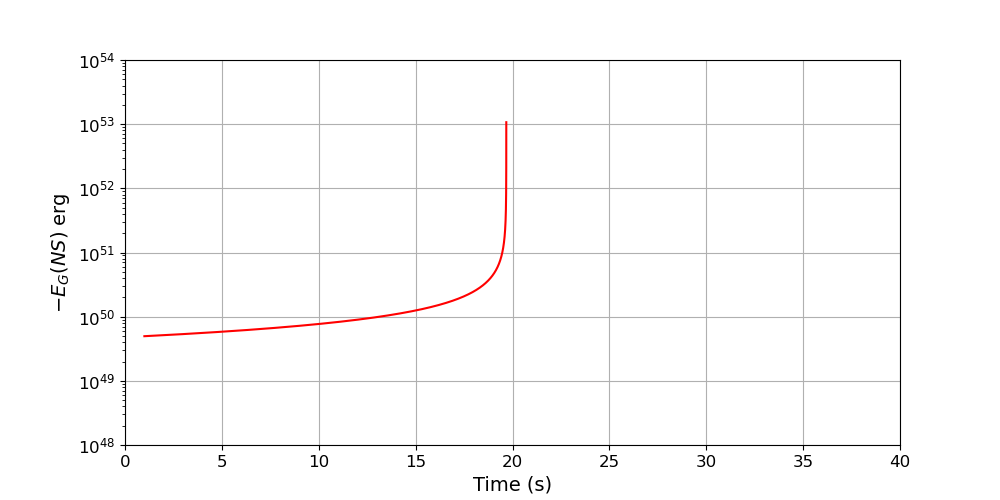}
\caption{Dipole luminosity ($L_{\nu \rm NS}$), and released gravitational energy ($E_g$ ($\nu$ NS) of 1.4 $M_\odot$ WD collapsing into a $\nu$NS vs. collapse time.}
\label{fig:specra3}
\end{figure*}

To quantitatively investigate the formation of the newborn neutron star during Episode~2, we model the gravitational collapse of the WD by assuming a free-fall radial evolution from the onset of dynamical instability until the compact remnant reaches its final degenerate configuration with a radius of $1.2\times10^{6}$~cm. Throughout the collapse, both angular momentum and magnetic flux are assumed to be conserved, providing a self-consistent framework for following the evolution of the stellar angular velocity, rotational period, magnetic field, gravitational binding energy, and electromagnetic luminosity. The dramatic reduction in the stellar radius leads to a corresponding decrease in the moment of inertia and, consequently, a rapid increase in the angular velocity through angular momentum conservation. Simultaneously, magnetic flux freezing amplifies the surface magnetic field by several orders of magnitude as the stellar surface contracts. This treatment enables us to determine the temporal evolution of the principal physical quantities governing the collapse and establishes a direct connection between the dynamical formation of the neutron star and the observed high-energy emission. As shown in Figure \ref{fig:specra2} and \ref{fig:specra3}, the collapse releases a total gravitational binding energy of $1.07\times10^{53}$~erg, in excellent agreement with the isotropic energy emitted during Episode~2, Slice~2, $(2.1\pm0.9)\times10^{53}$~erg, thereby providing strong evidence that this emission originates from the collapse of the WD and the birth of the $\nu$NS.

The collapse alone produces a rapidly rotating neutron star with a final spin period of approximately $78$~ms as a direct consequence of angular momentum conservation. At the same time, magnetic flux freezing amplifies the surface magnetic field to $3.9\times10^{14}$~G, substantially exceeding the quantum critical magnetic field, $B_{\rm c}=4.4\times10^{13}$~G, and placing the newborn object in the ultra-strong magnetic-field regime. Such conditions naturally provide the physical environment required for the subsequent high-energy evolution of the system.

\vspace{1em}
\noindent \textbf{Episode III: Ultra-relativistic Prompt Emission (UPE; Slices 4 and 6)} The two short-duration pulses observed in Slices~4 and 6 correspond to UPE~I and UPE~II. These emissions arise during the accretion process onto the $\nu$NS, when the system develops overcritical electromagnetic fields. Under these conditions, an $e^{+}e^{-}$ pair plasma is created via the QED Euler-Heisenberg process and undergoes repeated transparency episodes, producing the observed CPL+BB spectral structure. Slice~4 ($15.15$--$16.04$~s) and Slice~6 ($18.72$--$19.61$~s) last $0.89$~s each and release $(2.4 \pm 5.1)\times10^{53}$ erg and $(4.6 \pm 1.7)\times10^{53}$ erg, respectively. This represents one of the most significant analogies between the traditional BdHN model and the new, most energetic sources. In all cases, this leads to a fractal behavior, as described in detail in \citet{2021PhRvD.104f3043M,2022EPJC...82..778R}.

\vspace{1em}
\noindent \textbf{Episode IV: BH rise, growth, and GeV afterglow (Slice 11)}  
The formation of the BH is traced by the onset of the GeV emission. In the interval from $\sim 10$ to $\sim 20$~s, the continued accretion onto the companion NS of the first SN ejecta leads to the rise and growth of the BH, releasing an energy of $(2.1 \pm 0.3)\times10^{53}$ erg. Subsequently, from $\sim 20$ to $100$~s, the source enters the GeV afterglow phase, characterized by a power-law spectrum and powered by the spin-down of the BH, with an additional energy release of $(4.6 \pm 1.4)\times10^{53}$ erg.

\begin{figure}
\includegraphics[angle=0, scale=0.35]{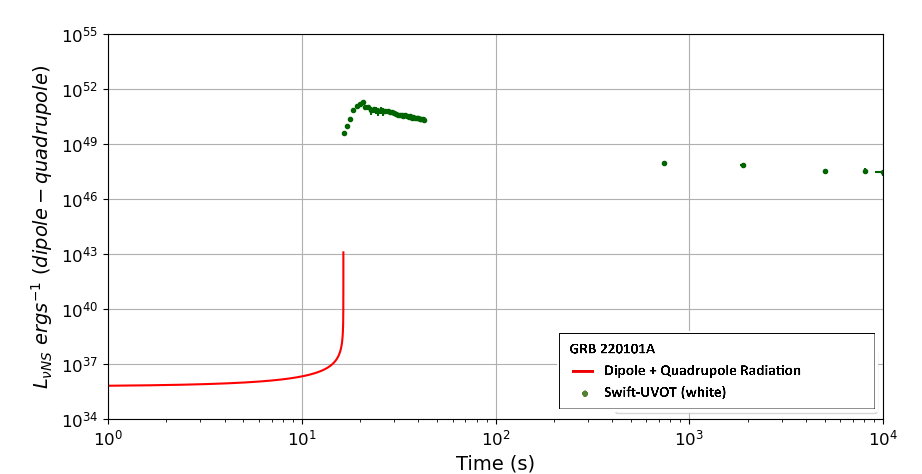}
\caption{Dipole and quadrupole luminosity of the pulsar during its formation (Event I) from Maclaurin ellipsoid with a magnetic field of $10^6$ G to a NS with $B = 10^{15} $ G. Green dots are Swift-UVOT observation. }
\label{fig:collapse}
\end{figure}

\vspace{1em}
\noindent \textbf{Episode V: $\nu$NS spin-up (Slice 10)}  
A second SN starting from $3.57$~s after the initial supernova. This explosion is triggered when the first SN ejecta destabilize the WD, leading to its collapse and the formation of a $\nu$NS. The following process corresponds to Slice~10, from $\sim 17$ to $21$~s, with an energy of $(2.6 \pm 0.5)\times10^{51}$ erg. During this phase, the $\nu$NS undergoes rapid spin-up due to accretion, reaching millisecond periods, as illustrated by the collapse luminosity and multipole-emission diagnostics in Figures \ref{fig:specra2}, \ref{fig:specra3} and ~\ref{fig:collapse}

The rapid collapse of WD to $\nu$NS marks only the beginning of the rotational evolution. As the ejecta from the first supernova continue to accrete onto the newly formed $\nu$NS, a substantial amount of mass and angular momentum is transferred to the compact object. This accretion-driven torque efficiently spins up the neutron star, progressively reducing its rotational period from $78$~ms to approximately $1.3$~ms. In our model, this transition is completed about $20$~s after the first supernova explosion, coinciding with the epoch of the intense prompt high-energy emission. The combined effects of gravitational collapse, magnetic-field amplification, and sustained angular momentum accretion therefore provide a coherent physical scenario that links the energetics of Episode~II with the subsequent evolution of the central engine in Episode~V. The resulting millisecond, highly magnetized neutron star constitutes the energy reservoir responsible for powering the later phases of the burst through continued rotational energy extraction and interaction with the expanding supernova ejecta. The dynamics of the neutralization process by the SN I ejecta are calculated in Ruffini et al. (2026, submitted); see Figures \ref{fig:collapse}.

\vspace{1em}
\noindent \textbf{Episode VI: X-ray afterglow (Slices 12 and 13)}  
The X-ray afterglow originates from the interaction of the expanding SN ejecta with the surrounding medium. The early rising phase (Slice~9) and the long decay phase (Slice~10) extend from about $16$~s to $\sim 2\times10^{5}$~s in the rest frame. This emission is well described by a power-law spectrum and carries a total isotropic energy of $(4.0 \pm 1.3)\times10^{53}$ erg.

\begin{table*}
    \centering
    \begin{tabular}{c|c|c|c|c|c} \hline
        & $L_{bol} (erg s^{-1}) $ & $L_X (erg s^{-1}) $ & $L_{opt} (erg s^{-1})$ & $Period \  (ms)$ & $d \Omega/dt$ \\
       \hline
        Crab (today) & $7.4 \times 10^{38} $ (a)& $1.5 \times 10^{38}$ (d) & $9 \times 10^{32}$ (b) & $33$ (a) & $-2.43 \times 10^{-9}$ (a)\\
       220101A (extraploated to $3 \times 10^{10}$ s & $1.2 \times 10^{38}$ (c) & $1.1 \times 10^{38}$ (c)  & -  & $56.7$  (c) & $-8.78 \times 10^{-10}$ (c) \\
        \hline
    \end{tabular}
    \caption{Comparison of 220101A extrapoloted to the crab age ($ 3 \times 10^{10} $ s) with crab nebula today. Refs: (a) \cite{1969ApJ...158L..71F}, (b) \cite{Quadir01}, (c) this study, (d) \cite{2008ARA&A..46..127H}. {\bf We consider this results as cosmic relevance.}}\label{crab}
\end{table*}

\begin{figure*}
\centering
\includegraphics[angle=0, scale=0.8]{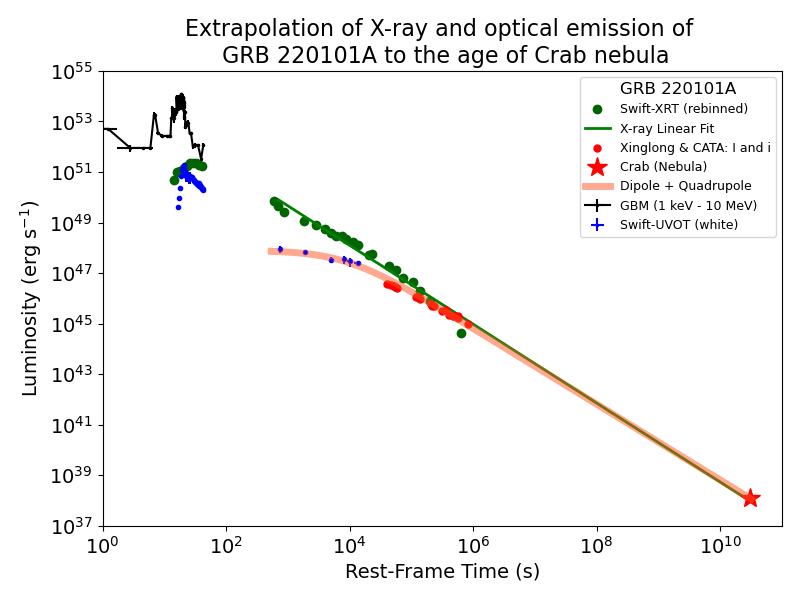}
\caption{X-ray and optical emission fit of 220101A extrapolated to $3 \times 10^{10}$ second (age of crab pulsar). from the spin-down energy from the $\nu$NS. The central compact object formed in GRB 220101A is predicted to evolve into a pulsar with a spin period of $56.7$ ms after $3 \times 10^{10}$ second.}\label{magplot}
\end{figure*}

\vspace{1em}
\noindent \textbf{Episode VII: Optical afterglow from the $\nu$NS/pulsar (Slice 11)}  
The optical afterglow is powered by the newly formed $\nu$NS. This emission, corresponding to Slice~13, extends from about $21$~s up to $\sim 3\times10^{5}$~s and releases an energy of $(5.3 \pm 1.8)\times10^{52}$ erg. After the initial spin-up phase, the emission is sustained first by magnetic energy release and then by rotational spin-down via dipole and multipole radiation, resulting in a smooth power-law decay in the light curve. The spin evolution and optical fit are shown in Figure~\ref{magplot}.

The $\nu$NS continues to inject energy, contributing to the X-ray, optical and radio afterglows. It carries substantial rotational energy, given by
\begin{equation}
E_{\text{rot}} = \frac{1}{2}I \Omega^2,
\end{equation}
where $\Omega$ is the angular velocity and $I$ is the moment of inertia. Assuming the emitted energy is driven by both dipole and quadrupole magnetic fields, the luminosity, powered by the rotational energy loss, is expressed as 
\begin{align}
L_{\text{NS}} &= -\frac{dE_{\text{rot}}}{dt} = -I \Omega \dot{\Omega} \\
&= \frac{2}{3c^3} \Omega^4 B_{\text{dip}}^2 R_{\text{NS}}^6 \sin^2 \chi_1 \left(1 + \eta^2 \frac{16}{45} \frac{R_{\text{NS}}^2 \Omega^2}{c^2} \right),
\end{align}
where
\begin{equation}
\eta^2 = (\cos^2 \chi_2 + 10\sin^2 \chi_2) \frac{B_{\text{quad}}^2}{B_{\text{dip}}^2}.
\end{equation}
where $B_{\text{dip}}$ and $B_{\text{quad}}$ are the dipole and quadrupole magnetic field strengths, and $R_{\text{NS}}$ is the $nu$NS radius. The angles $\chi_1$ and $\chi_2$ measure the inclinations of the magnetic moment components. The parameter $\eta$ quantifies the relative contribution of the quadrupole field to the total magnetic field, affecting the higher-order corrections in the luminosity. For details of the equations, we refer to  \citet{2018ApJ...869..101R,2019ApJ...874...39W,2022ApJ...936..190W,2022ApJ...939...62R,2023ApJ...945...95W,2022PhRvD.106h3002B, 2024ApJ...974...89W} and references therein.

We assume the $\nu$NS has $1.4 M_\odot$, radius $R_{\text{NS}} = 2 \times 10^6$~cm, and the inclination angles are chosen to have the maximal luminosity. From the fitting, we infer $B_{\text{dip}} = 6 \times 10^{13}$ G and $B_{\text{quad}} = 8 \times 10^{14}$ G. The spin period of the $\nu$NS evolves from $1.3$ ms to $56.7$ ms after $3 \times 10^{10}$ s comparable to the age of the crab nebula. The fitting in Figure \ref{magplot} shows the observed luminosity from Swift-UVOT, Xinglong, and CAHA telescopes aligns well with the combined dipole and quadrupole contributions.

As shown in Figure \ref{magplot}, the extrapolation of both the dipole and quadrupole spin-down models to an age of $3 \times 10^{10}$ s reproduces remarkably well the present-day luminosity of the Crab Nebula. This agreement provides an important consistency check on the long-term evolution predicted by our model. Table \ref{crab}  compares the extrapolated bolometric, X-ray, and optical luminosities of GRB 220101A at $10^{10}$ s with the corresponding observed luminosities of the Crab Nebula at its current age. The comparison shows that the predicted bolometric and X-ray luminosities are of the same order of magnitude as those measured for the Crab system, indicating that the late-time evolution of GRB 220101A converges toward the luminosity regime of a young pulsar-powered nebula.

The central compact object formed in GRB 220101A is predicted to evolve into a pulsar with a spin period of $56.7$ ms, substantially longer than the current spin period of the Crab pulsar. This difference is naturally explained by the lower rotational energy reservoir available in the remnant of GRB 220101A. In particular, the extrapolated bolometric luminosity of GRB 220101A at an age of $3 \times 10^{10}$ s is approximately a factor of six lower than that of the Crab Nebula. Since the spin-down luminosity is directly powered by the loss of rotational kinetic energy, a lower bolometric luminosity implies a reduced spin-down power and consequently a more slowly rotating neutron star. Therefore, the predicted longer spin period of the GRB 220101A remnant is a natural consequence of its lower late-time energy output and is fully consistent with the expectations of pulsar spin-down evolution.

\section{Black Hole Mass and GeV emission}\label{sec:upe-bh-gev}

\subsection{BH mass and spin growth: rising GeV emission}\label{sec:bh-angular-momentum}

Once the angular momentum excess is released, the BH can form and grow by accreting the matter still bound. We identify the BH formation instant with the starting time of the GeV emission, which occurs at the rest-frame time $t_0 \approx 10$ s (see Figure \ref{fig:lat-photon}).

In the initial BH growth phase, the evolution of the BH mass and spin is driven by
\begin{subequations}
    \begin{align}
        &\dot M_{\rm BH} = \epsilon\,\dot m,\label{eq:mdot}\\
        &\dot \alpha = \left(\frac{l}{\epsilon} - 2 \alpha\right)\,\frac{\dot M_{\rm BH}}{M_{\rm BH}},\label{eq:alphadot}
    \end{align}
\end{subequations}
where $\epsilon$ and $l$ are the energy and angular momentum per unit mass of the accreted material, and $\dot m$ is the accretion rate. On the other hand, the energy release from the gravitational energy gain of the inflowing matter until the inner accretion radius powers the luminosity, so
\begin{equation}\label{eq:Lacc}
    L = (1-\epsilon)\dot m c^2.
\end{equation}
Using Eq. (\ref{eq:Lacc}), we can rewrite Eq. (\ref{eq:mdot}) as
\begin{equation}\label{eq:Mdot2}
    \dot M_{\rm BH} = \frac{\epsilon}{1-\epsilon} \frac{L}{c^2}.
\end{equation}

For $\epsilon$ and $l$, we adopt the values of the last circular co-rotating orbit around a Kerr BH
\citep{1971ESRSP..52...45R,1972ApJ...178..347B,1974bhgw.book.....R}
\begin{subequations}\label{eq:EcLc}
\begin{align}
    \epsilon &= \frac{r^{3/2} -2 r^{1/2} + \alpha}{r^{3/4}(r^{3/2} - 3 r^{1/2} + 2 \alpha )^{1/2}},\\
    l &= \frac{r^2 - 2 \alpha r^{1/2} + \alpha^2}{ r^{3/4}(r^{3/2} - 3 r^{1/2} + 2 \alpha )^{1/2}},
\end{align}
\end{subequations}
with $r = R_{\rm lco}$, the radius of the last circular orbit
\begin{subequations}\label{eq:rlco}
    \begin{align}
    R_{\rm lco} &=3 + Z_2 - \sqrt{(3-Z_1)(3+Z_1+2 Z_2)},\\
    Z_1 &= 1 + \left(1-\alpha^2\right)^{1/3} \left[\left(1+\alpha\right)^{1/3} + \left(1- \alpha\right)^{1/3}\right],\\
    Z_2 &= \sqrt{3 \alpha^2 + Z_1^2},
\end{align}
\end{subequations}
where the radius is units of $G M_{\rm BH}/c^2$. Thus, by replacing Eq. (\ref{eq:rlco}) into Eqs. (\ref{eq:EcLc}), we have analytic (though cumbersome) functions $\epsilon = \epsilon(\alpha)$ and $l = l(\alpha)$. Therefore, given a luminosity evolution $L = L(t)$, we can integrate (numerically) the system of differential equations given by Eqs. (\ref{eq:alphadot}) and (\ref{eq:Mdot2}).

During this phase, the BH mass and spin grow. To set the initial conditions of the BH, we first introduce the BH mass-energy formula \citep{1970PhRvL..25.1596C,1971PhRvD...4.3552C,1971PhRvL..26.1344H}
\begin{equation}\label{eq:bhmass}
M_{\mathrm{BH}}^2 = \frac{\alpha^2 M_{\mathrm{BH}}^4}{4 M_{\mathrm{irr}}^2} +M_{\mathrm{irr}}^2,
\end{equation}
which relates the BH mass and spin with the BH irreducible mass, $M_{\rm irr}$. We can solve Eq. (\ref{eq:bhmass}) to express the BH spin in terms of $M$ and $M_{\rm irr}$:
\begin{equation}\label{eq:Mirr}
    \alpha = 2 \frac{M_{\rm irr}}{M} \sqrt{1-\left( \frac{M_{\rm irr}}{M}  \right)^2}.
\end{equation}
The BH extractable energy
\begin{equation}\label{eq:Eext}
    E_{\rm ext} = (M_{\rm BH} - M_{\rm irr}) c^2,
\end{equation}
also increases during this initial phase of BH growth.

As mentioned above, to integrate the equations, we must supply the luminosity, which we can approximate in this phase as (see Figure \ref{fig:lat-photon}, lower panel)
\begin{equation}\label{eq:Lrise}
    L = L_{\rm GeV,rise} = A_{\rm rise} t^\beta,
\end{equation}
with $A_{\rm rise}\approx 3.11 \times 10^{48}$ erg s$^{-1}$ and $\beta \approx 2.79$, valid from a rest-frame time $t_0 = t_{\rm BH} \approx 10$ s, to the maximum at $t_{\rm max} \approx 20$ s, where we indicate as $t_{\rm BH}$ the time of BH formation. The energy released in this phase is
\begin{equation}
    E_{\rm GeV,rise} = \frac{A_{\rm rise}}{1+\beta} \left(t_{\rm max}^{1+\beta}-t_0^{1+\beta}\right) \approx 2 \times 10^{53}\,\,{\rm erg}.
\end{equation}
Having the luminosity (\ref{eq:Lrise}), we can solve the system of differential equations (\ref{eq:alphadot}) and (\ref{eq:Mdot2}), in the time interval $[t_0,t_{\rm max}]$, given initial conditions $M_{\rm BH, 0} \equiv M_{\rm BH} (t_0)$ and $\alpha_0 \equiv \alpha (t_0)$, or replace one of them by $M_{\rm irr, 0} \equiv M_{\rm irr} (t_0)$. We adopt $M_{\rm irr, 0}=(2.35\pm 0.17) M_\odot$, the lower limit to the NS critical mass constrained by the highest NS mass measured, that of PSR J0952--0607 \citep[see][for details]{2022ApJ...934L..17R}. The value of $M_{\rm BH, 0}$ [and so of $\alpha_0$ via Eq. (\ref{eq:Mirr})], is set to satisfy the boundary condition, at $t = t_{\rm max}$, of matching the BH parameters consistent with the subsequent BH evolution, i.e., during the decreasing GeV emission, which we describe below.

\subsection{BH energy extraction: decaying GeV emission}\label{sec:momentum-extraction-by-gev}

When the accretion ceases, at $t=t_{\rm max} \approx 20$ s, the BH has reached a maximum mass, say $M_{\rm BH,max}$, spin $\alpha_{\rm max}$, and irreducible mass $M_{\rm irr,max}$. The extractable energy of this Kerr BH powers the subsequent long-term, power-law GeV emission, as we describe below.

It has been shown in \citet{2021MNRAS.504.5301R} that BdHNe I show a characteristic power-law decreasing GeV emission luminosity observed by the Fermi-LAT  \citep{2019ApJ...878...52A} (see also \citealp{2023ApJ...955...93A, 2019ApJ...878...52A}) of the form
\begin{equation}\label{eq:Lgev}
    L_{\rm GeV,dec} = A_{\rm dec}\,t^{-n},
\end{equation}
with the power-law index spanning a narrow range of values $n\approx 1.2$--$1.3$ (see Figure 3 in \citealp{2021MNRAS.504.5301R}).

The BdHN I scenario explains such a power-law GeV emission by the radiation of electrons accelerated by the electric field induced by the external magnetic field and the BH rotation \citep{2019ApJ...886...82R, 2020EPJC...80..300R, 2021A&A...649A..75M}. This electrodynamical mechanism is responsible for extracting the rotational energy of the Kerr BH \citep{2023EPJC...83..960R,2024EPJC...84.1166R}, which is the energy reservoir of the GeV emission. Therefore, in this phase, the mass and spin of the BH decrease. 

The modeling of the above GeV emission process has been done via the Wald solution \citep{1974PhRvD..10.1680W} of a Kerr BH immersed in an asymptotically uniform, aligned magnetic field, shows that electrons are accelerated outward and emit synchrotron radiation of GeV energy in a conical region with semi-aperture angle $\theta_{\rm GeV}\approx 60^\circ$ 
around the polar axis \citep{2021A&A...649A..75M,2022ApJ...929...56R}. 

Figure \ref{fig:lat-photon} (lower panel) shows that indeed the GeV emission luminosity of GRB 220101A shows a decreasing part at $t> t_{\rm max}$ that follows an approximate power-law of the form (\ref{eq:Lgev}), with index $n\approx 1.26$. It is worth noting that there are only a handful of observational data points from the peak onward, likely because the detector's line of sight moved outside the emission region. Thus, in our analysis, we assume that the LAT line of sight at times when data points exist is within the cone of semi-aperture angle $\theta_{\rm GeV}$ and that, within this cone, the energy distribution is homogeneous.

With the above, the intrinsic energy release in GeV photons is $\tilde{E}_{\rm GeV} = E_{\rm GeV, iso} (1-\cos\theta_{\rm GeV})$. For GRB 220101A, this equation implies $\tilde{E}_{\rm GeV} = 2.3 \times 10^{53}$ erg, where we have used $E_{\rm GeV} = 4.6 \times 10^{53}$~erg, the observed isotropic energy of the GeV emission during the BH spindown period (see Table \ref{tab:Summary1}).

We proceed to infer the evolution of the BH parameters during this phase. The time integral of Eq. (\ref{eq:Lgev}) gives the minimum value of the extractable energy of the BH, i.e.:
\begin{equation}\label{eq:Eextconstraint}
    E_{\rm ext} = \tilde{E}_{\rm GeV} = \int_{t_{\rm max}}^\infty L_{\rm GeV,dec}\,dt,
\end{equation}
where $t_{\rm max}$ is the rest-frame time at which the GeV emission luminosity peaks. The above equation implies
\begin{equation}\label{eq:Adec}
    A_{\rm dec} = \frac{\tilde{E}_{\rm GeV} (n-1)}{t_{\rm max}^{1-n}}.
\end{equation}
Using this power-law index $n=1.26$, we obtain using Eq. (\ref{eq:Adec}), $A_{\rm dec} \approx 1.24\times 10^{53}$ erg s$^{-1}$. With the above, Eq. (\ref{eq:Lgev}) sets the intrinsic GeV luminosity in this phase. 

From the constraint (\ref{eq:Eextconstraint}) on the extractable energy, the BH mass and irreducible mass at the peak time fulfill
\begin{equation}\label{eq:BHenergy}
(M_{\rm BH,max} - M_{\rm irr,max}) c^2 = \tilde E_{\rm GeV},
\end{equation}
assuming that the BH irreducible mass remains constant during the emission process. Gravitational processes to extract the BH energy, like the Penrose process \citep{1969NCimR...1..252P,1971NPhS..229..177P}, have recently been shown to cause an enormous increase of $M_{\rm irr}$ \citep{2025PhRvL.134h1403R,2025PhRvR...7a3203R}, rendering the process very inefficient. Instead, electrodynamical energy-extraction processes can efficiently limit the increase in the irreducible mass \citep{2023EPJC...83..960R,2024EPJC...84.1166R}. Thus, we adopt that in this phase, the BH irreducible mass remains constant at the value $M_{\rm irr, max}$. The latter, via Eq. (\ref{eq:Mirr}), is expressed in terms of $M_{\rm BH, max}$ and $\alpha_{\rm max}$, and replacing it into Eq. (\ref{eq:BHenergy}) leads to
\begin{equation}\label{eq:Mbhmax}
    M_{\rm BH,max} = \frac{\tilde E_{\rm GeV}}{1 - \sqrt{\frac{1 + \sqrt{1-\alpha^2_{\rm max}}}{2}}}.
\end{equation}

From the above, we can obtain the decreasing BH mass and spin in this phase, and the matching of this phase at $t = t_{\rm max} \approx 20$ s with the increasing BH mass and spin in the rising part of the GeV emission, i.e., $M_{\rm BH, max} = M_{\rm BH}(t_{\rm max})$, with $M_{\rm BH, max}$ fulfilling the constraint (\ref{eq:Mbhmax}), is used to set the initial BH spin $\alpha_0$.

\begin{table}
    \centering
    \begin{tabular}{c|c|c|c|c}
       $t$  & $M_{\rm BH}$ & $\alpha$ & $M_{\rm irr}$ & $E_{\rm ext}$ \\
       \hline
        $t_0 \approx 10$ & $2.40\pm 0.17$ & $0.40\pm 0.01$ & $2.35\pm 0.17$ & $0.05$ \\
        $t_{\rm max} \approx 20$ & $2.62\pm 0.17$ & $0.59\pm 0.02$ & $2.49\pm 0.17$ & $0.13$\\
        \hline
    \end{tabular}
    \caption{BH parameters at the beginning and peak of the GeV emission. We report the uncertainty in the parameters propagated by the assumed value of the irreducible mass $M_{\rm irr}(t_0) = (2.35 \pm 0.17) M_\odot$. We note that, since $M_{\rm BH}$ and $M_{\rm irr}$ decrease and increase by the same amount relative to the central value, the extractable energy, $E_{\rm ext} = (M_{\rm BH} - M_{\rm irr}) c^2$, remains fixed at its central value. Times are reported in seconds, masses in $M_\odot$, and $E_{\rm ext}$ in $M_\odot c^2$. }\label{tab:BHparameters}
\end{table}

\begin{figure}
    \centering
    \includegraphics[width=\hsize,clip]{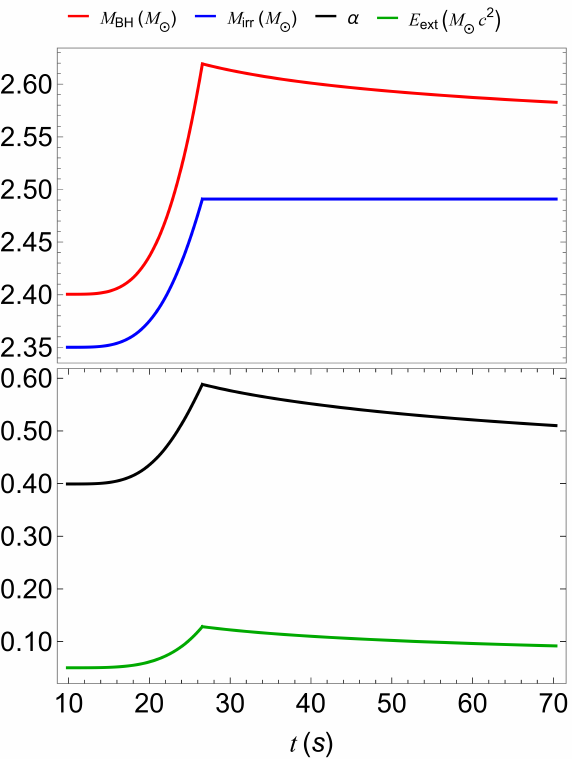}
    \caption{BH parameters evolution during the GeV emission phase, for an initial irreducible mass $M_{\rm irr,0} = 2.35 M_\odot$.}
    \label{fig:BHevolution}
\end{figure}

Following the above procedure, we obtained the BH parameters presented in Table \ref{tab:BHparameters}. In agreement with Eq. (\ref{eq:Eextconstraint}) and the constancy of the irreducible mass at the value $M_{\rm irr,max}$, the BH mass evolution fulfills
\begin{equation}\label{eq:BHenergyconserved}
    \dot M_{\rm BH} = -L_{\rm GeV,dec},
\end{equation}
whose integration leads to
\begin{equation}\label{eq:Mbh2}
    M_{\rm BH} = M_{\rm BH, max} - \frac{\tilde E_{\rm GeV}}{c^2}\left[ 1 - \left( \frac{t}{t_{\rm max}} \right)^{1-n} \right],
\end{equation}
and the BH spin is obtained from Eq. (\ref{eq:Mirr}). Figure \ref{fig:BHevolution} shows the resulting entire evolution of the BH parameters during the rise and decrease of the GeV emission.

Finally, by integrating Eq. (\ref{eq:mdot}), we can obtain the accreted mass onto the BH during the rising phase. We obtained $m_{\rm acc} = \int_{t_0}^{t_{\rm max}} \dot m\,dt \approx 0.24 M_\odot$.

The values we have obtained fall within a reasonable parameter range, indicative of the model's self-consistency.

\section{Overcritical field and the UPE phase}\label{sec:details}

\begin{figure}
    \centering
    \includegraphics[width=\hsize,clip]{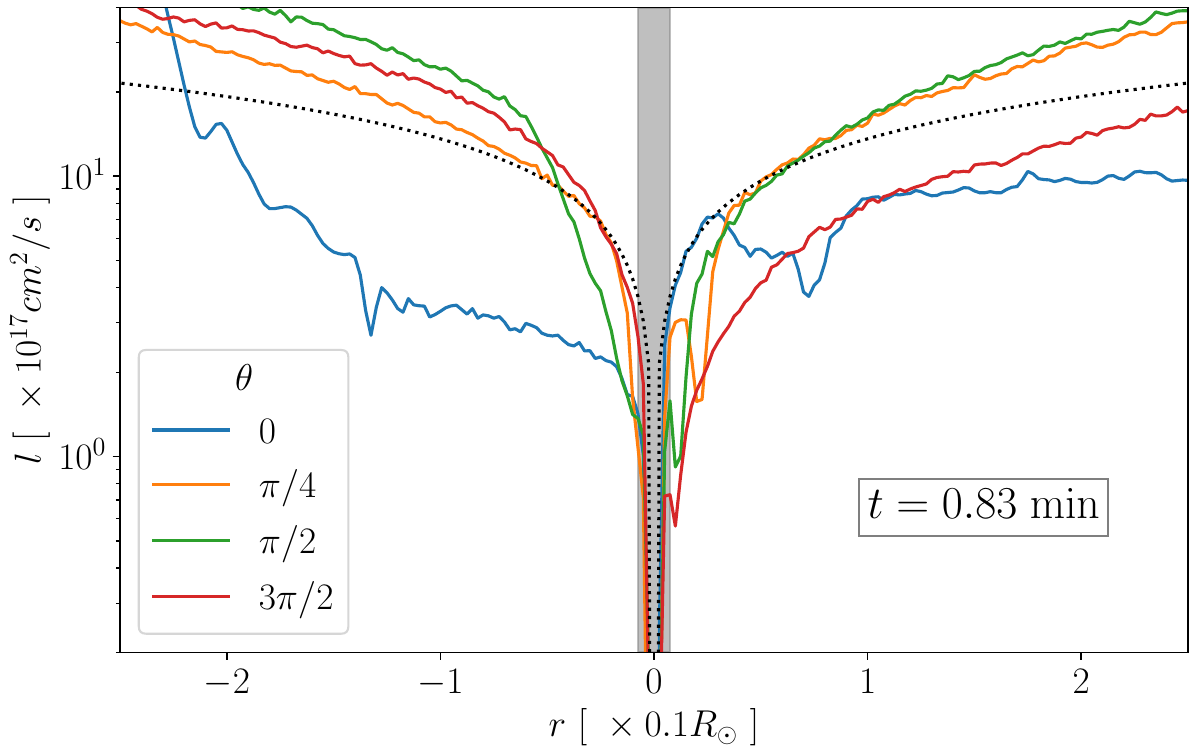}
    \caption{Angular momentum per unit mass of the matter bound to the NS companion, as a function of the radial distance, on selected planes $\theta=0$ (equatorial plane), $\pi/4$, $\pi/2$, and $3\pi/2$, at time $0.83$ min ($49.8$ s), corresponding to the time when the NS reaches the critical mass. The gray-shaded region shows the gravitational capture region of the NS companion. The dotted black curve shows the behavior of the specific angular momentum of material in Keplerian motion, i.e., $l \propto \sqrt{r}$ (see also Fig. \ref{fig:omega}).}
    \label{fig:LoverM}
\end{figure}

The white arrows in Fig. \ref{fig:simulation} are velocity field lines, which show the vortex motion around the NS companion. Figure \ref{fig:LoverM} shows the specific angular momentum (i.e., per unit mass), $l$, of the matter bound to the NS companion at the time it reaches the critical mass. The gray region shows the gravitational capture region. We find $l \approx (2$--$3) \times 10^{17}$ cm$^2$ s$^{-1}$ (or $l/c \approx (0.7$--$1) \times 10^7$ cm, in geometric units), and it has a mass $\Delta M \approx 0.5 M_\odot$. Therefore, the total angular momentum of the bound matter is $L \sim l \Delta M = (2$--$3) \times 10^{50}$ g cm$^2$ s$^{-1}$.

\begin{figure}
  \centering
  \includegraphics[width=\hsize,clip]{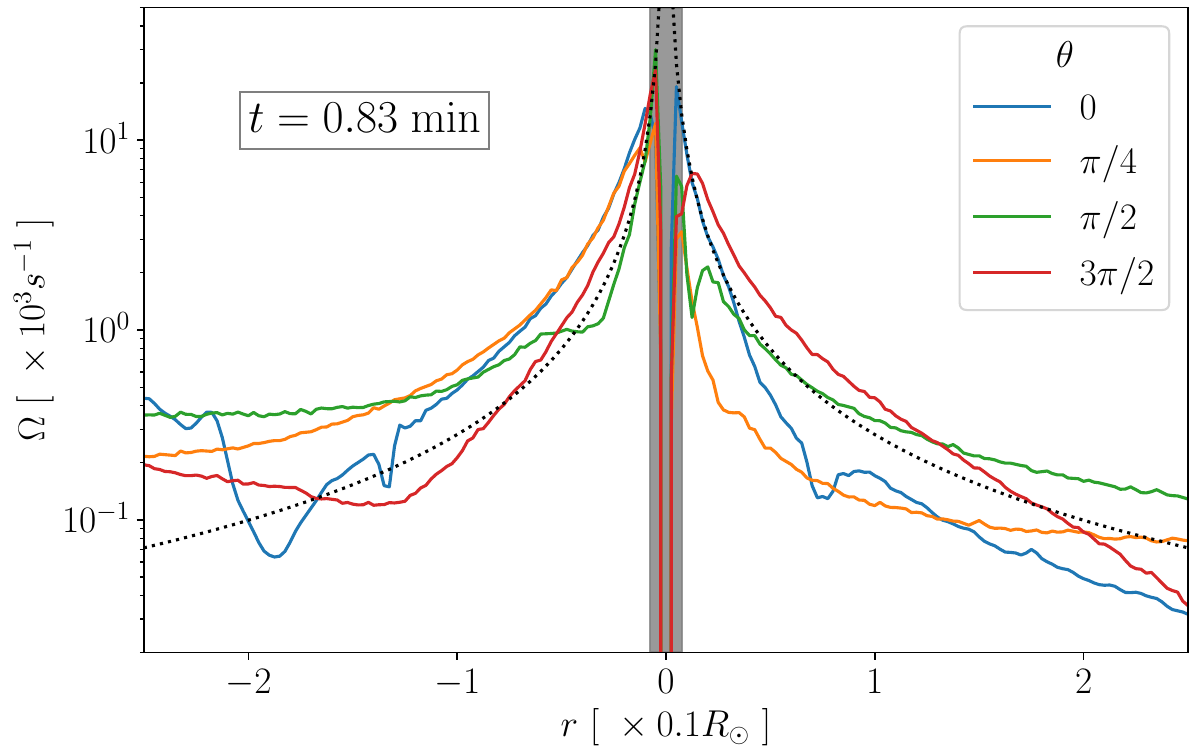}
  \caption{Corresponding angular velocity profiles of the SN ejected material close to the NS companion. The dotted black curve is the Keplerian angular velocity, $\Omega \propto r^{-3/2}$.}
  \label{fig:omega}
\end{figure}

The maximum spin parameter of a Kerr BH is $a_{\rm max}=M$ (geometric units). For example, a $3 M_\odot$-BH can have a maximum spin of $a_{\rm max}\approx 4.5\times 10^5$ cm, which is lower than the specific angular momentum of the matter bound to the accreting NS companion estimated above. This suggests that such bound material cannot participate in the BH formation; it must lose angular momentum to be incorporated into forming the Kerr BH.

The total rotational energy to be ejected should lead to the observed value of a few $\sim 10^{53}$ erg, characterizing the UPE Episode of the GRB. To assess this possibility, we plot in Figure \ref{fig:omega} the corresponding angular velocity profiles of the bound, rotating matter. The material inside the gravitational capture region (gray-shaded area) has $\Omega \sim 10^4$ rad s$^{-1}$. Therefore, it has a rotational energy $E_{\rm rot} = (1/2) \Omega L \sim 10^{54}$ erg, which covers the UPE energetics request.

We have seen in Table~\ref{tab:Summary1} that UPE~I lasting $0.89$~s ($15.15$--$16.04$~s) has an energy $E_{\rm iso}=(2.4\pm5.1)\times10^{53}$~erg and UPE~II of equal duration starts at $18.72$ and ends at $19.61$ separated by $1.68$ s occurs with the energy $E_{\rm iso}=(4.6\pm1.7)\times10^{53}$~erg. The UPE emission is concomitant with the GeV radiation, which begins at $10$~s and peaks at $20$~s, coinciding with the end of UPE~II. At $20$~s, the GeV afterglow begins extracting energy from the Kerr BH. In the previous section, we have already illustrated the rise of BH mass formation and its following decay by the extraction of its rotational energy. In this section, we return to the processes that characterize the UPE~I and UPE~II phases.  

We have shown in Section \ref{sec:simulation} that the accreting material onto the NS has a large amount of angular momentum, which does not allow the BH formation since, by definition, the BH formation must necessarily occur when $a/M$ is smaller than one. The rotational energy in the vortexing material is sufficient to power the UPE emission through the mechanism we discuss below.

The UPE phase is explained by the emission from the transparencies of the $e^+e^-$ pair plasma, which is reached through its ultra-relativistic self-accelerating expansion \citep{1999A&A...350..334R,2000A&A...359..855R,2021PhRvD.104f3043M,2022EPJC...82..778R,Ruffini2026UPE}. Here, based on our numerical simulations in Section \ref{sec:simulation}, we propose that the pairs are created by the overcritical electric field induced by the rotating magnetic field vortex, around the accreting NS, before it collapses to form the BH. We describe the structure of the electromagnetic field at this stage as that produced by a rotating magnetic dipole, described in Appendix A of \citet{2026JHEAp..5000464R}. The region where the electric field creates pairs is called the \textit{dyadoregion}, its surface radius and width are $r_d(\theta)$ and $\Delta_d = r_d - R$, which depend on the NS surface radius $R$, angular velocity $\Omega$, and polar surface magnetic field, $B$ (see Eqs. A.15--A.19 in \citealp{2026JHEAp..5000464R}). Next, we set their values.

In our picture, the start of the GeV emission marks the birth of the BH. That subsequent emission occurring after the UPE phase is powered by a different mechanism, based on the acceleration of charged particles by an undercritical electric field \citep{2019ApJ...886...82R, 2020EPJC...80..300R, 2021A&A...649A..75M,10.1093/mnras/stab724,2022ApJ...929...56R, Ruffini2026UPE}. This scenario implies that the UPE, so the pair creation, should end by $t_{\rm BH} = t_0 \approx 20$ s, at which the GeV emission utilizing the rotational energy of the Kerr BH starts to be effective. Hence, for the self-consistency of the model, we request that the condition for pair creation by the induced electric field ceases to be fulfilled at BH formation. In Appendix A of \citet{2026JHEAp..5000464R}, it has been shown that the minimum strength (at the pole) of the rotating magnetic field that induces an electric field able to create $e^+e^-$ pairs, $B_{p,\rm min} = 3 B_c/(4 \tilde\Omega)$, which depends only on $\tilde \Omega = \Omega R/c$, the dimensionless rotation parameter, which we now estimate for the present case.

The value of the angular velocity relevant in this estimate is at the moment when the electric field becomes undercritical, i.e., when the BH forms. From the analysis of the GeV emission, we infer the initial BH spin parameter $\alpha_0 = c J_0/(G M_0^2) = 0.4$ (see Table \ref{tab:BHparameters}), so the initial BH angular momentum is $J_0 = 0.4 G M_0^2/c$. Thus, we can assume that the mass and angular velocity of the NS at its critical mass point were $M_{\rm NS, crit} \approx M_0 = 2.4 M_\odot$, and $\Omega_{\rm crit} \approx 0.4 G M_{\rm NS,crit}^2/(c I)$, where $I$ is the NS moment of inertia. Using $R = 12$ km and $I \approx (2/5) M_{\rm NS, crit} R^2 = 2.75 \times 10^{45}$ g cm$^2$ (which agrees with numerical calculations, e.g., \citealp{2015PhRvD..92b3007C}), we obtain $\Omega_{\rm crit} \approx 7.37 \times 10^3$ rad s$^{-1}$.

Therefore, for the self-consistency of this picture, we set the UPE magnetic field for which the condition for pair creation is no longer fulfilled at $t=t_{\rm BH}$. Hence, we assume it to be the minimum magnetic field, $B_{p, \rm min}$, calculated with $\Omega = \Omega_{\rm crit}$, i.e.:
\begin{equation}\label{eq:Bminvalue}
    B_p = B_{p, \rm min} = \frac{3}{4}\left(\frac{\Omega_{\rm crit} R}{c}\right)^{-1} B_c = 1.12 \times 10^{14}\,\,\text{G}.
\end{equation}
Summarizing, at times $t < t_{\rm BH}$, where $\Omega > \Omega_{\rm crit}$, the dyadoregion exists ($\Delta_d > 0$), and when the angular velocity has decreased to $\Omega = \Omega_{\rm crit}$, the dyadoregion vanishes ($\Delta_d \to 0$). Accordingly, the energy available for the pairs decreases and becomes zero at $t = t_{\rm BH}$. The different pair-plasma initial conditions as a function of time explain the different transparencies, i.e., the various blackbodies observed in the time-resolved spectra of the UPE.

From the simulations presented in Sec. \ref{sec:simulation}, we estimated the angular velocity of the vortex around the accreting NS to be $\Omega\approx 10^4$ rad s$^{-1}$, which is indeed larger than $\Omega_{\rm crit}$. For this angular velocity, the dyadoregion extension along the pole is $r_d(\theta = 0) = 1.092 R = 13.11$ km, so its width is $\Delta_d (\theta = 0) = 0.092 R = 1.11$ km. 

Having set the radius, angular velocity, and magnetic field, we now estimate the energy available to the pairs (per emission event). Consistent with the present scenario, we estimate it as the electromagnetic energy stored in the dyadoregion (see Appendix A of \citealp{2026JHEAp..5000464R})
\begin{multline}\label{eq:Epairsvalue}
    {\cal E}_{e^+e^-} = \frac{1}{4} \int_0^\pi\int_R^{r_d(\theta)} (E^2 + B^2) r^2 \sin\theta dr d\theta \\= 4.62\times 10^{43}\,\,\text{erg}.
\end{multline}
The integrated isotropic energy of the UPE is given by the sum of the $E_{\rm iso}$ of Slices $4$ and $6$, so we obtain $E_{\rm UPE, iso} \approx 6.8\times 10^{53}$ erg (see Section \ref{sec:time-resolved} and Table \ref{tab:Summary1}). To compare it with the energy of the dyadoregion, we must reduce the former by an effective beaming factor, $f_b$, as the dyadoregion energy is contained within two polar lobes defined by $r = r_d(\theta$); see Figure A.5 in \citet{2026JHEAp..5000464R}. We can define such an effective beaming factor by requesting that the electromagnetic energy stored in a double-cone of height $r_d(0) = r_d(\theta =0)$ and semi-aperture angle $\theta_b$, equals the dyadoregion energy, i.e.:
\begin{equation}
    f_b \equiv 1-\cos\theta_b = \frac{{\cal E}_{e^+e^-}}{\frac{1}{6} B_p^2 r_d(0)^3} \approx 0.01,
\end{equation}
which implies an effective beaming angle $\theta_b\approx 10^\circ$. Applying this beaming-factor correction, the energy of the UPE to be explained is $E'_{\rm UPE} = f_b E_{\rm UPE} = 6.8\times 10^{51}$ erg. This result implies the UPE consists of $N \approx E'_{\rm UPE}/{\cal E}_{e^+e^-} \approx 1.5\times 10^8$ impulses or events.   

We can estimate some of the physical parameters of the pair plasma and its transparency using the blackbodies inferred from the spectral analysis of Slice 4 and Slice 6 in Sec. \ref{sec:time-resolved}. We follow the treatment in \citet{2021PhRvD.104f3043M,2022EPJC...82..778R,2026JHEAp..5000464R}, considering the dynamics of a spherically symmetric ultrarelativistic self-accelerated $e^+e^-$-photon plasma in a medium with a low baryonic-matter contamination. The latter is measured by the baryon load parameter ${\cal B} \equiv M_B c^2/{\cal E}_{e^+e^-}$, being $M_B$ the mass of the baryonic matter loaded into the plasma during its expansion. The transparency of the plasma leads to a blackbody spectrum in the MeV regime \citep{1999A&A...350..334R,2000A&A...359..855R,2021PhRvD.104f3043M,2022EPJC...82..778R,2026JHEAp..5000464R}.

The time-resolved spectral analysis of the UPE of GRB 190114C \cite{2021PhRvD.104f3043M} and GRB 180720B \cite{2022EPJC...82..778R} evidenced a \textit{hierarchical} (\textit{self-similar}) structure. The UPE spectra in rebinned time intervals exhibit composite blackbody-plus-cutoff power-law spectra, interpreted as the result of a repetitive microphysical process occurring on ever shorter timescales: successive transparencies of the pair plasma. The time-resolved spectra showed similar temperatures and blackbody-to-total flux ratios \citep[see, e.g., Table I in][]{2021PhRvD.104f3043M}. Therefore, although the observational data are limited to a time interval longer than the typical timescale of the pair-plasma transparencies, the similarity of the spectra allowed us to assume that those observed properties are shared by the $\sim 10^8$ events.

Having the above in mind, we estimate the Lorentz factor and radius of transparency, as well as the baryon load parameter by \citep{1999A&A...350..334R,2000A&A...359..855R,2021PhRvD.104f3043M,2022EPJC...82..778R,2026JHEAp..5000464R}
\begin{subequations}\label{eq:transparency}
    \begin{align}
    \Gamma &\approx \left( \frac{a T_{\rm obs}^4 \sigma_T \Delta_d}{16 m_N c^2}\frac{1-\epsilon_{\rm BB}}{\epsilon_{\rm BB}} \right)^{1/3},\\
    {\cal B} & = \frac{1-\epsilon_{\rm BB}}{\Gamma-1}\\
    R_{\rm tr} &= \sqrt{\frac{\sigma_T}{8\pi}\frac{{\cal B}{\cal E}_{e^+e^-}}{m_N c^2}},
\end{align}
\end{subequations}
where $\epsilon_{\rm BB} \equiv E_{\rm BB}/{\cal E}_{e^+e^-}$, being $E_{\rm BB}$ the energy of the blackbody component, $m_N$ is the nucleon mass, $\sigma_T$ is the Thomson cross-section, and $a = 4\sigma/c$, being $\sigma$ the Stefan-Boltzmann constant. Table \ref{tab:UPE220101A} shows the parameters inferred from Eqs. (\ref{eq:transparency}), using the blackbody temperature and blackbody-to-total flux ratio of Slices 4 and 6, which comprise the observational UPE data. From these parameters, we infer the typical $\Gamma$, ${\cal B}$, and $R_{\rm tr}$ of each of the transparency events.

The values of the above parameters are similar to the ones inferred in the UPE of GRB 190114C \citep{2021PhRvD.104f3043M} and GRB 180720B \citep{2022EPJC...82..778R}, with the difference that in the latter two references, the UPE was assumed to be produced by an $e^+e^-$ plasma formed around a Kerr BH in an external magnetic field. The parameters are similar because the size and energy of the dyadoregion of the NS approaching the critical mass are quantitatively similar to those of the dyadoregion of a Kerr BH in an external magnetic field with comparable mass, angular momentum, and magnetic field strength. We refer to Sec. 4 of \citet{2026JHEAp..5000464R} for further discussion on this comparison. The parameters are even more similar to those of the recent UPE analysis of the short GRB 090510 \citep{2026JHEAp..5000464R}. The closeness of the inferred parameters is due to the UPE of GRB 090510 being explained by the transparencies of the pair plasma produced by the induced overcritical electric field from the rotating magnetized NS-NS merged core, before it reaches the critical mass for BH formation. Such a rotating, highly magnetized merged core has mass, angular momentum, and a magnetic field similar to those of the accreting NS companion in GRB 220101A.

\begin{table}
    \centering
    \begin{tabular}{c|c|c|c|c}
    $i$-th event & $k_B T_{{\rm obs},i}$ (keV) & $\Gamma_i$ & ${\cal B}_i$ & $R_{{\rm tr},i}$ ($10^{10}$ cm)\\
     \hline 
    Slice 4 & 21.9 & 6.10 & 0.14 & 1.06 \\
    Slice 6 & 32.2 & 10.19 & 0.08 & 0.79 \\
 \hline 
    \end{tabular}
    \caption{Inferred properties of the $e^+e^-$ pair plasma transparencies observed in the UPE of GRB 220101A. The value of $k_B T_{{\rm obs},i}$ and $\epsilon_{{\rm BB},i}$ are obtained from the analysis of slices $4$ and $6$ in Sec. \ref{sec:time-resolved}. We estimate the latter as $\epsilon_{{\rm BB},i} = E_{{\rm BB},i}/{\cal E}_{e^+e^-,i} \approx E_{\rm BB}/E_{\rm UPE} \approx 0.3$.}
    \label{tab:UPE220101A}
\end{table}

\begin{table*}
\centering
\caption{Redshift, isotropic energy, time of Swift first observation, period and first derivative od period  of the selected GRBs.}
\label{tab:GRB_redshift_energy}
\begin{tabular}{lcccccl}
\hline
GRB & $z$ & $E_{\rm iso}$  & Swift first & $P(t=10^{10})$  & $\frac{dP}{dt}(t=10^{10})$ & Reference \\
& & (erg) & Observation (s) & (ms) & & \\
\hline
GRB~130427A & 0.34  & $8.5\times10^{53}$ & 104.4 & 56.5 & $7.52 \times 10^{-11}$ & Ruffini, et al. (2015), Ruffini, et al. (2018)\\
& & & & & &  Aimuratov, Ruffini, et al. (2022)\\
& & & & & & Ruffini, et al. (2019) \\
GRB~160509A & 1.17  & $8.56\times10^{53}$ & 3358.2 & 41.2 & $7.33 \times 10^{-11} $ & Liang, Ruffini, et al. (2019),  \\
& & & & & &  Ruffini, et al. (2021), Ruffini, et al. (2023) \\
& & & & & & Rueda, Ruffini, et al. (2020) \\
GRB~160625B & 1.41 & $5.0\times10^{54}$ & 550.3 & 22.4 & $5.61 \times 10^{-11} $ & Liang, Ruffini, et al. (2019),  \\
& & & & & & Ruffini, et al. (2021), Ruffini, et al. (2023) \\
& & & & & & Rueda, Ruffini, et al. (2020) \\
GRB~180720B & 0.65 & $5.92\times10^{53}$ & 53.4 & 2645.4 & $4.97 \times 10^{-11} $& Ruffini, et al. (2021) \\
& & & & & &  Rastegarnia, Ruffini, et al. (2022) \\
& & & & & & Rueda, Ruffini, et al. (2022) \\
GRB~220101A & 4.61  & $4.0\times10^{54}$ & 14.4 & 56.7 & $4.98 \times 10^{-11} $ & Ruffini, et al. (2022a), Ruffini, et al. (2022b) \\
& & & & & & Ruffini, et al. (2025), Ruffini, et al. (2026) \\
GRB~221009A & 0.15 & $1.2\times10^{54}$ & 79.6 & 720.7 & $9.93 \times 10^{-11} $ & Aimuratov, Ruffini, et al. (2022a) \\
& & & & & & Aimuratov, Ruffini, et al. (2022b) \\
& & & & & &  Ruffini, et al. (2022) \\
GRB~240825A & 0.66 & $1.7\times10^{53}$ & 50.1 &  4326.6 & $5.02 \times 10^{-11}$ & Ruffini, et al. (2024), Ruffini, et al (2026). \\
GRB~090423 & 8.2  & $3.5 \times 10^{52}$ & 8.42 &  -  &   -  & Bianco, Ruffini et al. (2023)    \\ 
\hline
\end{tabular}
\end{table*}

\section{ Summary }\label{sec:Summary}

GRB 220101A, at a redshift of ($z=4.61$) and with a total isotropic energy of approximately $4\times10^{54}$ erg, typical of a peta nova, provides an exceptional laboratory for investigating the most energetic long GRBs. Its unprecedented multi-wavelength coverage reveals a sequence of temporally and spectrally distinct emission episodes that cannot be naturally described as different manifestations of a single emission mechanism. The observations instead support an extended Binary Driven Hypernova (BdHN) framework, here applied to a new class of Binary-driven Peta-Nova (BdP-N) events.

\begin{enumerate}

\item The principal extension of the traditional BdP-N scenario is the introduction of a third compact object: a white dwarf of approximately $1.4,M_\odot$, in addition to a massive, highly magnetized CO core of approximately $10,M_\odot$ and a neutron-star companion. The central CO core is considered to possess a relatively small rotational angular momentum, while the surrounding NS and WD carry substantial orbital angular momentum. This configuration represents a significant conceptual departure from the classical picture of a rapidly rotating central object and provides the basis for the multi-stage evolution inferred for GRB 220101A.

\item The collapse of the magnetized CO core produces the first, unusually energetic supernova and initiates Episode I, releasing approximately $(1.2\pm1.0)\times10^{53}$ erg. The ejecta from this explosion subsequently interact with the WD. At approximately $3.6$ s, the WD undergoes induced gravitational collapse, producing a second supernova and a newly born neutron star $\nu$NS. The gravitational binding energy calculated for the WD collapse, $1.07\times10^{53}$ erg, is consistent with the energy observed in Episode II, providing a direct energetic connection between the WD collapse and the second emission episode.

\item The collapse of the WD initially produces a neutron star with a spin period of approximately 78  ms and amplifies its magnetic field to approximately $3.9\times10^{14}$ G through magnetic-flux conservation. Continued accretion of the first-supernova ejecta transfers both mass and angular momentum to the $\nu$NS, reducing its spin period to approximately 1.3 ms within about 20 s. Thus, the formation and subsequent spin-up of the $\nu$NS provides a continuous physical link between the second supernova, the prompt emission, and the later pulsar-powered emission.

\item The observational evidence is organized into seven episodes: (I) the first supernova from CO-core collapse; (II) WD collapse and accretion of the first SN ejecta; (III) the ultra-relativistic prompt emission (UPE); (IV) formation and growth of the BH followed by the GeV afterglow; (V) spin-up of the newly formed $\nu$NS; (VI) the X-ray afterglow associated with the expanding SN ejecta; and (VII) the optical afterglow powered by the newly born pulsar. Their temporal ordering and energetics provide a coherent evolutionary sequence from the initial stellar collapse to the long-term compact-object remnant.

\item The high redshift is not merely a limitation for detecting a distant GRB; in this exceptional event it becomes an observational advantage. The extreme intrinsic energy compensates for the large luminosity distance, allowing Fermi-LAT to detect high-energy photons, including photons reaching approximately 5 GeV in the rest frame. More importantly, cosmological time dilation stretches short intrinsic processes, allowing the early X-ray emission to become observable after the Swift-XRT slewing time. For GRB 220101A, XRT began observing at 87.47 s in the observer frame, corresponding to only 15.59 s in the source frame. This provides an unusually direct view of the early stages of the evolving central engine.

\item The UPE occurs while the accreting NS is still present and is associated with overcritical electromagnetic fields and the production of an ($e^+ \ e^-$) pair plasma. The subsequent GeV emission begins with the formation of the BH and is interpreted as the extraction of its rotational energy. This distinction is essential: the UPE represents the final energetic stage immediately preceding BH formation, whereas the GeV afterglow traces the electrodynamical evolution of the newly formed Kerr BH.

\item The synchrotron emission originated by $\nu$NS has three different X-ray, optical and radio components. See figure \ref{fig:xrt-lightcurve}. The synchrotron emission originate by $\nu$NS has three different X-ray, optical and radio components. The X-ray emission leads to the X-ray afterglow while the optical synchrotron component participates in the optical emission of the $\nu$NS.  The fitted pulsar evolution from an initial period of approximately 1.3 ms  would approach after $10^{10}$ second, namely 1000 years to a period of 56.7 ms, and bolometric luminosity of $1.2 \times 10^{38} \ ergs^{-1}$, very close to the period and luminosity of the Crab pulsar. This point to the conclusion that indeed the SN explosion of 1054 can explain perfectly all the observed features discussed in \cite{Ruffini2024Crab} using GRB 190114C, see table \ref{crab}. We call this cosmic relevance.

\end{enumerate}

The combined observational and theoretical analysis of GRB 220101A supports a picture in which the extraordinary energy release is not produced by a single event, but by a sequence of dynamically connected processes occurring in a compact three-body progenitor system. The multi wave length observation from the ground and space observatories which has been reaching an  unprecedented level both in quality and quantity as well as  high-redshift observations resolve these processes with exceptional temporal and spectral detail, allowing the seven BdP-N episodes to be associated with distinct physical stages. GRB 220101A therefore represents a particularly important case for testing and extending the BdP-N model and for establishing a broader class of Binary-driven Peta-Nova explosions. We have applied this model to GRB 130472A, GRB 160509A, GRB160625B, GRB 180720B, GRB 221009A and GRB 240825A, see table \ref{tab:GRB_redshift_energy}.

\section{Main Results}\label{sec:mainresults}

Following the discovery of GRB 220101A we called attention on the exceptionality of this source in two GCN reports \citep{2022GCN.31465....1R, 2022GCN.31648....1R} pointing out some immediate consequence of the measured red shift of $z= 4.61$. This redshift corresponds to distance of $7.6 \times 10^9$ pc and  age of universe equal to 1.3 billion years. Based on the observations of FERMI and SWIFT satellites, in order to estimate the $E_{iso}$ of the source which we estimated larger than $10^{54}$ erg, typical of a Peta-Nova. We were in presence of the most powerful GRB ever observed and at very high red shift. We soon realized the importance of this result and the possibility to use cosmological principle in the study of this source. Truly overwhelming were the unprecedented quality and quantity of the data by Swift and Fermi satellites. We chose to exploit the unique nature of this source to probe the depths of the comprehension of these new extreme energetic sources. We did expect to address a very strong and exciting commitment, enriched by a continuous injection of new ideas, of additional complementary observations.  We started the definition of the new paradigm of the BDP-N, to approach both the GRBs understanding and their cosmic implications. This procedure has continued without interruption for five years, duly adjourning our observational data and theoretical understanding, presented in accompanying publications and here, finally, completed.

We have identified the seven independent Episodes characterizing the BdP-N model summarized in Section \ref{sec:episodes}. Particular important where the observations of James Web telescope \citep{2024ApJ...968L..18T} and the IXPE satellites \citep{2024arXiv240712779W}, including the observations of synchrotron radiation in the Crab Nebula. 

Equally fundamental has been the conceptual breakthrough introduced by Michel. Mayor, \citep{Mayor2020, Mayor2024Plurality} following O. Struve \citep{1952Obs....72..199S}, and F. Hoyle \citep{Hoyle1947}, addressing non rotating massive systems surrounded by binary fast rotating binary companions, which has become leading in our formulation of the BdP-N.
Among the many results obtained. we like to recall in particular

\begin{enumerate}
   
\item The detailed spectral analysis across various time slices has revealed
the complex and here firstly explored nature of GRB 220101A, including the initial collapse of the CO core, the explosion of the HB Supernova, the observation of UPE emission, indicating the appearance of high-energy QED phenomena of energies up to $10^{54}$ erg, manifested in the occurrence of a most powerful jet observed normal to the GRB equatorial plane. See episode III in Fig 1 and Fig 9, the new physics observed in this UPE has been presented in \citep{Ruffini2026UPE} addressed in where the many new physical process some presenting a fractal structure has been presented opening a new domain of quantum electrodynamics processes conceptually different from the non-linear general relativistic processes dominating the Black Hole formation.

\item The detection of high-energy GeV emissions of quantum-electrodynamic processes occurring around the newly born BH originating from the gravitational collapse of the NS companion induced by the accretion of the ejecta of the HB SN. This has allowed to follow, since the onset at $T= 10$ s. The mass and spin of the newly born BH at the moment formation from  $M =  2.40 \pm 0.17 \ M_\odot$    and spin $\alpha = 0.40 \pm 0.01$  and reaching  $2.62 \pm 0.17  \ M_\odot$  and spin $\alpha = 0.59$ at its peak value.

\item The accretion of the ejecta on the white dwarf companion leads to the
second SN. The gravitational collapse of the white dwarf core leads to
the emission of the energy observed in Episode II due to the increase
of gravitational energy leading to a new NS ($\nu$NS) formation the
formation of the $\nu$NS occurs at 17 seconds. The subsequent spin up of
the $\nu$NS by the ejecta reaching a 1.3 msec period occurs at 20 seconds.
The observed optical afterglow provides, as well, the evolutionary
picture of the $\nu$NS created by the collapse of the accreting white dwarf originating the second SN. The X-ray afterglow provides a comprehensive understanding of the synchrotron process in the X-ray, optical and radio, created by the interaction of the $\nu$NS with the SN remnant. Finally the seventh Episode linked to the pulsar observation further emphasizes the versatility of the BdP-N model explaining emissions of different origins at each GRB Episode. 
\end{enumerate}

\section{New perspectives}\label{sec:perspectives}

Following the basic understanding of GRB 220101A,

\begin{enumerate}

\item we started to probe the validity of the model of the BdP-N to the seven sources we had previously identified with similar $E_{iso}$ of $10^{54}$ erg. A new understanding gradually emerged, differentiating the properties of the optical and X-ray afterglows allowing identifying in all these sources, the presence of a millisecond pulsar, see table \ref{tab:GRB_redshift_energy}.

\item The observational and physical properties of the Crab pulsar today are
remarkably consistent with those inferred for all above sources when extrapolated to $10^{10}$ seconds. From this we can conclude that indeed, the astrophysical system which exploded in 1054 in the Crab Nebula was indeed a source equal in physical processes  to GRB220101 and their seven equally energetic companion.

\item Extrapolating the observed optical flux of these pulsars to the present epoch yields an apparent magnitude of approximately $m \sim 28$, placing these sources within the sensitivity range of the James Webb Space
Telescope (JWST). This sign just the beginning of a new era of additional crucial observations of GRB. 

\end{enumerate}

\section{Conclusion}\label{sec:conclusion}

The extensive study of GRB 220101A  has provided unprecedented insights into the complex mechanisms of such cosmic phenomena. Our analysis, leveraging multi-wavelength data from Fermi-GBM, Fermi-LAT, Swift, and other satellites, as well as many ground-base observatories, has outlined a comprehensive picture of the seven different Episodes occurring during this GRB event. These observations have been pivotal in validating the BdP-N model, which posits  binary systems origin for long GRBs all the way to the most energetic ones up to $E_{iso} \approx 10^{54}$ erg. We open new directions of understanding historical events in our galaxy, including the supernova 1054 event and observation of the pulsar (PSR J0534+2200) in 1967 \citep{1968Sci...162.1481S}. In this work, we explain both events in terms of observation of GRB 220101, as well as in terms of the seven equally energetic GRBs reported in table \ref{tab:Summary1}.

The fact that an equally energetic GRB, had been observed at $z = 9.4$  \citep{2011ApJ...736....7C,Bianco...2024ApJ...966..219B}, brought to our attention an unexpected coincidence: that the earliest GRB we have studied are coeval with the observations of the Red Dots.

This has  been a major topic of our research in the last two years \citep{Jiang_2025, RuffiniWang2026, WangRuffini2026}. Indeed, in these papers, we have evidenced that the earliest  Red Dots originate at redshift $z = $10 - 20. The crucial new point is that  Red Dots  originate  in the early universe from  black holes of mass larger than  $\sim 5 \times 10^6  M_\odot$. There are mounting evidence that they are composed of Dark matter of X-fermions of approximately 380 keV \citep{RuffiniVereshchagin2025}. 

These BHs are very different from  the baryonic matter BH of 2.3 solar masses observed in GRB 220101, which occurs at the end point of thermonuclear evolution of stars currently observed.  

The further occurrence of the cosmological nucleosynthesis at cosmological redshift $10^8$ where the light element begins to form, all the way to $z = $10-20 when the iron forms, see, e.g., \citep{OhanianRuffini2013} indicates a third fundamental process. 

 The earliest GRB occurring at $z = 9.4$,  the birth of red dots occurring in $z = $10-20, and  the cosmological neucleu-synthesis  between $z= 10^8$ and $z=$10-20, all occur in comparable interval in cosmological time.
 
 It is clear that we are witnessing a new era of scientific research which includes:

 \begin{enumerate}
     \item the study of the  physics and astrophysics of baryonic matter genesis starting in the early universe and evolving to the current formation of baryonic matter black holes we are observing in GRBs, and
     
     \item the dark matter genesis starting also in the early Universe from neutral fermions interacting only gravitationally and leading at the end of their evolution to the formation of the quasars \citep{Jiang_2025, RuffiniWang2026, WangRuffini2026}, 
     
     \item  the fundamental contributions of George Gamow, Enrico Fermi  and Anthony Turkevich \citep{Gamow1946, AlpherBetheGamow1948, AlpherHerman1950} to be extended to the consideration not only of Baryonic matter but also to the dark matter component of the universe, following the latest observations. 
 \end{enumerate}

 For these reasons, in the book \citet{Ruffini2027}, dedicated to the  principle physics and astrophysics processes in the early Universe, the Fermi-Turkevich-Gamow work was presented in all technical details so opening the way to a new approach, ready to be extended both to baryonic matter and dark matter components, in view of the current observational evidence.

\section*{Data Availability}

This research uses publicly available multi-wavelength observations of GRB~220101A obtained from space-based and ground-based observatories. The data used in this work are accessible through the following archives and public data repositories.

The \textit{Neil Gehrels Swift Observatory} data are available through the NASA Swift Data Archive and the HEASARC at \url{https://swift.gsfc.nasa.gov/archive/}. The GRB~220101A Swift observation page is also available at \url{https://swift.gsfc.nasa.gov/archive/grb_table/fullview/220101A/}. The \textit{Fermi Gamma-ray Space Telescope} data are publicly available from the Fermi Science Support Center (FSSC) at \url{https://fermi.gsfc.nasa.gov/ssc/data/}. 
The \textit{Xinglong 2.16,m Telescope} observations are reported in GCN Circular 31353. The \textit{CAHA 2.2,m Telescope} data  are described in \citep{2023ApJ...959..118Z,2022GCN.31388....1C}. The radio observations include the \textit{Karl G. Jansky Very Large Array (VLA)} are accessible through the NRAO Data Archive at \url{https://science.nrao.edu/facilities/vla/archive}.  The corresponding ALMA science data are available through the ALMA Science Archive at \url{https://almascience.eso.org/alma-data}. 

\vspace{5mm}
\facilities{{\it Fermi}/GBM}
\software{3ML\citep{Vianello2015}}  

\newpage

\appendix

\begin{deluxetable}{cccccccc}
\tabletypesize{\scriptsize}
\tablecaption{Results of the time-resolved spectral fits of GRB 220101A}
\label{tab:initialchecking}
\tablehead{
\colhead{$t_{1}$$\sim$$t_{2}$}
&\colhead{$t_{1}$$\sim$$t_{2}$}
&\colhead{$S$}
&\colhead{Band}
&\colhead{CPL+BB}
&\colhead{DIC}
&\colhead{pDIC}
&\colhead{UPE}\\
\hline
(s)&(s)\\Obs&Rest} 
\colnumbers
\startdata
\hline\noalign{\smallskip}
\multicolumn{8}{c}{Iteration I}\tabularnewline
\hline\noalign{\smallskip}
64$\sim$160&11.40$\sim$28.52&77.81&Constrained&Unconstrained&\nodata&\nodata&\nodata\\
64$\sim$80&11.40$\sim$14.26&26&Constrained&Unconstrained&\nodata&\nodata&No\\
80$\sim$90&14.26$\sim$16.04&42.3&Constrained&Constrained&\nodata&\nodata&Yes(weak)\\
90$\sim$100&16.04$\sim$17.83&65.27&Constrained&Unconstrained&\nodata&\nodata&No\\
100$\sim$110&17.83$\sim$19.61&63.57&Constrained&Constrained&3805&-3.6&Yes\\
110$\sim$120&19.61$\sim$21.39&55.26&Constrained&Unconstrained&1908&1909&No\\
\hline\noalign{\smallskip}
\multicolumn{8}{c}{Iteration II}\tabularnewline
\hline\noalign{\smallskip}
80$\sim$85&14.26$\sim$15.15&19.23&Constrained&Unconstrained&2967&-53&No\\
85$\sim$90&15.15$\sim$16.04&41.29&Constrained&Constrained&2807&-227&Yes(weak)\\
100$\sim$105&17.83$\sim$18.72&43.03&Constrained&Unconstrained&3007&-56&No\\
105$\sim$110&18.72$\sim$19.61&49.37&Constrained&Constrained&3107&-9&Yes\\
110$\sim$115&19.61$\sim$20.50&49.42&Constrained&Unconstrained&2375&-752&No\\
\hline\noalign{\smallskip}
\multicolumn{8}{c}{Iteration III}\tabularnewline
\hline\noalign{\smallskip}
85$\sim$87.5&15.15$\sim$15.60&19.22&Constrained&Unconstrained&937&-1328&No\\
87.5$\sim$90&15.60$\sim$16.04&39.5&Constrained&Unconstrained&1838&-510&No\\
105$\sim$107.5&18.72$\sim$19.16&35.00&Constrained&Unconstrained&2348&-57&No\\
107.5$\sim$110&19.16$\sim$19.61&35.88&Constrained&Unconstrained&2389&-28&No\\
\hline\noalign{\smallskip}
\multicolumn{8}{c}{Iteration IV}\tabularnewline
\hline\noalign{\smallskip}
105$\sim$106.25&18.72$\sim$18.94&24.02&Constrained&Unconstrained&1557&-68&No\\
106.25$\sim$107.5&18.94$\sim$19.16&25.89&Constrained&Unconstrained&...&-49&No\\
107.5$\sim$108.75&19.16$\sim$19.39&24.73&Constrained&Unconstrained&-798&-160&No\\
108.75$\sim$110&19.39$\sim$19.61&26.42&Constrained&Unconstrained&1645&-29&No\\
\enddata
\end{deluxetable}

\startlongtable
\begin{deluxetable}{ccccccccccccc}
\tabletypesize{\scriptsize}
\tablecaption{Spectral fit results of GRB 220101A with the Band model.}
\tablehead{
\colhead{GRB}
&\colhead{$t_{1}$$\sim$$t_{2}$}
&\colhead{$S$}
&\colhead{$K$}
&\colhead{$\alpha$}
&\colhead{$E_{\rm p}$}
&\colhead{$\beta$}
&\colhead{$F$}
&\colhead{$S_\gamma$}
&\colhead{$k_c$}
&\colhead{$E_{\rm p,z}$}
&\colhead{$L_{\gamma,\rm iso}$}
&\colhead{$E_{\gamma,\rm iso}$}\\
\hline
&(s)
&
&(10$^{-2}$)
&
&(keV)
&
&(10$^{-6}$)
&(10$^{-6}$)
&(keV)
&
&(10$^{52}$)
}
\colnumbers
\startdata
\hline
220101215&64$\sim$80&27.0&0.84$^{+0.12}_{-0.12}$&-1.04$^{+0.09}_{-0.09}$&437$^{+139}_{-152}$&-3.98$^{+2.12}_{-3.37}$&0.60$^{+0.20}_{-0.16}$&9.6$^{+3.1}_{-2.6}$&0.73&2456$^{+779}_{-853}$&9.9$^{+3.2}_{-2.7}$&157.8$^{+51.9}_{-42.5}$\\
220101215&80$\sim$96&64.5&3.57$^{+0.22}_{-0.23}$&-0.61$^{+0.04}_{-0.04}$&223$^{+12}_{-12}$&-3.33$^{+0.84}_{-0.38}$&1.12$^{+0.14}_{-0.14}$&18.0$^{+2.3}_{-2.2}$&0.91&1251$^{+68}_{-67}$&23.0$^{+2.9}_{-2.9}$&368.5$^{+47.0}_{-45.9}$\\
220101215&96$\sim$112&78.0&3.26$^{+0.15}_{-0.15}$&-0.70$^{+0.03}_{-0.03}$&334$^{+21}_{-21}$&-2.08$^{+0.08}_{-0.07}$&2.05$^{+0.18}_{-0.16}$&32.7$^{+2.9}_{-2.6}$&0.55&1878$^{+117}_{-117}$&25.6$^{+2.3}_{-2.0}$&410.1$^{+36.7}_{-32.7}$\\
220101215&112$\sim$128&45.1&1.73$^{+0.11}_{-0.11}$&-0.95$^{+0.05}_{-0.04}$&269$^{+21}_{-21}$&-6.11$^{+2.61}_{-2.61}$&0.73$^{+0.09}_{-0.07}$&11.7$^{+1.4}_{-1.1}$&1.01&1514$^{+118}_{-118}$&16.6$^{+1.9}_{-1.6}$&266.2$^{+31.2}_{-25.0}$\\
220101215&128$\sim$144&13.6&0.49$^{+0.12}_{-0.13}$&-1.21$^{+0.15}_{-0.16}$&221$^{+67}_{-74}$&-6.05$^{+2.67}_{-2.67}$&0.19$^{+0.11}_{-0.06}$&3.0$^{+1.7}_{-1.0}$&1.05&1242$^{+374}_{-418}$&4.5$^{+2.5}_{-1.5}$&71.5$^{+40.6}_{-23.4}$\\
\hline
\enddata 
\vspace{3mm}
\tablecomments{Col.(6) lists erg cm$^{-2}$ s$^{-1}$, Col.(10) lists fluence, erg cm$^{-2}$, Col.(12) lists in units of erg s$^{-1}$, Col.(9) lists in units of erg.}
\end{deluxetable}\label{tab:Band}

\appendix

\section{Pair creation region around the accreting NS}\label{app:A}

We define here the region where pair creation occurs around the accreting NS, the \textit{dyadoregion}. We follow the treatment in \cite{2026JHEAp..5000464R} hereafter. We adopt the solution of the Maxwell equations for a perfectly conducting sphere of radius $R$, endowed with a magnetic dipole of strength $B_p$ at the pole, rotating with angular velocity $\Omega$, and a uniformly distributed co-rotating charge that minimizes the electromagnetic energy of the configuration, obtained by \citet{1973ApL....13..109R}. The components of the electric and magnetic field, up to first order in the dimensionless rotation parameter $\tilde \Omega \equiv \Omega R/c$, are given by 
\begin{subequations}\label{eq:EB}
    \begin{align}
    &(E_r,E_\theta,E_\phi) = \begin{cases}
        \displaystyle\left(-r \sin^2\theta \frac{\Omega}{c},-r \sin\theta\cos\theta \frac{\Omega}{c}, 0\right) B_p,& r<R\\
        \\
        \displaystyle\left(-\frac{1}{3} -P_2(\cos\theta)\frac{R^2}{r^2}, -2\sin\theta\cos\theta \frac{R^2}{r^2} ,0\right)\frac{\tilde\Omega B_p R^2}{r^2},& r\geq R
    \end{cases}\label{eq:E}\\
    &(B_r,B_\theta,B_\phi) = \begin{cases}
        \displaystyle\left(\cos\theta,-\sin\theta, 0\right) B_p,& r<R,\\
        \\
        \displaystyle\left(2\cos\theta,\sin\theta, 0\right) \frac{B_p}{2}\frac{R^3}{r^3},& r\geq R.
    \end{cases}\label{eq:B}
\end{align}
\end{subequations}
where $P_2 (\cos\theta) = (3\cos^2\theta-1)/2$ is the Legendre polynomial.

The Schwinger pair production rate ($e^+e^-$ pairs per unit volume, per unit time) is
\begin{equation}\label{eq:rate2}
	\dot{n} =\frac{e^2 \tilde E \tilde B}{4 \pi^2 \hbar^2 c}\sum_{n=1}^\infty\frac1{n}\coth\left(n\pi \frac{\tilde B}{\tilde E}\right)e^{-\frac{n\pi E_c}{\tilde E}},
\end{equation}
where ${\bf \tilde E}$ and ${\bf \tilde B}$ are the moduli of the electric and magnetic field in the frame where they are parallel, which can be obtained from the Maxwell invariants: $\tilde E=[({\mathcal F}^2+{\mathcal G}^2)^{1/2}-{\mathcal F}]^{1/2}$ and $ \tilde B=[({\mathcal F}^2+{\mathcal G}^2)^{1/2}+{\mathcal F}]^{1/2}$, being ${\cal F}\equiv\frac14F_{\mu\nu}F^{\mu\nu}$, $
{\cal G}\equiv\frac14F_{\mu\nu}{}^*F^{\mu\nu}$, with $F_{\mu \nu}$ and $^*F_{\mu \nu}$ the electromagnetic field tensor and its dual. From the exponential cutoff of the pair creation rate (\ref{eq:rate2}), it follows the natural definition of dyadoregion
\begin{equation}\label{eq:dyadoregion}
    \tilde{E}(r,\theta) = E_c,
\end{equation}
which determines the dyadoregion equation, $r = r_d(\theta)$. Using $\tilde E\approx E |\cos\theta|$, one obtains the solution of Eq. (\ref{eq:dyadoregion})
\begin{equation}\label{eq:rdya}
  r_d(\theta) = \sqrt{\frac{{\cal R} + {\cal D}}{2}},
\end{equation}
where ${\cal R} = \sqrt{z + 2a_2/3}$, ${\cal D} = \sqrt{-{\cal R}^2 + 2 a_2 + 2 a_1/{\cal R}}$, $z = -2 \sqrt{p/3} \sinh{\left[\frac{1}{3}\text{arcsinh}\left(\frac{3 q}{2 p} \sqrt{3/p}  \right)\right]}$, being $p = (3 c - b^2)/3$, $q = (2 b^3-9 b c+27 d)/27$, $b = a_2$, $c=4 a_0$, $d = 4 a_0 a_2 - a_1^2$, with $a_0 = \frac{1}{4}(\beta \tilde\Omega)^2 [(3 \cos^2\theta-1)^2 + \sin^2 2\theta] \cos^2\theta$, $a_1 =\frac{1}{3}(\beta \tilde\Omega)^2 (3 \cos^2\theta-1)^2 \cos^2\theta$, $a_2 = \frac{1}{9}(\beta \tilde\Omega)^2 \cos^2\theta$, and we introduced the dimensionless magnetic field parameter $\beta\equiv B_p/B_c$. The dyadoregion width is $\Delta_d \equiv r_d(\theta)-R$. The dyadoregion extension is largest along the polar axis
\begin{equation}\label{eq:rdpole}
    r_d(\theta = 0) = R \sqrt{\frac{\beta \tilde\Omega}{6}} \left( 1 + \sqrt{1+\frac{36}{\beta \tilde\Omega}} \right).
\end{equation}
Demanding that $r_d(\theta = 0) = R$, so a non-zero width $\Delta_d > 0$, we obtain the minimum magnetic field strength to produce pairs in the exterior of the rotating dipole
\begin{equation}\label{eq:Bmin}
    \beta_{\rm min} = \frac{B_{p,\rm min}}{B_c} = \frac{3}{4 \tilde\Omega}.
\end{equation}
The energy available for the pair plasma is given by the electromagnetic energy in the dyadoregion volume
\begin{equation}\label{eq:Epairs}
    {\cal E}_{e^+e^-} = \frac{1}{4} \int_0^\pi\int_R^{r_d(\theta)} (E^2 + B^2) r^2 \sin\theta dr d\theta.
\end{equation}
%

\section{Sample of BdHN I with UPE Episodes}\label{app:B}

One of the main new results has been the identification of the UPE~I and UPE~II Episodes and their energy source. In Tables~\ref{tab:episodes190114C}, \ref{tab:episodes180720B}, and \ref{tab:episodes240825A}, we show the Episodes of additional examples of GRBs from a BdHN I sample under investigation.

\begin{table*}
\centering
\caption{The Episodes and afterglows of GRB 190114C. This table reports the event name, duration (start -- end in seconds), best-fit spectrum, isotropic energy ($E_{\rm iso}$ in erg), and underlying astrophysical processes for GRB 190114C. The redshift is $z=0.424$, and the total isotropic energy is $E_{\rm iso}=(2.48\pm0.22)\times 10^{53}$ erg. Unless specified, energy values are isotropic.}
\label{tab:episodes190114C}
\small
\setlength{\tabcolsep}{4pt}
\begin{tabular}{c l c c}
\toprule
Event       & Duration (s)      & Spectrum & $E_{\rm iso}$ (erg)  \\
\midrule
    SN-rise & 0.79 (0.00--0.79) & CPL    & $(3.52 \pm 0.15) \times 10^{52}$ \\
    UPE I   & 0.39 (0.79--1.18) & CPL+BB & $(1.00 \pm 0.11) \times 10^{53}$ \\
    BH-rise & 0.72 (1.18--1.9)  & CPL    & $(3.75 \pm 0.11) \times 10^{52}$ \\
    UPE II  & 2.09 (1.9--3.99)  & CPL+BB & $(1.47 \pm 0.20) \times 10^{53}$ \\
\bottomrule
\end{tabular}
\end{table*}

\begin{table*}
\centering
\caption{The Episodes and afterglows of GRB 180720B. This table reports the event name, duration (start -- end in seconds), best-fit spectrum, isotropic energy ($E_{\rm iso}$ in erg), and underlying astrophysical processes for GRB 180720B. The redshift is $z=0.654$, and the total isotropic energy is $E_{\rm iso}=5.92\times 10^{53}$ erg. Unless specified, energy values are isotropic.}
\label{tab:episodes180720B}
\small
\setlength{\tabcolsep}{4pt}
\begin{tabular}{c l c c}
\toprule
Event & Duration (s)& Spectrum & $E_{\rm iso}$ (erg) \\
\midrule
    SN-rise & 4.84 (0.00--4.84)   & Band   & $(1.53 \pm 0.09) \times 10^{53}$ \\
    UPE I   & 1.21 (4.84--6.05)   & CPL+BB & $(6.37 \pm 0.48) \times 10^{52}$ \\
    BH-rise & 3.02 (6.05--9.07)   & CPL    & $(1.13 \pm 0.04) \times 10^{53}$ \\
    UPE II  & 1.82 (9.07--10.89)  & CPL+BB & $(1.60 \pm 0.95) \times 10^{53}$ \\
\bottomrule
\end{tabular}
\end{table*}

\begin{table}
\centering
\caption{The Episodes and afterglows of GRB 240825A. This table reports the event name, duration (start -- end in seconds), best-fit spectrum, isotropic energy ($E_{\rm iso}$ in erg), and underlying astrophysical processes for GRB 240825A. The redshift is $z=0.658$.}
\label{tab:episodes240825A}
\begin{tabular}{cccc}
\toprule
  Event & Duration (s)  & Spectrum & $E_{\rm iso}$ (erg) \\
\midrule
  SN-rise & -0.3--0.5 & CPL     & $(5.09\pm0.87) \times 10^{50} $   \\
  UPE-1   & 0.7--1.1  & Band+BB & $(7.09\pm1.24) \times 10^{52}$    \\
          &            & BB      & $(2.05 \pm 1.30) \times 10^{51}$ \\
  UPE-2   & 1.1--2.1  & Band+BB & $(8.93\pm0.73) \times 10^{52}$    \\
          &            & BB      & $(8.11\pm3.38) \times 10^{50}$   \\
\bottomrule
\end{tabular}
\end{table}

\end{document}